\documentclass[12pt]{article}
\usepackage{mathtext}
\usepackage[english]{babel}

\usepackage{graphicx,color}
\usepackage{amscd}
\usepackage{amsmath}

\newcommand{\cM}{\cal M}
\usepackage[pdftex]{hyperref}

\usepackage{amsmath,amssymb,amsthm,latexsym}
\usepackage{stmaryrd,wasysym,upgreek,mathrsfs,dsfont}

\newtheorem{lemma}{Lemma}[section]
\newtheorem{remark}{Remark}[section]
\newtheorem{example}{Example}[section]

\newtheorem{definition}{Definition}[section]
\newtheorem{theorem}{Theorem}[section]
\newtheorem{corollary}{Corollary}[section]
\newcommand{\be}{\begin{equation}}
\newcommand{\ee}{\end{equation}}
\newcommand{\bea}{\begin{eqnarray}}
\newcommand{\eea}{\end{eqnarray}}

\newcommand{\om}{\Omega}

\newcommand{\prf}{\noindent{\bf Proof}\ }

\newcommand{\al}{\alpha}

\newcommand{\bt}{\beta}

\newcommand{\nn}{\nonumber}

\newcommand{\tr}{\mbox{tr}}

\newcommand{\og}{{\omega}}
\newcommand{\Om}{{\Omega}}

\newcommand{\Tr}{{\rm Tr}}
\newcommand{\oj}{{\omega(j)}}

\newcommand{\bQ}{{\bf Q}}
\newcommand{\bV}{{\bf V}}
\newcommand{\bX}{{\bf X}}
\newcommand{\bA}{{\bf A}}

\newcommand{\bP}{{\bf P}}

\newcommand{\mR}{\mathbb {R}}

\newcommand{\cT}{{\cal{T}}}
\newcommand{\cO}{{\cal{O}}}
\newcommand{\cV}{{\cal{V}}}
\newcommand{\cC}{{\cal{C}}}
\newcommand{\cA}{{\cal{A}}}
\newcommand{\cB}{{\cal{B}}}
\newcommand{\cH}{{\cal{H}}}

\newcommand{\cI}{{\cal{I}}}

\newcommand{\cR}{{\cal{R}}}
\newcommand{\cG}{{\cal{G}}}
\newcommand{\cW}{{\cal{W}}}
\newcommand{\cP}{{\cal{P}}}
\newcommand{\cF}{{\cal{F}}}
\newcommand{\cS}{{\cal{S}}}
 \newcommand{\cJ}{{\cal{J}}}
\newcommand{\bbone}{{\bf 1}}

\begin{document}

\title{Constructive Renormalization of the
 $2$-dimensional Grosse-Wulkenhaar Model}


\author{\Large{Zhituo Wang} \footnote{Institute for Advanced Study in Mathematics Science,  Harbin Institute of Technology, Xidazhi Street 92, Harbin, China, 150006, Email: wzht@hit.edu.cn
\ }
\footnote{Research Center for Operator Algebras, East China Normal University, 3663 ZhongShan North Road, Shanghai, China. 200062} }

\maketitle

%


%

\begin{abstract}
We study a quartic matrix model with partition function $Z=\int d\ M\exp\Tr\ (-\Delta M^2-\frac{\lambda}{4}M^4)$.
The integral is over the space of Hermitian $(\Lambda+1)\times(\Lambda+1)$ matrices, the matrix $\Delta$, which is not a multiple of the identity matrix, encodes the dynamics and $\lambda>0$ is a scalar coupling constant. We proved that the logarithm of the partition function is the Borel sum of the perturbation series, hence is a well defined analytic function of the coupling constant in certain analytic domain of $\lambda$, by using the multi-scale loop vertex expansions. All the non-planar graphs generated in the perturbation expansions have been taken care of on the same footing as the planar ones.

This model is derived from the self-dual $\phi^4$ theory on the 2 dimensional Moyal space, also called the 2 dimensional Grosse-Wulkenhaar model. This would also be the first fully constructed matrix model which is non-trivial and not solvable.

\end{abstract}

\medskip


\medskip

\section{Introduction}

Quantum field theories on noncommutative space time became popular
after the discovery that they may arise as effective regimes of string theory either due
to the compactification \cite{Connes} or due to the presence of the
Green-Schwarz $B$ field for open strings \cite{Sho, SW}. The simplest
non-commutative space is the Moyal space, which could
be considered also as a low energy limit of open string theory.
However the usual quantum field theories on the Moyal space, defined by
simply replacing the commutative algebra of scalar fields by the 
noncommutative Moyal algebra, are non-renormalizable due to a phenomenon called the $UV/IR$ mixing, namely the amplitudes for the non planar graphs are infrared-divergent and non-renormalizable after integrating out the ultraviolet degree of freedoms \cite{Minwala}.

Several years ago H. Grosse and R. Wulkenhaar made a breakthrough by introducing
a harmonic oscillator term to the ill-defined $\phi^{\star 4}_4$ Lagrangian such that the 
resulting Lagrangian obeys a symmetry called the Langmann-Szabo duality \cite{LaSz}. 
They proved  in a series of papers  \cite{GWp,GW0,GW1} (see also \cite{RVW}) that this new model, also called the Grosse-Wulkenhaar model or $GW_4$ for short, is perturbatively renormalizable to all orders. Based on this idea many other noncommutative QFT models with Langman-Szabo symmetry \cite{LsSZ2,Fabien1,GS1,Wang1, GF} have been shown to be {\emph{perturbatively}} renormalizable to all orders. More details could be found in
\cite{Bourbaphy}.

The Grosse-Wulkenhaar model is not only perturbatively renormalizable but also
asymptotically safe \cite{beta1}, \cite{MagRiv}, \cite{DGMR}, namely, the
renormalization flow of the coupling constant is bounded. So it is even better behaved than many of the commutative quantum field theory models, for example, there are the Landau ghost problems in the 4-dimensional $\phi^4$ theory and $QED$, and infrared confinement problem in non Abelian gauge theory. Based on this result Grosse and Wulkenhaar proved that this model
can be exactly solved \cite{gws1,gws2} in the limit $\theta\rightarrow\infty$, where $\theta$ characterizes the volume of the noncommutative Moyal space, by using methods from the integrable model and the fixed point theorem. 

But this is not the whole story. The Feynman graphs generated in the perturbation expansions are ribbon graphs \cite{hooft, tutt} which include both planar graphs \cite{hooft} and the nonplanar ones, and only planar graphs are non-vanishing after taking the limit $\theta\rightarrow\infty$. In other words, the method is not suitable for the case when $\theta$ is finite, for which all the non-planar graphs survive. Different method should be considered.  

Constructive renormalization theory \cite{Riv, GJ}
builds the exact Green's functions whose Taylor
expansions correspond to perturbative quantum field theory.
The traditional techniques for Bosonic constructive
theories are the cluster and Mayer expansions \cite{GJS, BK}, which requires division of the Euclidean space into cubes and test the localization of the interaction vertices. However they are not suitable for noncommutative quantum field theories due to the non-commutativity of the space coordinates and 
the non-locality of the interaction vertex.

About one decade ago a new method in constructive renormalization theory, called the Loop Vertex Expansions \cite{RivLve, RivMag, RivW1, RivW2, Rivhigh}
(LVE for short), has been invented exactly for overcoming these difficulties. The LVE is a combination of the Hubbard-Stratonovich intermediate field technique with the BKAR forest formula \cite{BK, AR1} and has been proved to be very useful for constructing models that doesn't require renormalizations. In order to implement the multi-scale analysis one has to generalize this method to the Multi-scale Loop Vertex Expansions \cite{MLVE} (MLVE for short), which consist of three major steps: the intermediate field representation of the partition function, which is the same as LVE, the slice-testing expansions, which play the role of renormalizations, and the two-level forest expansions, which rewrite the partition function and the density functions as convergent perturbation series. This step plays the role of cluster and Mayer expansions. The MLVE has been applied successfully to the construction of the commutative $\phi^4_2$ model \cite{RivW3} as well as tensor field theory models \cite{ Delepouve:2014bma, DR, fabien}.

In this paper we shall construct the Grosse-Wulkenhaar model on the $2$ dimensional
Moyal plane ($GW_2$ for short) of finite volume (namely $\theta<\infty$), with the method of MLVE, as a first step towards building constructively the four dimensional Grosse-Wulkenhaar model. One can choose suitable basis on Moyal space under which the scalar field becomes Hermitian matrix, and the $GW_2$ model becomes an Hermitian matrix model with non-trivial covariance. Recall that the Hermitian matrix model is a probability measure on the space of Hermitian matrices $M$ of the form
$$\frac{1}{Z_\Lambda}e^{\Tr\ [-\Delta M^2-V(\lambda, M)]}\ dM,$$
where $Z_\Lambda:=\int\exp{\Tr[-\Delta M^2-V(\lambda, M)]}\ dM$ is the partition function,
 $V(\lambda, M)$ is a polynomial function of the Hermitian matrices $M$ and $\lambda$ is the coupling constant. The matrix elements are not identically distributed if the Laplacian $\Delta$, which is now a matrix, is not proportional to the identity matrix.

\begin{figure}[!t]
\begin{center}
{\includegraphics[width=0.2\textwidth]{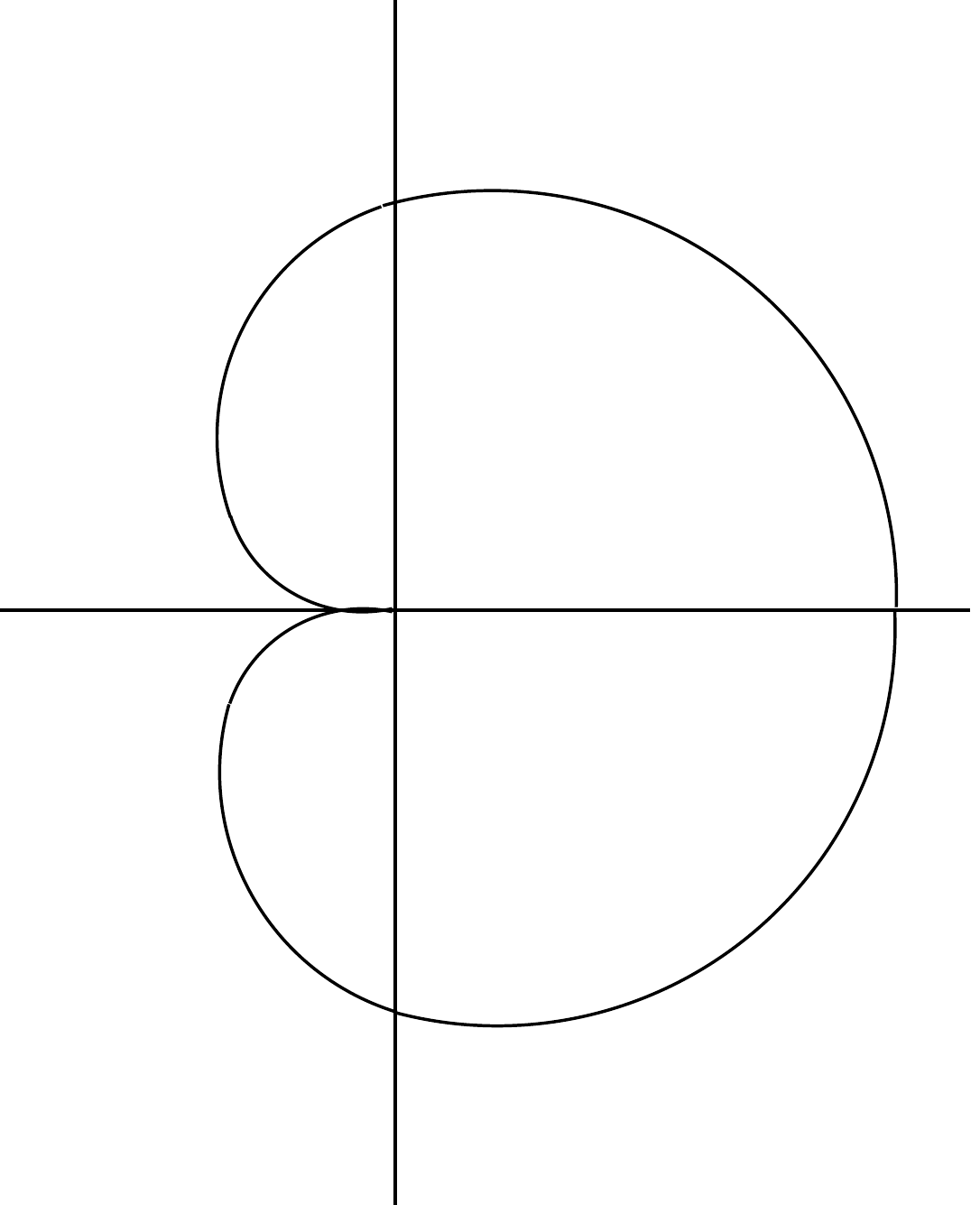}}
\end{center}
\caption{A Cardioid Domain}
\label{cardio}
\end{figure}

We will study the asymptotic behavior of the partition function when the size of the matrix $\Lambda$ tends to infinity, with the method of MLVE. The main result of this paper is (see also Theorem \ref{thetheorem}):
\begin{theorem}[{\bf The Main Theorem}]\label{main1}
Let $V_\theta$ be the volume of the Moyal space and $\rho>0$ a fixed constant that is small enough. The vacuum correlation function $\lim_{\Lambda\rightarrow\infty}\frac{1}{V_\theta}\log Z_\Lambda(\lambda)$ for the 2 dimensional Grosse-Wulkenhaar model exists and is an analytic function of $\lambda$
in the Cardioid domain $\lambda\in{\cal{C}}ard_\rho:=\{\lambda\in\mathds{C}\ \vert\ |{\arg}\ \lambda|<\pi, |\lambda|<\rho\cos^2(\frac{1}{2}{\arg}\ \lambda ) \}$ (see Figure \ref{cardio} for an illustration). Furthermore it is the Borel sum of its perturbation series in $\lambda$.
\end{theorem} 

The plan of this paper is as follows: in Section 2 we provide the mathematical definition of this model and state the main theorem. In Section 3 we introduce the intermediate field representation of this model, at the heart of MLVE, which rewrites the partition function as a functional integral over a Hermitian intermediate field. This is the first key step for performing the MLVE. In Section 4 we introduce the slice-testing expansions (STE), which is second step of the MLVE. As briefly introduced before, the STE adapted to the current model is a conditional expansion playing two roles: on the one hand it is a perturbative renormalization procedure, in which the tadpole graphs are compensated in a multi-scale way; on the other hand it generates enough convergent power-counting factors from the marked propagators (see Definition \ref{markedp} in that section), so as to compensate the non-perturbative bound. In Section 5 we briefly recall the combinatorial definition of forests and the forest formulas of different level as well as perform the two level forest formula, one for the Bosonic intermediate fields and the other for the Fermionic fields, so as to write the partition function and the vacuum correlation function as $convergent$ perturbation series. One single forest formula would not be enough, as remarked in \cite{RivW3}, since otherwise one would generate unbounded factors from the contraction of the intermediate fields. The Fermionic variables play the role of the hard-core constraints as in the cluster expansion and the second forest formula play the role of a Mayer expansion.
Section 6 is devoted to the proof of both the perturbative  and the non-perturbative bounds of the perturbation series and finally the Borel summability \cite{Sok, EMS} of the perturbation series. In the Appendix we perform the second order slice-testing expansion.



\section{Moyal space and the Grosse-Wulkenhaar Model}
\subsection{Basic properties of the Moyal space}
The $D$-dimensional Moyal space $\mathbb {R}^{D}_{\Theta}$ for $D$ even is generated by the
non-commutative coordinates $x^{\mu}$ that obey
the commutation relation $[ x^{\mu},x^{\nu}]=i\Theta^{\mu\nu}$, where
$\Theta$ is a $D\times D$ non-degenerate skew-symmetric
matrix such that $\Theta^{\mu\nu}=-\Theta^{\nu\mu}$.
It is the simplest and best studied model of non-commutative space (see
\cite{GPW, Bourbaphy} for more details).

The Moyal algebra of smooth and rapidly decreasing functions on $\mathbb {R}^{D}_{\Theta}$ is
the Schwartz space of functions equipped with the non-commutative 
\emph{Groenewold-Moyal} product defined by:
\begin{eqnarray}
  &&(f\star_{\Theta} g)(x)=\int_{\mathbb{R}^D} \frac{d^{D}k}{(2\pi)^{D}}d^{D}y\, f(x+{\textstyle\frac 12}\Theta\cdot
  k)g(x+y)e^{i k\cdot y}\\
  &=&\frac{1}{\pi^{D}|\det\Theta|}\int_{\mathbb{R}^D\times \mathbb{R}^D} d^{D}yd^{D}z\,f(x+y)
  g(x+z)e^{-2i y\Theta^{-1}z}\; ,\ \ \forall f,g\in\cS_{D}:=\cS(\mathbb{R}^{D}).\nonumber
\end{eqnarray}

We shall consider the case of $D=2$ in the rest of this paper, for which the symplectic matrix $\Theta=\begin{pmatrix}0&\theta\\-\theta&0\end{pmatrix}$ and $\theta\in\mathds{R}^+$. For better understanding the Moyal algebra it is convenient to define the creation and annihilation operators by:
\cite{GW0,GW1,GPW}:
\begin{align}
  a &= \frac{1}{\sqrt{2}}(x_1+i x_2)\;, & \bar{a} &=
  \frac{1}{\sqrt{2}}(x_1-{i} x_2)\;, \nonumber
  \\*
  \frac{\partial}{\partial a} &= \frac{1}{\sqrt{2}}(\partial_1 -
  {i} \partial_2 )\;, & \frac{\partial}{\partial \bar{a}} &=
  \frac{1}{\sqrt{2}}(\partial_1 + {i} \partial_2 )\;,
\end{align}
where $x_1, x_2$ be the coordinates of the Moyal plane $\mR^2_\Theta$.
Then for any function $f\in\cS_{D}$ we have:
\begin{align}
  (a \star f)(x) &= a(x) f(x) + \frac{\theta}{2} \frac{\partial
    f}{\partial \bar{a}}(x)\;, & (f \star a)(x) &= a(x) f(x) -
  \frac{\theta}{2} \frac{\partial f}{\partial \bar{a}}(x)\;, \nonumber
  \\*
  (\bar{a} \star f)(x) &= \bar{a}(x) f(x) - \frac{\theta}{2}
  \frac{\partial f}{\partial a}(x)\;, & (f \star \bar{a})(x) &=
  \bar{a}(x) f(x) + \frac{\theta}{2} \frac{\partial f}{\partial
    a}(x)\;.
\end{align}

With the creation and annihilation operators we can define the matrix basis $f_{mn}(x)$:
\begin{eqnarray}\label{matrixbasis}
  f_{mn}(x) &:=&\frac{1}{\sqrt{n! m! \,\theta^{m+n}}} \, \bar{a}^{\star m}
  \star f_0 \star a^{\star n}(x)\\
& = &\frac{1}{\sqrt{n! m! \,\theta^{m+n}}} \sum_{k=0}^{\min(m,n)}
(-1)^k \binom{m}{k} \binom{n}{k} \,k! \,2^{m+n-2k}\, \theta^k
\,\bar{a}^{m-k} \,a^{n-k} f_0(x)\;, \nonumber
\end{eqnarray}
with $f_0(x)= 2 \mathrm{e}^{-\frac{1}{\theta}(x_1^2+x_2^2)}$. 

The matrix basis is orthogonal and complete in the sense that 
\begin{equation}
  (f_{mn} \star f_{kl})(x) = \delta_{nk} f_{ml}(x),\quad \quad \int_{\mathbb{R}^2} d^2x f_{mn}(x)=2\pi\theta\delta_{mn},
\label{fprod}
\end{equation}
and that for any square integrable function
 $\phi(x)$ defined on $\mathbb {R}^{D}_{\theta}$ there exists a sequence $\phi_{mn}\in\mathds{C},\forall\ m,n\in\mathds{N}$ such that:
\begin{equation}
\phi(x)=\sum_{m,n\in \mathds{N}}\phi_{mn}f_{mn}(x).
\end{equation}
Remark that matrices $\phi_{mn}$ carries the topology of the function space of $\phi$ \cite{gayral}, due to nice properties of the Moyal plane.
\subsection{The 2-dimensional Grosse-Wulkenhaar Model}
The 2-dimensional Grosse-Wulkenhaar model ($GW_2$ for short) is
defined by the action:
\begin{eqnarray}\label{mainaction}
S&=&\int_{\mathbb{R}^2} d^2x \big [\frac{1}{2}\sum_{\mu={1,2}}\partial_\mu\phi\star
\partial^\mu\phi+\frac{\Omega^2}{2}\sum_{\mu={1,2}}(\tilde x_\mu\phi)
\star(\tilde x^\mu\phi)+\frac{\kappa^2}{2}\phi\star\phi\nonumber\\
&+&\frac{\lambda}{4}:\phi\star\phi\star\phi\star\phi
:\big],
\end{eqnarray}
where $\phi:\mathds{R}^2\rightarrow\mathds{R}$ is a real scalar field, $\tilde
x_\mu=2(\Theta^{-1})_{\mu\nu}x^\nu$; $\kappa\in \mathds{R}^+\cup {0}$, $\lambda\in\mathds{R}^+$ are the mass and coupling constant, respectively, and the Euclidean signature has
been used. Wick ordering is taken to the interaction so that the
tadpoles are renormalized. The explicit form of the Wick ordered interaction is given by \eqref{Wick1}. 

The main feature of this model is the introduction of the harmonic potential term, namely the second term in (\ref{mainaction}). This term relates the infrared dynamics (large $\tilde x$) with the ultraviolet behavior (large derivations in the Laplacian) and is essential for curing the UV/IR mixing problem and proving the renormalizability of this model \cite{GW1, RVW}. Another interesting feature of the action is a symmetric property called the Langmann-Szabo duality \cite{LaSz}.

The scalar fields become Hermitian matrices $[\phi_{mn}]$, $m, n\in \mathds{Z}$, under the matrix basis \eqref{matrixbasis} and integration over $\mathds{R}^2$ is replaced by summing over the matrix indices. The action becomes:
\begin{eqnarray}
S[\phi] = 2\pi\theta \sum_{m,n,k,l\in \mathds{N}} \Big[ \frac{1}{2}
\phi_{mn} \Delta_{mn;kl} \phi_{kl} + \frac{\lambda}{4}
:\phi_{mn} \phi_{nk} \phi_{kl} \phi_{lm}:\Big],\label{action1}
\end{eqnarray}
where $2\pi\theta$ is the volume of the Moyal space, $\Delta$ is the Laplacian (remark that the mass term is also included) with elements $\Delta_{mn,kl}$:
\begin{eqnarray}
&&\Delta_{mn,kl}=[\kappa^2+\frac{2(1+\Omega^2)}{\theta}(m+n+1)]\delta_{mk}\delta_{nl}-\frac{2}{\theta}(1-\Omega^2)\nonumber\\
&\times&\big[\sqrt{(m+1)(n+1)}\delta_{m+1,k}\delta_{n+1,l}+\sqrt{mn}\delta_{m-1,k}\delta_{n-1,l} \big].
\end{eqnarray}

Introducing an ultraviolet cutoff $\Lambda\in \mathds{N}^+$, then the Laplacian becomes an $\Lambda^2\times\Lambda^2$ dimensional matrix. The covariance $C_{sr,kl}$ is defined by the following equation: \cite{GW0}:
\begin{equation}
\sum_{r,s=0}^{ \Lambda}\Delta_{mn,rs} C_{sr,kl}=\delta_{ml}\delta_{nk}.
\end{equation}

Notice that the Laplacian becomes diagonal at $\Omega=1$:
\begin{equation}
\Delta_{mn,kl} =2\pi\theta\Big[\kappa^2+ \frac{4}{\theta}(m{+}n{+}1)
\Big]\delta_{ml}\delta_{nk} ,\label{cova}
\end{equation}
so does the covariance matrix:
\begin{equation} \label{covmatrix}
C_{mn,kl}=\frac{1}{2\pi\theta}\frac{1}{\kappa^2+ \frac{4}{\theta}(m{+}n{+}1)}\delta_{ml}\delta_{nk}:=\frac{1}{2\pi\theta}C_{mn}\delta_{ml}\delta_{nk}.
\end{equation}
Setting $\theta=4$ and forgetting inessential factor
$\frac{1}{8\pi}$ for simplicity, we have $C_{mn}=\frac{1}{m+n+1+\kappa^2}$, where the term $1$ in the denominator plays the role of infrared regulator for $m=0=n$. In order to further simplify the calculation we set $\kappa^2=0$. Remark that this setting doesn't mean that the theory is massless; it is massive and the mass term does receive renormalization. Setting $\kappa^2=0$ makes the propagator in the simplest form. We forget also the volume factor $2\pi\theta=8\pi$ and the density function becomes $\log Z(\lambda)$. 

Remark that $\Omega=1$ is a fixed point of the 4 dimensional Grosse-Wulkenhaar model \cite{MagRiv}, at which the $\beta$ function is vanishing for all orders \cite{beta1, MagRiv} and the flow of the coupling constant is bounded \cite{DGMR}. So it is important to take $\Omega=1$ for the $GW_4$ model.
But is is not the case for $GW_2$ model since it is super-renormalizable. We set $\Omega=1$ in the present paper only for simplicity.

The Wick-ordered interaction with ultraviolet cutoff $\Lambda$ reads:
\begin{equation}\label{Wick1}
\sum_{mnkl}:\phi_{mn} \phi_{nk} \phi_{kl} \phi_{lm}:=\sum_{mnkl}\phi_{mn} \phi_{nk} \phi_{kl} \phi_{lm}-\sum_{mp}4\phi_{mp}\phi_{pm}T^\Lambda_m+2\sum_m (T^\Lambda_m)^2.
\end{equation}
where
\begin{equation}\label{fish}
T^\Lambda_m=\sum_q C_{mq,qm}=\sum_{q=0}^{\Lambda}\frac{1}{q+m+1}\sim\log\frac{\Lambda+m}{m+1}\sim\log\Lambda,\  {\rm for}\ 0\leqslant m\ll\Lambda,
\end{equation}
is the counter-term for the tadpole.
Let $\Pi^\Lambda:=\sum_{m=0}^{\Lambda}(T^\Lambda_m)^2$, we have
\begin{equation}\label{glasses}
\Pi^\Lambda=\sum_{m=0}^{\Lambda}(\sum_{p=0}^{\Lambda}\frac{1}{m+p+1})(\sum_{q=0}^{\Lambda}\frac{1}{m+q+1})\sim\sum_m\log^2\frac{\Lambda+m}{m+1}\sim c_1\Lambda+c_2\ln^2\Lambda+c_3\ln\Lambda,
\end{equation}
where $c_1, c_2, c_3$ are positive constants such that $1<c_1< 3$. $\Pi^\Lambda$ is also called the $vacuum$ $tadpole$.
\begin{figure}[!htb]
\centering
\includegraphics[scale=0.3]{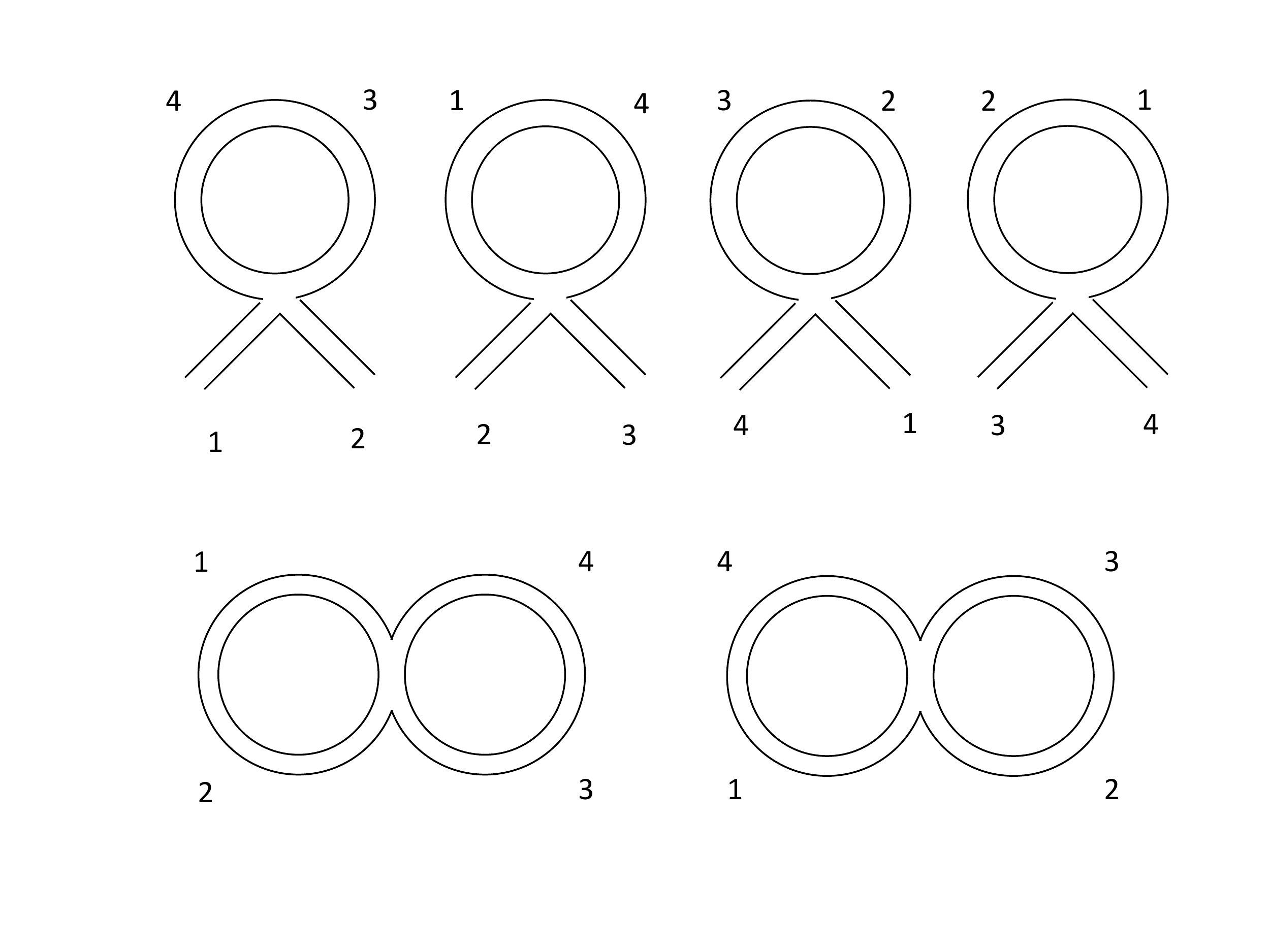}
\caption{The counter-terms $\Tr (\phi^2 T^\Lambda)$ and $\Pi^\Lambda$ from the Wick ordering.}
\label{wick1}
\end{figure}
\vskip.5cm

The partition function reads:
\begin{equation}\label{part0}
Z(\lambda, \Lambda)=\int d\mu^\Lambda(\phi)e^{-S[\phi]},
\end{equation}
where
\begin{equation}\label{gauphi}
d\mu^\Lambda(\phi)=\pi^{-\frac{\Lambda^2}{2}}\ \det C^{-1}\ e^{-\frac{1}{2}\Tr\ \phi\Delta\phi}\prod_{m\le n}^\Lambda d{\rm Re}(\phi_{mn})\prod_{m< n}^\Lambda d {\rm Im}(\phi_{mn})\ ,
\end{equation}
is the normalized Gaussian measure for the Hermitian matrices $[\phi_{mn}]$ with covariance $C=\Delta^{-1}$ (see Formula \eqref{covmatrix}) and $S[\phi]$ is the Wick ordered interaction term. Before proceeding some remarks on the Gaussian integrations are in order: Gaussian integrations over $(\Lambda+1)\times(\Lambda+1)$ dimensional Hermitian matrices $[A_{mn}]_{(\Lambda+1)\times(\Lambda+1)}$ are considered as integrations over the  $(\Lambda+1)^2\times 1$ dimensional vectors $\vec A$, which is formed by listing the elements of the matrices $[A_{mn}]$ as 
$\vec A=(A_{01}, A_{02}\cdots, A_{0\Lambda}, \cdots,A_{\Lambda0},\cdots,A_{\Lambda\Lambda})$, one row after another, and result in a determinant of a $(\Lambda+1)^2\times(\Lambda+1)^2$ dimensional matrix.

The rest of this paper is devoted to proving the main theorem, namely proving the well-definedness  the density function $\log Z(\Lambda,\lambda)$ (recall that $V_\theta=1$) in the cardioid domain ${\cal{C}}ard_\rho$ (see Section 6) in the limit $\Lambda\rightarrow\infty$. The main difficulty for proving this theorem lies in the fact that the action in \eqref{action1} is not positive. Observe that for $\lambda>0$,
\be
e^{-\frac{1}{4}\Tr\ [\phi^4-4\phi^2\ T^\Lambda+2\ (T^\Lambda)^2 ]}=e^{-\frac{1}{4}\Tr\ [(\phi^2-2\ T^\Lambda)^2-2\ (T^\Lambda)^2 ]}\le e^{\frac{\lambda}{2}\Pi^\Lambda},\label{nelson}
\ee
reproducing the Nelson's divergent factor \cite{nelson} (also called the Nelson's bound):
\be
\vert\ Z(\lambda, \Lambda)\ \vert\le e^{\lambda O(1)\Lambda}.\label{lid}
\ee
Remark that the divergent behavior is much worse than the $\phi^4_2$ model, for which we have
 $\vert Z_{\phi^4_2}(\lambda, \Lambda)\vert\le e^{\lambda O(1)\ln^2\Lambda}$, when $\Lambda\rightarrow\infty$.
 
The Nelson's factor will be compensated in a multi-scale way, step by step following the flow of the renormalization group, by convergent power counting factors generated in the slice-testing expansions. This is the content of the Section 4. Before discussing this point in more detail we shall first of all introduce the intermediate field technique of the partition function, a key step for the MLVE and an important preparation for the slice-testing expansions.

\section{The Multi-scale loop vertex expansions, Step one: The intermediate field representation}
\subsection{The intermediate field representation}

The multi-scale loop vertex expansions (MLVE) combine an intermediate field representation of the partition function with a replica trick, a slice-testing expansion and a two-level forest formula to express the connected function of a Bosonic field theory with quartic interaction as a convergent perturbation series indexed by two-level trees. As a first step of applying MLVE we shall introduce the $(\Lambda+1)\times(\Lambda+1)$ dimensional Hermitian matrices $\sigma=[\sigma_{mn}]$ as intermediate fields and write the partition function (using \eqref{nelson})
$\int d\mu^\Lambda(\phi)e^{-\frac{\lambda}{4}\Tr[(\phi^2-2 T^\Lambda)^2]+\frac\lambda2\Pi^\Lambda}$
as
\be
\int d\mu^\Lambda(\phi)e^{-\frac{\lambda}{4}\Tr(\phi^2-2 T^\Lambda)^2}e^{\frac\lambda2\Pi^\Lambda}=\int d\phi e^{-\frac{1}{2}\Tr\phi C^{-1}\phi}\int d\sigma e^{-\frac{1}{2}\Tr\sigma^2}e^{-\frac{i\sqrt{2\lambda}}{2}\Tr[\sigma(\phi^2-2T^\Lambda)]}e^{\frac\lambda2\Pi^\Lambda}.\label{inta}
\ee
Remark that the term $-\frac{i\sqrt{2\lambda}}{2}\Tr\sigma\phi^2$ in the above formula should be written as $-\frac{i}{4}\sqrt{2\lambda}\Tr(\sigma\phi^2+\phi^2\sigma)$, as $\sigma$ can be added to different sides of the matrix $\phi^2$ to make a full square $\phi^4$. 

Now we consider the Gaussian integral over $\phi$ and we have 
\be
\int d\phi e^{-\frac{1}{2}\phi C^{-1}\phi} e^{-\frac{i}{2}\sqrt{2\lambda}\sigma\phi^2}=e^{-\frac12\Tr\log(1+i\frac{\sqrt{2\lambda}}{2}C(1\otimes\sigma+\sigma^t\otimes1))}.
\ee

So the partition function can be written as:  
\begin{equation}\label{partitionf}
Z(\lambda)=\int d\nu(\sigma)\ e^{\frac{ i\sqrt{2\lambda}}{2}\ \Tr\ T^\Lambda(I\otimes \sigma+\sigma^t\otimes I)-
\frac{1}{2}\Tr\log[1+i\frac{\sqrt{2\lambda}}{2} C(I\otimes \sigma+\sigma^t\otimes I )\ ]+\frac{1}{2}\lambda \Pi^\Lambda},
\end{equation}
where \begin{equation}
d\nu(\sigma)=\pi^{-\Lambda^2/2}e^{-1/2\ \Tr\ \sigma^2}\prod_{m\le n} d{\rm Re}(\sigma_{mn})\prod_{m< n} d {\rm Im}(\sigma_{mn}),
\end{equation}
is the normalized Gaussian measure with covariance
$ <\sigma_{mn},\sigma_{kl}>=\int d\mu(\sigma)\sigma_{mn}\sigma_{kl}=\delta_{nk}\delta_{ml}$. 
The term
$\exp\{-\Tr\log[1+i\frac{\sqrt{2\lambda}}{2} C(I\otimes \sigma+\sigma^t\otimes I )]\}$ means the $(\Lambda+1)^2\times(\Lambda+1)^2$ dimensional determinant resulted from the Gaussian integration and $\sigma^t$ means the transpose of the matrix $[\sigma_{mn}]$. 
Let $\hat\sigma=I\otimes \sigma+\sigma^t\otimes I$, then the partition function can be written as
\be\label{lv1}
Z(\lambda)=\int d\nu(\sigma)\ e^{\frac{ i\sqrt{2\lambda}}{2}\ \Tr\ T^\Lambda\hat\sigma-
\frac{1}{2}\Tr\log[1+i\frac{\sqrt{2\lambda}}{2} C\hat\sigma\ ]+\frac{1}{2}\lambda \Pi^\Lambda}\ ,
\ee
which is called the intermediate field representation for the partition function and the interaction term $\Tr\log[1+i\frac{\sqrt{2\lambda}}{2}C\hat\sigma]$, which is also called the $loop\ vertex$ \cite{RivLve}, is defined in the operator sense. Formula \eqref{lv1} is the starting point for the future expansions.

The main message of this intermediate field technique is that it makes the constructive renormalization of this model much easier (the cumulants can be expressed in terms of ordinary propagators and the resolvents operators (see Formula \eqref{symrev}), for which we have the nice "resolvent bound" (see Lemma \ref{keybound})).

\begin{figure}[!htb]
\centering
\includegraphics[scale=0.3]{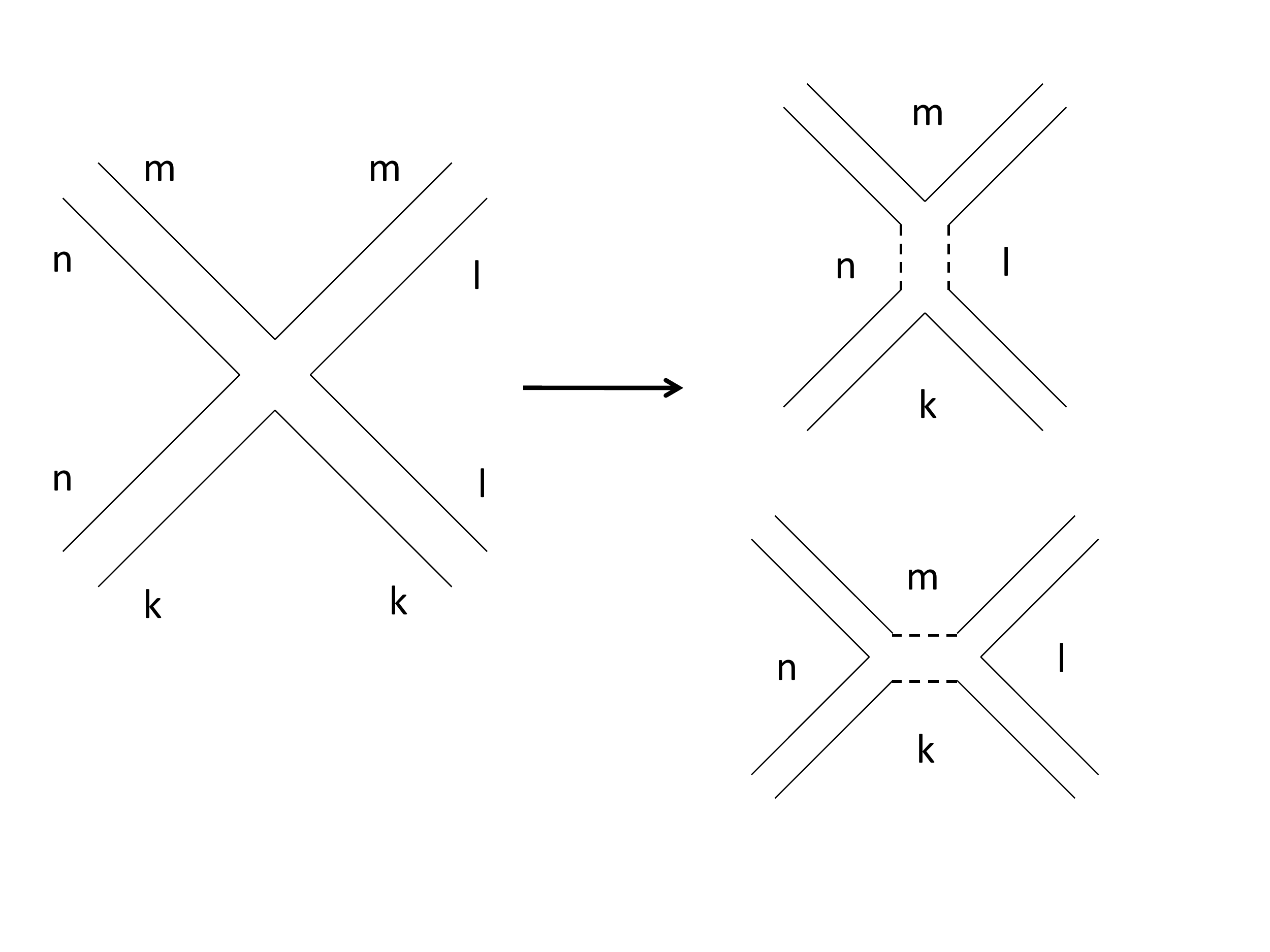}
\caption{Two different ways of decomposing a $\phi^4$ vertex. The dotted lines stand for the $\sigma$ propagators.} \label{decomp}
\end{figure}

The matrix elements of $C\hat\sigma$ read:
 \bea\label{matrixrule}
(C\hat\sigma)_{mn,kl}=\sum_{pq}C_{mn,pq}(I\otimes \sigma+\sigma^t\otimes I)_{qp,kl}
=C_{mn}\sigma_{nk}\delta_{lm}+\sigma_{lm}C_{mn}\delta_{nk},
\eea
and the linear counter-term reads: 
\be\frac{ i\sqrt{2\lambda}}{2}\ \Tr\ T^\Lambda((I\otimes \sigma+\sigma^t\otimes I))=2\times\frac{ i\sqrt{2\lambda}}{2}\sum_m T^\Lambda_m\sigma_{mm},\label{lict}\ee 
where the factor $2$ reflects the fact that there are two different ways of decomposing the quartic term $\Tr\phi^4$ into the cubic term $\Tr\ \phi^2\sigma$, see Figure \ref{decomp}.

Let $e_m\otimes e_n$, $m,n=0, 1\cdots\Lambda$, be the basis of the Hilbert space $\cH^\Lambda\otimes\cH^\Lambda$ and define the matrix operator $\hat{\sigma}=I\otimes \sigma+\sigma^t\otimes I$ with \be\hat\sigma_{mn,pq}= [I\otimes\sigma+\sigma^t\otimes I]_{mn,pq}=\sigma_{np}\delta_{qm}+\sigma_{qm}\delta_{np}\label{int}\ ,\ee then 
$\hat\sigma$ acts on $\cH^\Lambda\otimes\cH^\Lambda$ with the following rule:
\be
\hat\sigma:\ e_m\otimes e_n\rightarrow\sum_{k}\sigma_{mk}e_k\otimes e_n+e_m\otimes\sum_k\sigma_{kn} e_k\ \label{hbasis}.
\ee
The Feynman graphs generated in the perturbation expansion are Ribbon graphs \cite{GW1, tutt,tutt2}. Each edge of a ribbon graph comprises of two borders. Ribbon graphs can be embedded into a surface of genus $g\ge 0$ in such a way that the edges do not intersect and dissecting the surface along the edges decomposes it into faces each of which is homeomorphic to a disc. 

It is useful to label the two Hilbert spaces in $\cH^\Lambda\otimes\cH^\Lambda$ as $\cH^\Lambda_1\otimes\cH^\Lambda_2$ and identify the two borders of a ribbon with the two Hilbert spaces ${\cal{H}}^\Lambda_1$ and ${\cal{H}}_2^\Lambda$, respectively. Define the border of an edge on which the operator $I\otimes\sigma$ acts, namely the border corresponding to the Hilbert space $\cH^\Lambda_2$, as the inner border and the border on which $\sigma^t\otimes I$ acts as the outer border. Graphically \eqref{hbasis} means that the intermediate field $\sigma$ can hook to both the inner border and the outer border of a ribbon.

In order to simplify the combinatorial factors in the perturbation expansion we'd like to expand the loop vertex term as \cite{RivW3}:
\begin{eqnarray}\label{killleaf}
- \frac{1}{2}\Tr\log[1+i\frac{\sqrt{2\lambda}}{2}C\hat\sigma]=- \frac{i}{2}\Tr(\frac{\sqrt{2\lambda}}{2}C\hat\sigma )-\frac{1}{2}\Tr\log_2[1+i\frac{\sqrt{2\lambda}}{2}C\hat\sigma],
\end{eqnarray}
where $\log_n(1+x)$ for $n\geq 2$ is defined as the $n$-th Taylor
remainder of the function $\log(1+x)$:
$\log_n(1+x)=\log(1+x)-[x-x^2/2+x^3/3\cdots+(-1)^{n+1}x^n/n]$\ .

Using \eqref{matrixrule} we can easily find that $-i/2\ \Tr(\sqrt{2\lambda}/2\ C\hat\sigma )=-i\sqrt{2\lambda}/2\sum_m T^\Lambda_m\sigma_{mm}$, which partially cancels the counter-term (see Formula \eqref{lict}). The resulting interaction vertex reads:
\begin{eqnarray}\label{fulver}
V(\sigma)=-\frac{1}{2}\Tr\log_2\ [1+i\frac{\sqrt{2\lambda}}{2}C\hat\sigma]
+i\frac{\sqrt{2\lambda}}{2}\sum_m\ T^\Lambda_m\sigma_{mm}+\frac{1}{2}\lambda \Pi^\Lambda\ .
\end{eqnarray}

Define the resolvent matrix $R$ by
\be\label{res0}
R=: \frac{1}{1+i\frac{\sqrt{2\lambda}}{2}C\hat\sigma}\ ,
\ee
which originates from the derivation over the loop vertex w.r.t. $\sigma$:
\bea\label{cder1}
&&\frac{\partial}{\partial\sigma_{\alpha\beta}}\ \Tr\log[ 1+i\frac{\sqrt{2\lambda}}{2}C\hat\sigma]=
\frac{\partial}{\partial\sigma_{\alpha\beta}} \sum_{mn,nm}\big[\sum_N\frac{(-1)^{N+1}}{N}(i\frac{\sqrt{2\lambda}}{2})^N(C\hat\sigma)^N\ \big]_{mn,nm}\nonumber\\
&&= 
i\frac{\sqrt{2\lambda}}{2}\ [\sum_{m} R_{m\beta,\alpha m}C_{m\alpha}+\sum_{n}R_{\alpha n,n\beta}C_{\beta n}],
\eea
where $R_{mn,pq}=\big[\ \frac{1}{1+i\frac{\sqrt{2\lambda}}{2}C\hat\sigma}\ \big]_{mn,pq}$
is an element of $R$. The resolvent matrices appear frequently in the next section when we consider the forest expansions and play a very important role for constructing the $GW_2$ model. 

We can easily find that the resolvent matrix obeys the following equation:
\be\label{diffr}
\frac{\partial}{\partial\sigma_{\alpha\beta}}R_{mn,pq}=-i\frac{\sqrt{2\lambda}}{2}\sum_{s}[R_{mn,\alpha s}C_{s\alpha} R_{s\beta,pq}+R_{mn,s\beta}C_{\beta s}R_{\alpha s,pq}\ ],
\ee
which could be symbolically written as
\be
\frac{\partial}{\partial\sigma}R=\ -i\frac{\sqrt{2\lambda}}{2}\ RCR.\label{opeq1}
\ee

Remark that since the operators $C$ are positive, we can define naturally the operator $C^{1/2}$ with elements $[C^{1/2}]_{mn,kl}=\big(\frac{1}{m+n+1}\big)^{1/2}\delta_{ml}\delta_{nk}$ and equivalently write
\eqref{fulver} as:
\bea\label{fulverb}
V(\sigma)&=&-\frac{1}{2}\Tr\log_2\big[1+i\frac{\sqrt{2\lambda}}{2}\ C^{1/2}\ \hat\sigma\ C^{1/2}\ \big]\nonumber\\
&+&i\frac{\sqrt{2\lambda}}{2}\sum_m\ T^\Lambda_m\sigma_{mm}+\frac{1}{2}\lambda \Pi^\Lambda\ .
\eea

Similarly we can define the symmetric resolvent by the following equation:
\bea\label{cder2}
&&\frac{\partial}{\partial\sigma_{\alpha\beta}}\ \Tr\log[1+i\frac{\sqrt{2\lambda}}{2}C^{1/2}\hat\sigma C^{1/2}]\nonumber\\
&&=
i\frac{\sqrt{2\lambda}}{2}[\sum_m\  C^{1/2}_{m\beta}\hat R_{m\beta,\alpha m}C^{1/2}_{m\alpha}+   \sum_n\ C^{1/2}_{\alpha n}\hat R_{\alpha n,n\beta}C^{1/2}_{\beta n}],
\eea
where $\hat R$ is called the symmetric resolvent matrix with elements:
\be\label{symrev}
\hat R_{mn,pq}=\big[\ \frac{1}{1+i\frac{\sqrt{2\lambda}}{2} C^{1/2}\hat\sigma C^{1/2}}\ \big]_{mn,pq}\ .
\ee
We have
\be\label{diffrs}
\frac{\partial}{\partial\sigma_{\alpha\beta}}\hat R_{mn,pq}=-i\frac{\sqrt{2\lambda}}{2}\sum_{s}[\hat R_{mn,\alpha s}C^{1/2}_{s\alpha}\ C^{1/2}_{\beta s}\hat R_{s\beta,pq}+\hat R_{mn,s\beta}C^{1/2}_{\beta s}\ C^{1/2}_{s\alpha}\hat R_{\alpha s,pq}\ ].
\ee

We shall use the resolvents and symmetric resolvents and formulas \eqref{diffr} and \eqref{diffrs} interchangeably in the rest of this paper. We have the following lemma for the bound of the symmetric resolvent:

\begin{lemma}\label{keybound}
Let $\lambda\in\mathds{C}\setminus\mathds{R}_-$, then the symmetric resolvent $\hat R$ is well defined 
and uniformly bounded:
\be
||\hat R||<\frac{1}{\cos(\frac{1}{2}\arg{\lambda})},\label{hsb}
\ee
where $||\cdot||$ is the operator norm for the matrix operator $\hat R$.
\end{lemma}
\begin{proof}
First of all remark that $C^{1/2}\sigma C^{1/2}$ is a self-adjoint operator acting on a Hilbert space, so that its eigenvalues are real. We can easily prove the following bound:
\be
||(I+i\frac{\sqrt{2\lambda}}{2} L)^{-1} ||<\frac{1}{\cos(\frac{1}{2}\arg\lambda)},
\ee
for any self-adjoint operator $L$ and $\lambda\in\mathds{C}\setminus\mathds{R}_-$,
using the spectral
mapping theorem \cite{reed-simon}.
\end{proof}

Remark that to each resolvent operator $R$ or $\hat R$ is associated with a factor $\sqrt\lambda$ (see Formulae \eqref{cder1} and \eqref{opeq1} or \eqref{cder2} and \eqref{diffrs}), as well as a propagator $C$, which can be simply bounded by a constant of order one.  In order to obtain the non-perturbative bound for the vacuum correlation function one needs to consider the H.S. norm of the operator $\Vert \sqrt\lambda \hat R\Vert$. It is easy to find that
\be
\Vert \sqrt\lambda \hat R\Vert =\Vert \frac{\sqrt\lambda}{\cos(\frac{1}{2}\arg\lambda)}\Vert<\sqrt{\rho},\ {\rm for}\ \lambda\in \cC ard_\rho,
\ee
which is the desired result, namely $ \sqrt\lambda \hat R$ can be bounded by a small constant $\rho$. And that is the reason why the analytic domain for $\lambda$ should be $\cC ard_\rho$ in order that Theorem \ref{thetheorem} being true. Since $\cC ard_\rho\subset \mathds{C}\setminus\mathds{R}_-$ for $\rho$ small enough, the resolvent $\hat R$ obeys the same bound as \eqref{hsb} for $\lambda\in\cC ard_\rho$. 

\section{The Multi-scale loop vertex expansions, Step two: The slice-testing expansions}

One of the main problems in the construction of the $GW_2$ model is to bound properly the divergent factors in \eqref{fish} and \eqref{lid} so as to render the partition function $Z$ and
the vacuum connected function $\log Z$ well defined. The compensation of these divergent terms is not performed at one stroke but is realized in multi-steps in a recursive way: these divergent terms are further decomposed into geometric series such that each term is indexed by an integer, called the scaling index, and is of order 1. The compensations of the divergent factors with the counter-terms then take place to these order one terms of the geometric series, one after another following the decreasing order of the scaling indices. This is the basic idea of the renormalization groups invented by K. Wilson. The renormalization procedure adapted to the current situation is called the slice-testing expansions. 

Let's first of all introduce the scaling indices.
%
%
%
%

\subsection{Sliced propagators and the index sets}\label{slice1}
Let $I_\Lambda=\{1, 2,\cdots, \Lambda\}$ be the set of matrix indices.
Define $S:=\{ 0, 1, \cdots, j_{max}\}$ as the set of coarse  scaling indices, where
for a fixed integer $M\ge 2$, $j_{max}$ is defined as the integer part of the number $\tilde j_{max}$ such that $M^{\tilde j_{max}}=\Lambda$. Then $S$ divides $I_\Lambda$ into $j_{max}+1$ slices: 
\be I_\Lambda=I_0\cup I_1\cup I_2\cup\cdots\cup I_{j_{max}},\label{slice0}\ee
each of which is made of the integers $\oj\in I_j=[M^j, M^{j+1}-1]$, called the refined scaling indices. Define also $I_{\le j}=\cup_{k=0}^{j}\ I_k$. 
Let $J\subseteq S$ be subset of $S$, we can define the union of index sets $I_J=\cup_{j\in J} I_j$. Clearly $I_J\subseteq I_\Lambda$.

Define the propagator of refined scaling index $\oj$ as:
\bea
C^{\omega(j)}_{mn}&:=&C_{mn}\cdot \mathds{I}_{m+n=\oj \in I_j},
\eea
where $\mathds{I}_x$ is the characteristic function for the event $x$:
\be
\mathds{I}_x=\bigg\{\begin{matrix}1,\quad {\rm if\ x\ is\ true},\\
0,\quad {\rm otherwise},
\end{matrix}
\ee
we set
\be
C_{mn}=\sum_{\og\in I_\Lambda}\ C^\og=\sum_{j=0}^{j_{max}}\sum_{\omega(j)\in I_j}C_{mn}^{\omega(j)}.
\ee

It is easy to find that
\be\label{cog}
O(1)M^{-j-1}\le\vert C^\oj_{mn}\vert\le O(1)M^{-j},
\ee 
where $O(1)$ is a general name for the inessential constants of order 1.

Then the tadpole term can be written as $T^\Lambda_m=\sum_{j=0}^{j_{max}}\sum_{\omega(j)\in I_j}T^{\omega(j)}_m$, where
\bea
T^{\omega(j)}_m&=&\sum_{n\in I_{\le j}} C^{\omega(j)}_{mn}\cdot\mathds{I}_{ m+n=\oj\in I_j}=\ \frac{1}{\oj+1}\label{bd1} .
\eea
Define $T^j_m=\sum_{\oj\in I_{j}}T^\oj_m$. Using the fact that $\oj\in I_j= [M^j, M^{j+1}-1]$
and $\vert I_j\vert=(M-1)M^j$, we have:
\be\label{bd1a}
T^j_m=\sum_{\oj=M^j}^{M^{j+1}-1}  \frac{1}{\oj+1} =O(1), 
\ee
which reproduces the bound for the tadpole (see Formula \eqref{fish}):
\be
T^\Lambda_m=\sum_j\  T^{j}_m=O(1)\ln\Lambda.\label{bd2}
\ee

Similarly we can decompose the vacuum tadpole (see Formula \eqref{glasses}) as
\be
\Pi^\Lambda=\sum_{j=0}^{j_{max}}\Pi^j,\label{vatad1}
\ee
where
\be
\Pi^j=\sum_{m\in I_j}(\sum_{p\in I_{\le j}}\frac{1}{p+m+1})(\sum_{q\in I_{\le j}}\frac{1}{q+m+1})\le\sum_{m\in I_j}O(1)\le
 M^j.\label{vatad2}
\ee
So we have
\be
\Pi^\Lambda=\sum_{j=0}^{j_{max}}\Pi^j\le\sum_{j=0}^{j_{max}} M^j\sim O(1)\Lambda,\label{bd3}
\ee
which reproduces the linear divergence of \eqref{glasses}.

%

\subsection{The slice-testing expansion}\label{slicetesting}
The slice-testing expansions play two roles in the renormalization process: on the one hand, it generates at each scale the counter-terms to cancel exactly the sliced tadpoles \eqref{bd1}, on the other hand, it generates the $marked$ propagators, which are propagators with scaling index $\oj$ that are not tadpoles, so as to compensate the Nelson's bound for the connected function . This compensation is more involved. Since the amplitude of each vacuum tadpole at coarse scale $j$ is bounded by $M^j$, the Nelson's bound at each coarse scale is given by $e^{\lambda O(1)M^j}$, while from each marked propagator of refined scaling index $\oj$ one can only gain a convergent factor $M^{-j}$ (see \eqref{cog}). Obviously one marked propagator at each coarse scale $j$ is not enough. Instead one has to generate $aM^j$, $a\in(0,1)$, marked propagators at each coarse scale $j$. This is possible since the cardinality of the set of the refined scaling indices $I_j$ is $(M-1)M^j$. Of course one can generate much more marked propagators at each scale $j$ hence cumulate a much better convergent factor. But this is also problematic since this would generate large combinatorial factor in the slice-testing expansion such that the convergent factor we gained is not enough to compensate it. So we have to set a stopping rule to the slice testing expansion. Namely the expansion should be terminated as long as the number of marked propagators at each scale $j$ reaches $aM^j$. Now we go to the details.

First of all we introduce inductively a set of interpolation parameters $t:=\{t^{\omega(j)}\}\in [0, 1]$, $\oj\in I_j$, $j\in S$, one for each propagator $C^\oj$. Define the interpolated propagator, the counter-term and the interaction vertex as:
\begin{eqnarray}
&&C(t)=\sum_j\sum_\oj t^\oj C^\oj,\quad T(t)=\sum_j\sum_\oj t^\oj T^\oj,\nonumber\\
&&V(\sigma)(t)=V_\Lambda(\sigma, t)=-\frac{1}{2}\Tr\log_2[1+i\frac{\sqrt{2\lambda}}{2}
\sum_{j=0}^{j_{max}}\sum_{\omega(j)=M^{j}}^{M^{j+1}-1}\ C^{\omega(j)}t^{\omega(j)}\hat\sigma\ ]\label{dp0}\\
&&+\ i\frac{\sqrt{2\lambda}}{4}\Tr\ [ \sum_{j=0}^{j_{max}}\sum_{\omega(j)=M^{j}}^{M^{j+1}-1}\ T^{\omega(j)}t^{\omega(j)}\hat\sigma]+\frac{1}{2}\lambda\Tr\ [ \sum_{j=0}^{j_{max}}\sum_{\omega(j)=M^{j}}^{M^{j+1}-1}\ T^{\omega(j)}t^{\omega(j)}]^2.\nonumber
\end{eqnarray}
Then the partition function is noted as $Z^{j_{max}}(\lambda, t^{\omega(1)},\cdots t^{\omega(j_{max})})\vert_{t^{\omega(j)}=1, j\in[0,j_{max}]}$. For each interpolation parameter $t^{\omega(j)}$
we shall use the fundamental theorem of calculus:
\bea\label{cal1}
&&Z^{j_{max}}(\lambda, \cdots, t^{\omega(j)}\cdots)\vert_{t^{\omega(j)}=1}\\
&&=Z^{j_{max}}(\lambda, \cdots, t^{\omega(j)}\cdots)\vert_{t^{\omega(j)}=0}+\int_{0}^1d t^{\omega(j)}\frac {\partial Z^{j_{max}}(\lambda, \cdots, t^{\omega(j)}\cdots)}{\partial{t^{\omega(j)}}}\nonumber\ .
\eea

Let $\om_j\subseteq I_j$ be a subset of refined scaling indices of coarse scale $j$. For $J\subseteq S=[0, j_{max}]$, we can define $\om_J=\cup_{j\in J} \om_j$ be a subset of $I_J=\cup_{j\in J} I_j$. 
Then the partition function can be written as:
\be\label{st0}
Z^{j_{max}}(\lambda)=\sum_{ J\subseteq  S}\sum_{\om_J\subseteq I_J}\int d\nu(\sigma)\ [\ \prod_{j\in  J}\prod_{\oj\in \om_j}\int_0^1 dt^\oj\frac{\partial}{\partial t^\oj} ]\ e^{V(\sigma)(t)}\ \vert\ _{t^\oj=0,\ {\rm for}\ j\notin{ J},\ \oj\notin \om_J}.
\ee
The above formula is also called the multi-variable interpolation formula, as we use the fundamental theorem of calculus \eqref{cal1} for each interpolation parameter $t^{(\omega(j))}$. We use the convention that when $J=\emptyset$, we have $t^{\omega(j)}=0,\ \forall \omega(j)\in I_j, \forall j\in[0,j_{max}]$, for which we have $Z^{j_{max}}(\lambda, 0\cdots, 0)=1$.

\begin{figure}[!htb]
\centering
\includegraphics[scale=0.4]{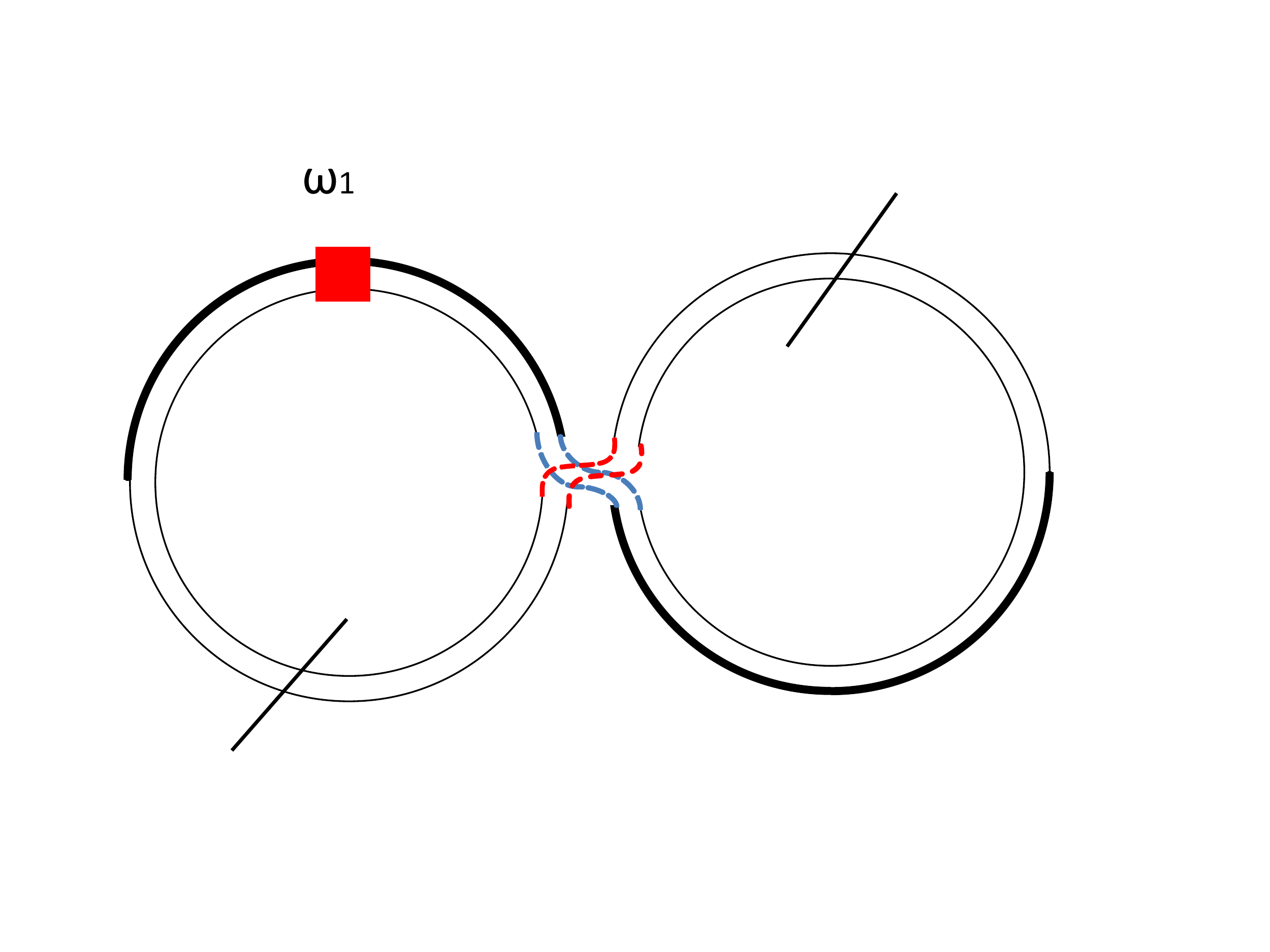}
\caption{A nonplanar graph in the first order slice-testing expansion. We used different colors for the $\sigma$ propagators.} \label{nonplanar2}
\end{figure}

\begin{figure}[!htb]
\centering
\includegraphics[scale=0.3]{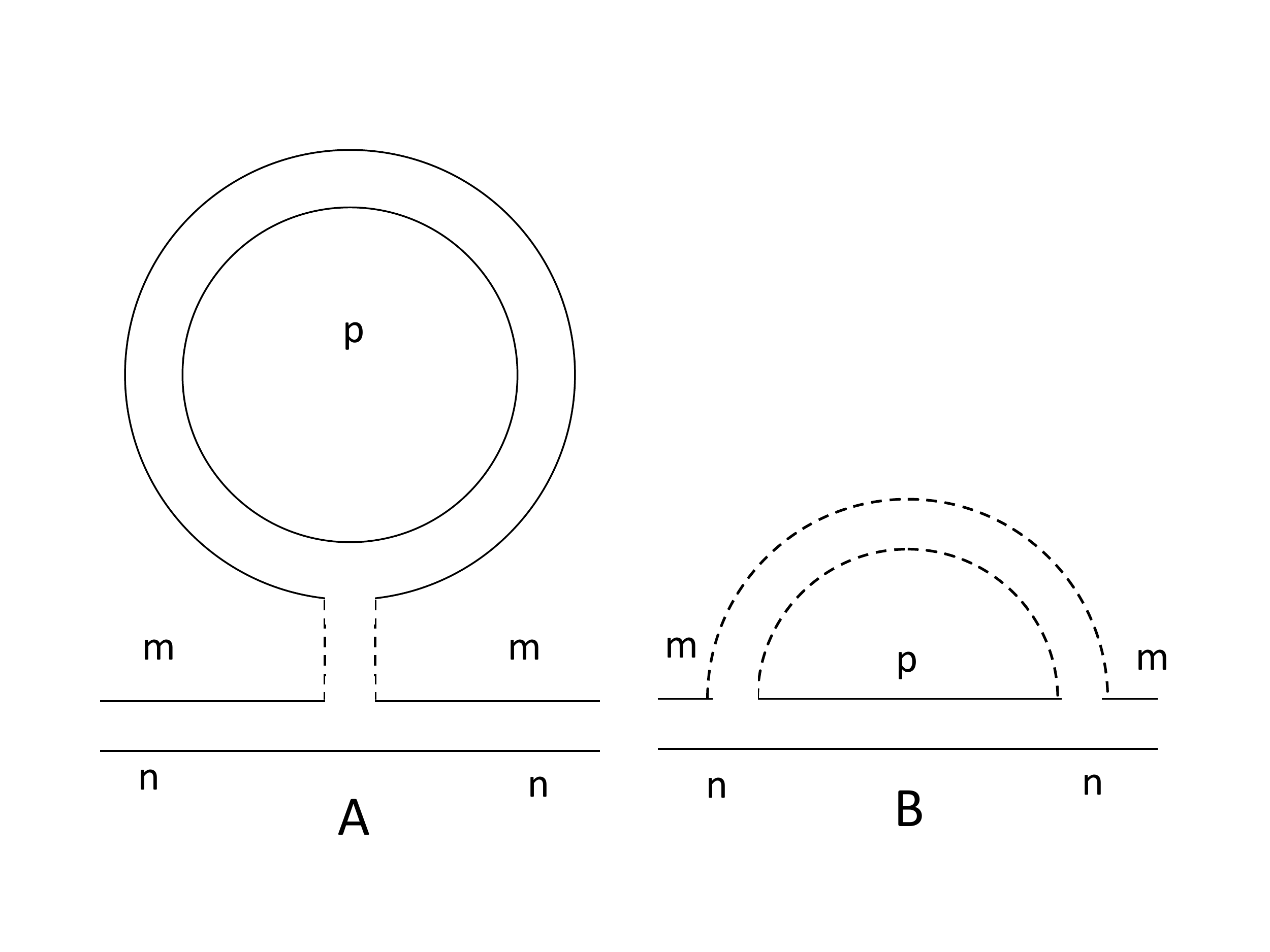}
\caption{Duality for the tadpoles.} \label{duality}
\end{figure}

Each derivation $\frac{\partial}{\partial t^\oj}$ on $ V(\sigma, t)$ results in:
\bea
&&\frac{\partial}{\partial t^\oj}\ V(\sigma, t) = -i\frac{\sqrt{2\lambda}}{4}\ \Tr\ [C^{\og(j)}(I\otimes\sigma+\sigma^t\otimes I)(R-1)]\nonumber\\
&+&i\frac{\sqrt{2\lambda}}{4}\ \Tr\ [T^{\og(j)}(I\otimes\sigma+\sigma^t\otimes I)]+\lambda\ \Tr\ [ T^{\og(j)}(\sum_{j}\sum_{\og(j)}T^{\og(j)}t^\oj)].
\eea
The linear terms $\sigma^t\otimes I$ and $I\otimes\sigma$ in the numerator will contract on the resolvents and the vertex $e^{V(\sigma, t)}$ that are functions of $\sigma$, due to the Gaussian integration. Tadpoles will be generated in this process and they are automatically compensated with the corresponding counter-terms and the resulting terms, which are called the renormalized amplitudes, are free of tadpoles and counter-terms. 
\begin{definition}[Marked propagators and tadpoles]\label{markedp}
A propagator $C^\oj$ of refined scaling index $\oj$ generated in the slice-testing expansion is called a marked propagator. A tadpole term of scaling index $\oj$ is defined by $T^\oj_{n}=(\tr\ C^\oj)_{nn}$,  where $n$ is the index for the border of the tadpole which doesn't form a closed face and the trace is defined in \eqref{ptrace} (see also Formula \eqref{fish}).
 \end{definition}
Before proceeding it is useful to consider the slice-testing expansions of the first order.
\begin{example}
Let $\cA_1$ be the amplitude of the first order slice-testing expansion, we have:
\bea
\cA_1&=&\int d\nu(\sigma)\int_0^1 dt^{\omega(j_1)}\frac{\partial}{\partial t^{\og(j_1)}}e^{V(t,\sigma)}\nonumber\\
&=&\int d\nu(\sigma)\int dt^{\omega(j_1)}e^V\ \big\{-i\frac{\sqrt{2\lambda}}{4}\Tr\ [(R-1)C^{\og(j_1)}(I\otimes\sigma+\sigma^t\otimes I)]\nonumber\\
&+&i\frac{\sqrt{2\lambda}}{4}\Tr\ [T^{\og(j_1)}(I\otimes\sigma+\sigma^t\otimes I)]+\lambda\Tr\  [T^{\og(j_1)}(\sum_{j}\sum_{\og(j)}T^{\og(j)}t^\oj)]\ \big\},
\eea
where we have used the fact that $\partial t^{\og(j)}/\ \partial t^{\og(j_1)'}=\delta_{\og(j)\og(j_1)'}$.

Writing explicitly the matrix indices we have:
\bea\label{exp1}
\cA_1&=&\int d\nu(\sigma)\int dt^{\omega(j_1)}e^{V(t,\sigma)}\ \big\{-i\frac{\sqrt{2\lambda}}{4}\sum_{mnpq}\ [(R-1)_{mn,pq}C^{\og(j_1)}_{qp,pq}(I\otimes\sigma+\sigma^t\otimes I)_{qp,nm}]\nonumber\\
&+&i\frac{\sqrt{2\lambda}}{4}\sum_{mn}\ [T^{\og(j_1)}_m(I\otimes\sigma+\sigma^t\otimes I)_{mn,nm}]+\lambda\Tr\  [T^{\og(j_1)}(\sum_{j}\sum_{\og(j)}T^{\og(j)}t^\oj)]\ \big\},
\eea
First of all we consider the first term in the above formula. Using the fact that $(I\otimes\sigma+\sigma^t\otimes I)_{qp,nm}=\sigma_{pn}\delta_{qm}+\sigma_{mq}\delta_{np}$ this term can be written as
\bea\label{exp1a}
&&\int d\nu(\sigma)dt^{\omega(j_1)}e^{V(t,\sigma)}\sum_{mnpq}\{-i\frac{\sqrt{2\lambda}}{4}\frac{\partial}{\partial{\sigma_{np}}}(R-1)_{mn,pq}C^{\omega(j_1)}_{qp,pq}\delta_{qm}\\
&&+(-i\frac{\sqrt{2\lambda}}{4})
[\frac{\partial}{\partial{\sigma_{np}}}V] (R-1)_{mn,pq}C^{\omega(j_1)}_{qp,pq}\delta_{qm}+derivations \ w.r.t. \ \sigma_{qm}\}.\nonumber
\eea
Graphically the terms resulted from derivations over $\sigma_{np}$ correspond to attaching and contracting the intermediate fields $\sigma$ to the border $n$ of the ribbon while the terms resulted from derivations over $\sigma_{qm}$ is attaching and contracting the intermediate fields $\sigma$ to the border $m$ of the ribbon. Due to the symmetric properties of ribbon graphs
the amplitude of the terms from derivations over $\sigma_{qm}$ is the same as that of derivations over $\sigma_{np}$ and we don't explicitly write these terms.

The first term in Formula \eqref{exp1a} reads:
\be
\int d\nu(\sigma)dt^{\omega(j_1)}e^{V(t,\sigma)}\sum_{mnps}\{-\frac{\lambda}{4}(R_{mn,ps}C_{sp,ps}R_{sn,pm}C^{\omega(j_1)}_{mp,pm}+R_{mn,sn}
C_{ns,sn}R_{ps,pm}C^{\omega(j_1)}_{mp,pm})\}.
\ee
It is easy to find that while the first term corresponds to a planar term, the second term corresponds to a non-planar graph (see Figure \ref{nonplanar2}), which is also the only non-planar term in the first order expansion.

The second term in Formula \eqref{exp1a} is equal to
\bea
&&\int d\nu(\sigma)dt^{\omega(j_1)}e^{V(t,\sigma)}\{-\frac{\lambda}{8}\sum_{mnps}[(R-1)_{sp,ns}C_{sn,ns}(R-1)_{mn,pm}C^{\omega(j_1)}_{mp,pm}\nonumber\\
&&+(R-1)_{ns,sp}C_{ps,sp}(R-1)_{mn,pm}C^{\omega(j_1)}_{mp,pm}]+\frac{\lambda}{4}\sum_{mn} T^{\Lambda}_{n}R_{mn,nm}C_{mn,nm}\}.
\eea
Here both of the first two terms correspond to planar graphs, namely, the $\sigma$ propagators are attached to the outer border for the first term and are attached to the inner border for the second term. The amplitudes for the two terms are equal and can be written as:
\be\label{exp1b}
\int d\nu(\sigma)dt^{\omega(j_1)}e^{V(t,\sigma)}\{-\frac{\lambda}{4}\sum_{mnps}(R-1)_{sp,ns}C_{sn,ns}R_{mn,pm}C^{\omega(j_1)}_{mp,pm}+\frac{\lambda}{4} \sum_{mn}T^{\Lambda}_{n}R_{mn,nm}C_{mn,nm}\}.
\ee
Taking into account the contributions from the derivations of $\sigma_{qm}$, Formula \eqref{exp1a} is equal to
\bea\label{exp1c}
&&\int d\nu(\sigma)dt^{\omega(j_1)}\big[-\frac{\lambda}{2}\sum_{mnps}(R_{mn,ps}C_{sp,ps}R_{sn,pm}C^{\omega(j_1)}_{mp,pm}+R_{mn,sn}C_{ns,sn}R_{ps,pm}C^{\omega(j_1)}_{mp,pm})\\
&&-\frac{\lambda}{2}\sum_{mnps}(R-1)_{sp,ns}C_{sn,ns}(R-1)_{mn,pm}C^{\omega(j_1)}_{mp,pm}+\frac{\lambda}{2}\sum_{mn} T^{\Lambda}_{n}R_{mn,nm}C_{mn,nm}\ \big].\nn
\eea

Now we consider the second term of \eqref{exp1}, which is equal to
\bea
&&i\frac{\sqrt{2\lambda}}{2}T^\oj\frac{\partial}{\partial\sigma_{mm}}V=\frac{\lambda}{4}\sum_{ms}T^\oj_m[(R-1)_{sm,ms}C_{sm,ms}\nonumber\\
&&+(R-1)_{ms,sm}C_{ms,sm}]-\frac{\lambda}{2}
\sum_{m}T^\oj_m T^\Lambda_m\ .\label{exp1d}
\eea

Again both of the first two terms correspond to planar graphs and have the same amplitude, we can write \eqref{exp1d} as
\be
\frac{\lambda}{2}\sum_{ms}T^\oj_m(R-1)_{sm,ms}C_{sm,ms}-\frac{\lambda}{2}
\sum_{m}T^\oj_m T^\Lambda_m.
\ee 

Combine the terms in \eqref{exp1c} and \eqref{exp1d} and write the first term in Formula \eqref{exp1b} as
\be
-\frac{\lambda}{2}\sum_{s}[(R-1)+1]_{mn,ps}C_{sp,ps}[(R-1)+1]_{sn,pm}C^{\oj}_{mp,pm},
\ee
in which we used the fact that $\tr\ C^\Lambda_{nn}=T^\Lambda_{nn}$, $\tr\ C^\oj_{nn}=T^\oj_{nn}$,
then formula (58) can be written as
\bea
&&\int d\nu(\sigma)de^{\oj}e^{V(t,\sigma)}\sum_{mnps}[-\lambda (R-1)_{mn,ps}C_{sp,ps}(R-1)_{sn,pm}C^{\oj}_{mp,pm}\nn\\
&&\ \ -\frac{\lambda}{2}R_{mn,sp}C_{ps,sp}R_{ns,pm}C^{\oj}_{mp,pm} ],\label{npl}
\eea
in which the first term corresponds to a planar graph while the second term corresponds to a non-planar graph. Remark that in order to obtain the above formula we have used the duality relation such that the tadpoles in Graph A and Graph B in Figure \ref{duality} have the same amplitude, which is a consequence of the fact that each $\phi^4$ interaction can be decomposed in two different ways into $\phi^2\sigma$ vertex (see Figure \ref{decomp}). 
\end{example}
In order to simplify the notation, it is useful to introduce the circle product as follows:
\begin{definition}\label{traceprod}
Let $A$ and $B$ be two $N^2\times N^2$ dimensional matrices with double indices, we can define the trace product $\tr A\circ\tr B$ as 
\be \tr A\circ\tr B:=\sum_n(\tr A)_{nn}(\tr B)_{nn}= \sum_{n=1}^N[\sum_{m=1}^N A_{nm,mn}][\sum_{k=1}^N B_{nk,kn}],\label{circ}\ee
where $\tr A$ means taking the partial trace to the matrix $A_{nm,mn}$, namely summing over the indices corresponding to the inner border of the ribbon, and 
\be(\tr A)_{nn}:=\sum_{m=1}^N A_{nm,mn}\label{ptrace}.\ee
Let A be an $N^2\times N^2$ matrix and B an $N\times N$ diagonal matrix, we can still define the produce as 
\be
\tr A\circ B=\sum_n (\tr A)_{nn} B_{nn}.
\ee
If both A and B are $N\times N$ diagonal matrices we can define the circle product as
\be
A\circ B=\sum_n A_{nn} B_{nn}.
\ee
It is easy to find that the definition of the circle product can be generalized to more matrices. The trace operator $\Tr$ means that taking the full trace for matrix $A_{nm,mn}$, namely summing over all matrix indices.
\end{definition}

Then the expression of $\cA_1$ can be written as:
\bea
\cA_1&=&\int d\nu(\sigma)\int dt^{\omega(j_1)}e^{V(t,\sigma)}\big\{-\frac{\lambda}{2}
\tr\ [(R-1)C^{\omega(j_1)}]\circ\tr\ [(R-1)C(t)]\nonumber\\
&+&\frac{\lambda}{2}\tr\ [(R-1)C^{\omega(j_1)}]\circ T{(t)}
+\frac{\lambda}{2}\tr\ [(R-1)C(t)]\circ T^{\omega(j_1)}\nonumber\\
&-&\frac{\lambda}{2}\tr\ [RC^{\omega(j_1)} ]\circ\tr \ [RC(t)]+{\rm non-planar\ term}\ \big\}\nonumber\\
&=&\int d\nu(\sigma)\int dt^{\omega(j_1)}e^{V(t,\sigma)}\big\{-\lambda\ \tr\ [(R-1)C^{\omega(j)}]\circ \tr\ [(R-1)C(t)]\nonumber\\
&+&{\rm non-planar\ term}\big\},\label{oda}
\eea
where the non-planar term is given in the second line of Formula \eqref{npl}.

\begin{figure}[!htb]
\centering
\includegraphics[scale=0.5]{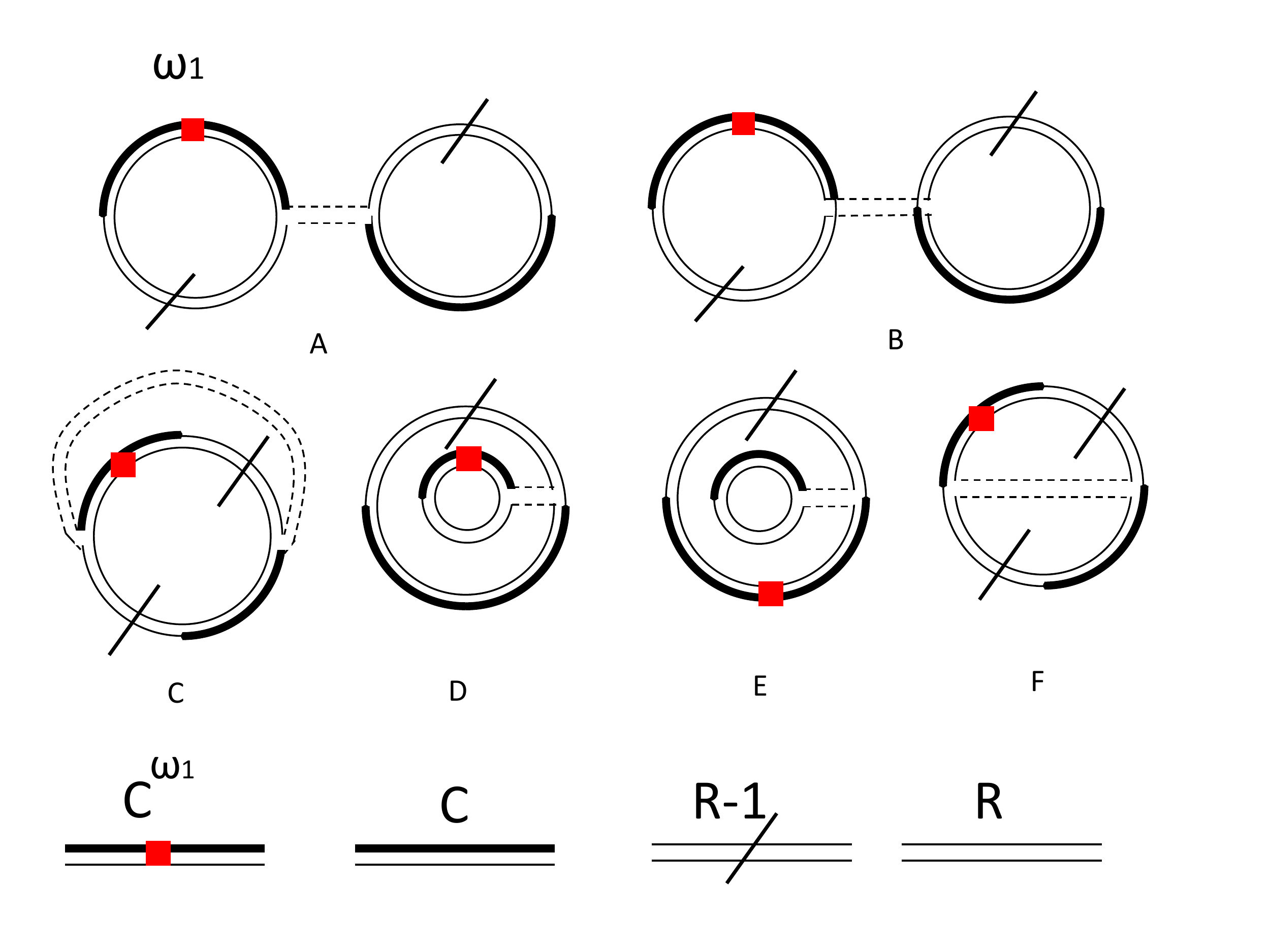}
\caption{Intermediate Feynman graphs for the first order slice-testing expansions. The dash line means the propagator for the intermediate fields $\sigma$, $C^{\og_1}$ represents a marked propagator of scale $\og$, $C$ is the full propagator $C(t)$; $R-1$ is the renormalized resolvent and $R$ is the full resolvent. Graph $A$ as well as its variants $B$, $D$ and $E$ are called dumbbell graphs.} \label{lve2}
\end{figure}

An illustration of the tadpole is given in Figure \ref{duality} and an illustration of the first order expansion is given in Figure \ref{lve2}.

The propagator $C^{\oj}$ is the source of the convergent factors. Another source of convergent factors is the crossings, for each of which of scale $\oj$ we can also gain a convergent factor $M^{-j}$ (see Formula \eqref{cog}).

An explanation of the numerical factors in \eqref{oda} is in order. We have $-\lambda=-\frac{\lambda}{8}\times 4- \frac{\lambda}{4}\times2 $, where $\frac{\lambda}{8}$ is the contribution from the graphs A, B, C and the total combinatorial factor is 4, \footnote{Remark that graph A represents two different graphs where the tadpole with marked propagator can appear on the left side or on the right side.} and $\frac{\lambda}{4}$ is the contribution from the graphs D and E for which the total combinatorial factor is 2. 

The the second order slice-testing expansion is given in the appendix. 
One can better understand the graphical representation of the slice-testing expansion amplitudes by the following definition:
\begin{definition} \label{resolg}
The ribbon graphs generated in the slice-testing expansions are also called the resolvent graphs, as they bear resolvents $R$. Let $\om_J$ be the set of refined scaling indices. An $\om_J-$ resolvent graph is defined as a resolvent graph in which exactly $\vert\om_J\vert$ marked propagators bear marks $\om_J$. An $\om_J$ resolvent graph is called minimal if any connected component of the graph bears at least one mark and the total perturbation order of the graph, i.e. the number of $\sigma$ propagators, is at most $\vert\om_J\vert$. The set of minimal $\om_J$ resolvent graphs is noted $\cG(\om_J)$ and we denote $\cG=\cup_J\cG(\om_J)$. 
\end{definition}
\begin{definition}\label{cprop}
A resolvent propagator is defined as a combination of a resolvent 
$R$ or $(R-1)$ with the propagator $C(t)$ or $C^\oj$. We call also the propagator $C(t)$ or $C^\oj$ the c-propagator.
\end{definition}

From the example of the second order slice-testing expansion we find that many different resolvent graphs can be generated. It is important to estimate the total number of graphs generated by an arbitrary order of expansion. We have the following lemma:
\begin{lemma}\label{comb}
The total number of resolvent graphs generated by the $N$-th order slice-testing expansion is bounded by $4^{N+1}N!$.
\end{lemma}
\begin{proof}
Observe that each derivation w.r.t. $t$ on the interaction $V$ generate $4$ $\sigma$ variables, which eventually acting on either the resolvent amplitude or on the interaction term again and generate new terms. Remark that due to the cyclic ordering the newly generated terms are not in arbitrary positions of the graph but still follow the cyclic ordering according to the previous position of the terms. The $N!$ are due to the combinatorial factors of the marked propagators.
\end{proof}
Remark that the bound is by no means optimal. The actual number of resolvent graphs can be much smaller, due to symmetric properties. But it will be enough for the purpose of this paper.

Now we have the following theorem for the renormalized partition function:

\begin{theorem}\label{generalslice}
For an arbitrary minimal $\om_J$ resolvent graph $G$ the partition function can be expressed
as follows in terms of the renormalized amplitude $A_G^R$ of the minimal resolvent graphs $G$:
\bea\label{partition}
Z^{j_{max}}(\lambda)=\sum_{J\subseteq S}\sum_{\om_{ J}\subseteq  I_J}\sum_{G\subset {\cal G}(\om_{J})}c(G)\int d\nu(\sigma)\prod_{\oj\in \om_{ J}}
\int_0^1 dt^\oj [e^{V(t)}A^R_G(t,\sigma)]|_{t^\oj=0,\ {\rm for}\ \oj\notin I_{ J}}\label{test1},\nonumber\\
\eea
where 
\be\label{gena}
A^R_G(t,\sigma)=(-\lambda)^{|\om_j|}\prod_{G^c}\Tr^{G^c}\{\prod_{\ell\in CP(G),\ \ell\ tadpole}[(R(\sigma)-1)\tilde C(\ell)]
\prod_{\ell\in CP(G), \ell\ not\  tadpole}[R(\sigma)\tilde C(\ell)]\},
\ee
is the renormalized amplitude, $c(G)=\prod c(G_c)$ is an order-one numerical factor characterizing the symmetric properties of a graph $G$, which factors over connected components $G_c$; the trace operator $\Tr^{G^c}$ means summing over the double indices for the matrices in $A^R_G(t,\sigma)$, namely it is taken over the $(\Lambda+1)^2\times(\Lambda+1)^2$ dimensional matrices, for the connected component $G^c$ in a general graph $G$, which could have many connected components; $CP(G^c)$ is the set of c-propagators of $G^c$, namely the propagators which are $C^\Lambda$ or $C^\oj$ and we use $\tilde C(\ell)$ to denote a c-propagator. The sum over ${\cal{G}}(\om_{J})$ runs over a set of minimal resolvent graphs. The index $\oj(\ell)$ specifies the markings, that is, restricts the c-propagator $\ell$ to be $C^\oj$ if that propagator bears the mark $\oj$. All propagators belonging to a single loop of the form $\Tr\ RC$  are renormalized, hence accompanied by an $R-1$ resolvent factor. The c-propagator which do not bear any mark are equal to $C(t)$.

\end{theorem}

\begin{proof}[Proof of Theorem \ref{generalslice}]
We shall prove this theorem inductively. Assume that Theorem \ref{generalslice} holds up to $n-1$ slices, which means that $|\om_J|=n-1$, and the renormalized amplitude is written as:
\bea
&&A^{R, n-1}_G(t,\sigma)\\
&&=(-\lambda)^{(n-1)}\prod_{G^c}\Tr^{G^c}\{\prod_{\ell\in CP(G^c),\ \ell\ tadpole}[(R(\sigma)-1)\tilde C(\ell)]
\prod_{\ell\in CP(G^c), l\ not\  tadpole}[R(\sigma)\tilde C(\ell)]\}.\nonumber
\eea
Let $\oj\notin\Omega_J$ and consider
\bea
\int d\nu(\sigma)\int dt^{\oj}\frac{\partial}{\partial t^{\oj}}[ A_G^{n-1,R}\ e^{V(\sigma, t)}].
\eea

Without losing generality (the number of such contributions is estimated in Lemma \ref{comb}) we can write
$$\frac{\partial}{\partial t^{\oj}}\ A_G^{n-1,R}=\tr\ (\frac{\partial}{\partial t^{\oj}}[ \tilde R\tilde C])\circ\tr A_{G,{rest}}^{n-1,R}(t, \sigma),$$ 
where $\tilde R$ is a general name for the resolvent $R$ or $(R-1)$, which depend on $t$ via \eqref{res0} and \eqref{dp0}, $\tilde C$ is a general name for $C^\Lambda$ or a a marked propagator $C^{\og}$, $\og\in\Omega_J$, and $A_{G, {rest}}^{n-1,R}(\sigma)$ corresponds to the other terms in $A_G^{n-1,R}(\sigma)$ such that the derivation $\partial/\partial t^{\oj}$ doesn't act on.

Each intermediate field in the numerator in the form of $\sigma\otimes I$ or $I\otimes\sigma$ will act either on the vertex $V$ or the resolvent $R$ by integration by parts. 
Remark that the derivation of $\sigma\otimes I$ w.r.t. $I\otimes\sigma$ will generate 
a non-planar graphs from which we gain convergent factors. 

We have:
\bea\label{proof}
&&\int d\nu(\sigma)\int dt^{\oj}\frac{\partial}{\partial t^{\oj}}[\ A_G^{n-1,R}\ e^{V(\sigma, t)}]\\
&=&\int d\nu(\sigma)\int dt^{\oj}\  e^{V(\sigma, t)}\ \big\{\tr\ A_{G, {rest}}^{n-1,R}(\sigma)\circ\tr (-i\frac{\sqrt{2\lambda}}{2}RC^{\oj}\hat\sigma R\tilde C\ )]\nn\\
&&\quad\quad\quad\quad+\tr[\tilde R C^{\omega(j)}]\circ\tr A^{n-1,R}_{G,rest}(t,\sigma)\nonumber\\
&+&A_G^{n-1,R}\big(  -i\frac{\sqrt{2\lambda}}{4}\Tr\ [(R-1)C^{\oj}\hat\sigma ]+i\frac{\sqrt{2\lambda}}{4}\Tr\ [T^{\oj}\hat\sigma]+\lambda\Tr\  [T^{\oj}T(t)]\ \big)  \ \big\}.\nn
\eea
The matrix elements of $\hat\sigma$ in the numerator will act on all other terms that are functions of $\sigma$ as differential operators, due to the Gaussian integrals and integration by parts. We shall consider all the possibilities and prove that each tadpole generated in this process is canceled exactly with the corresponding counter term.

We start the analysis with the term in the second line of \eqref{proof}:
\begin{itemize}
\item if $\hat\sigma$ acts on the adjacent resolvent, by using formula \eqref{diffr} and $\tr (RC(t))=\tr (R-1)C(t)+ T(t)$, a tadpole term with amplitude 
$(-i\frac{\sqrt{2\lambda}}{2})\times (-i\frac{\sqrt{2\lambda}}{2}) T(t)\times 2=-\lambda\ T(t)$ will be generated;

\item if $\hat\sigma$ acts on any other resolvents or on the $\log_2[\cdots]$ term in the interaction $V$, the resulting terms contain no tadpoles nor counter terms hence
belong to $ A_G^{n,R}$;

\item if $\hat\sigma$ acts on the term $ i\frac{\sqrt{2\lambda}}{4}\tr\ \hat\sigma T(t)$ from $V(\sigma, t)$, then the term
$(-i\frac{\sqrt{2\lambda}}{2})\times (i\frac{\sqrt{2\lambda}}{4} T(t))\times 4=\lambda T(t) $ will be generated. This term cancels exactly with the tadpole generated above.

\end{itemize}
So we find that, after contracting the $\sigma$ fields in in the second line of \eqref{proof},
the resulting terms contain neither tadpoles nor counter terms, hence belong to $ A_G^{n,R}$.

Now we consider the first term in the third line. 
\begin{itemize}
\item
If $\hat\sigma$ acts on the adjacent resolvent, then we get 
\bea\label{rnmain}
&&\tr\ [-i\frac{\sqrt{2\lambda}}{4}\ RC^{\oj}]\circ\tr\ [-i\frac{\sqrt{2\lambda}}{2}RC ]\times 2\nonumber\\
&&=-\frac{\lambda}{2}\big[\ \tr\ [ (R-1)C^{\oj}]\circ\tr\ [ (R-1)C]+ \tr\ [ (R-1)C]\circ T^{\oj}\nonumber\\
&&\quad +\ \tr\ [ (R-1)C^{\oj}]\circ T(t)+ T^{\oj}\circ T(t)\ \big].
\eea
Recall that each tadpole term $T(t)$ or $T^{\oj}$ is an $(\Lambda+1)\times(\Lambda+1)$ dimensional matrix and we have $T^{\oj}\circ T(t)=\sum_{n}T^{\oj}_{nn} T(t)_{nn}$,
\item If $\hat\sigma$ acts on the counter term in $V(\sigma, t)$, then we get
$\frac{\lambda}{2}\tr\ [(R-1)C^{\omega_n}]\circ T$, which cancels exactly the third term in \eqref{rnmain},
\item If $\hat\sigma$ acts on $ A_G^{n-1,R}$ or on the $\log_2(\cdots)$ term in $V(\sigma, t)$, then
no tadpole nor counter term will be generated.
\end{itemize}

Now we consider the last two terms in \eqref{proof}.
\begin{itemize}
\item if $\hat\sigma$ acts on $\log_2(\cdots)$, then the term $$(-i\frac{\sqrt{2\lambda}}{4}\tr\ T^{\oj})\circ\  (i\frac{\sqrt{2\lambda}}{4}\tr\ [(R-1)C]\ )\times 4=\frac{\lambda}{2}\tr\ [ (R-1)C]\circ  T^{\oj}$$ will be generated, which cancels with the second term in \eqref{rnmain},
\item if $\hat\sigma$ acts on the counter term in $V(\sigma, t)$, then the term $-\frac{\lambda}{2}T^{\oj}T$ will be generated, which, together with the last term in \eqref{rnmain}, cancel with the last term in
\eqref{proof}. 
\end{itemize}

So all tadpoles generated in the slice-testing expansions cancel exactly with the counter-terms and $A^{R,n}$ contains only terms that are finite.
 So we have proved this Theorem.
\end{proof}


\begin{remark}
From Lemma \ref{comb} we find that if we perform the slice-testing expansions at very high order, we may generate a large combinatorial factor which could be unbounded. Hence we need to control the number of the marked propagators by controlling the order of the slice-testing expansions so that it generates just enough marked propagators for getting the convergent factors while not generates unbounded combinatorial factors. For the resolvent graphs which are not generated we simply put $c(G)=0$ while for the graphs generated in the slice-testing expansions we put $c(G)=O(1)$.
\end{remark}
So we have the following stopping rule for the slice-testing expansions:
\begin{definition}[The Stopping Rule]
Let $n_j\in\mathds{Z}_+$ be the number of marked propagators of coarse scale index $j\in [0, j_{max}]$ in the renormalized amplitude \eqref{gena} and $a>0$ be a constant to be decided later. The stopping rule for the slice-testing expansions is defined as follows: the slice-testing expansion is to be performed recursively starting from $j=j_{max}$ in the decreasing order of $j$ until $j_{min}$ and at each coarse scale $j$ one should continue the expansion until $n_j=aM^j$; Then one restart the expansions at scale $j-1$.
\end{definition}
\begin{remark}
The merit of the stopping rule is as follows. Let the set of refined scaling indices for the marked propagators at each coarse scaling index $j$ be $\bar\Omega_j$,then $|\bar\Omega_j|=aM^j$ marked propagators would be generated in the slice-testing expansion, from which we gain a convergent power-counting factor $(M^{-j})^{|\bar\Omega_j|}\sim O(1)e^{-ajM^j}$ (see Formula \eqref{cog}), which could compensate the constructive bound.
If we didn't impose the stopping rule, the total number of marked propagators could be as large as $aM^{2j}$ and the corresponding combinatorial factor would be unbounded. 
\end{remark}
It is easy to find that the ribbon graphs labeling the resulting expressions of the partition function of slice-testing expansions are not connected in general (see the second order slice-testing expansions shown in the Appendix.). A general graph $G$ can be decomposed as $G=\cup G^a$, where $G^a$ is a connected component of $G$ and $a$ runs over all connected components.

Define $\bar\Omega_J:=\cup_{j\in J}\bar\Omega_j\subset\Omega_J$, as the set of refined scaling indices of the marked propagators generated in the slice-testing expansions. Then we have: $\bar\Omega_J=\cup_{a}\bar\Omega^a$, where $\bar\Omega^a$ is a set of refined scaling indices for the marked propagators in $G^a$. Obviously, $\bar\Omega^a\cap\bar\Omega^{a'}=\emptyset$ for $a\neq a'$.

So the partition function also factorizes as product of different amplitudes, each of which is labeled by a connect graph $G^a$ with refined
indices set $\bar\Omega^a$. In order to explicit this decomposition we should further slice the integrand and the Gaussian measure of the partition function, respectively. These are the contents of Section $4.3$ and Section $5$.

\subsection{Slicing the integrand}
In order to factorize the slice-testing expanded partition function one should first of all factorize the exponent $e^V$ in Formula \eqref{partition} over the set of refined indices $\{\bar\Omega^a\}$ . It is useful to define a new index set $\om^J$, by listing all the elements of $\bar\om_J$ in the increasing order and labeling these indices by ordinal natural numbers as: $$\om^J:=\{\og_1, \og_2,\cdots, \og_h\ \vert\og_j>\og_i,\ \forall j>i, \forall\og_j\in\bar\om_J\},\ h=\vert \bar\om_J \vert\ .$$ Clearly $\om^J=\bar\om_J$ as sets. Let $\om_a$ the set of refined scaling indices attributed to each connected graph $G^a$, where $a$ runs over all connected components in $G$, we have: $\om^J=\cup_a\om_a$. Again the sets $\{\om_a\}$ are disjoint. Define also the index set $\Omega_{stop}$ to be the set of refined scaling indices labeling the marked propagators that are not generated in the slice-testing expansions. Clearly we have $\Omega^J\cap\Omega_{stop}=\emptyset$.

Now we factorize the interaction $V$ according to the index sets $\{\om_a\}$.

\subsubsection{Slicing the interaction}

In order to perform this factorization we shall attribute to each loop vertex the index of its highest marked propagator 
and to each interpolation parameter $t^\oj$ we introduce an additional auxiliary parameter $u^\oj $ which multiplies $t^\oj$.
We can consider the multiplication  $t^\oj(u^\oj)=u^\oj t^\oj$ as a substitution of each $t^\oj(u^\oj)$ by $u^\oj t^\oj$.

So the interaction with cutoff $\oj$ reads:
\bea
V(\{t\},\sigma)_{\le \oj}
&=&-\frac{1}{2}\Tr\ \log_2[\ 1+i\frac{\sqrt{2\lambda}}{2}
\sum_{j'=0}^j\sum_{\omega'(j')=M^{j'}}^{M^{j'+1}-1} C^{\omega'(j')}t^{\omega'(j')}(u^{\omega'(j')})\hat\sigma\ ]|_{u^{\omega'(j')}=1}\nonumber\\
&+&i\frac{\sqrt{2\lambda}}{4}\Tr\ [\ \sum_{j'=0}^{j}\sum_{\omega'(j')=M^{j'}}^{M^{j'+1}-1} T^{\omega'(j')}t^{\omega'(j')}(u^{\omega'(j')})\hat\sigma\ ] |_{u^{\omega'(j')}=1}\nonumber\\
&+&\frac{1}{2}\lambda\Tr\ [\ \sum_{j'=0}^{j}\sum_{\omega'(j')=M^{j'}}^{M^{j'+1}-1} T^{\omega'(j')}t^{\omega'(j')}(u^{\omega'(j')})\ ]^2|_{u^{\omega'(j')}=1}.
\eea


The specific part of the interaction attributed to the scaling index $\oj$ is the sum over all loop vertices with at least one marked propagator at scale $\oj$ and all other c-propagators at scales $\le \oj$. We shall abbreviate $t^\oj(u^\oj)$ as $t^\oj$ and we have:
\bea  V_{\oj} &:=&  V_{\le \oj}  - V_{\le \oj-1} =  V_{\le \oj} \vert_{u^\oj = 1} - V_{\le \oj} \vert_{u^\oj = 0}
\nonumber\\ &=& \int_0^1  du^\oj \frac{d}{du^\oj}\ \bigg\{\  \frac{1}{2}\lambda \sum_{j_1=0}^{j}\sum_{p\in I_{\le j_1}}(\sum_{\omega_1(j_1)=M^{j_1}}^{M^{j_1+1}-1}T_p^{\omega_1(j_1)}t^{\omega_1(j_1)}\ )^2\nonumber\\
&+& \Tr\   \biggl[\ i \frac{\sqrt{2\lambda}}{4} [\ \sum_{j_1=0}^{j}\sum_{\omega_1(j_1)=M^{j_1}}^{M^{j_1+1}-1}T^{\omega_1(j_1)}t^{\omega_1(j_1)}] \hat\sigma\nonumber\\ 
&-&\frac{1}{2}\log_2[1+i\frac{\sqrt{2\lambda}}{2} \sum_{j_1=0}^{j}\sum_{\omega_1(j_1)=M^{j_1}}^{M^{j_1+1}-1} C^{\omega_1(j_1)} t^{\omega_1(j_1)} \hat\sigma ]\  \biggr] \bigg\}\nonumber \\
&=&  \int_0^1 t^\oj du^\oj\ \bigg\{ \lambda\sum_j \sum_{p\in I_{\le j}}(\sum_{\omega_1(j_1)=M^{j_1}}^{M^{j_1+1}-1} T_p^{\omega_1(j)}t^{\omega_1(j)}\ )\
T_p^\oj t^\oj \nonumber\\
&+&\Tr  \big[\ i \frac{\sqrt{2\lambda}}{4} T^{\oj}t^\oj \hat\sigma 
 - i \frac{\sqrt{2\lambda}}{4}  [ R_{\le \oj}(t) -1 ]  C^{\omega(j)} t^\oj\hat\sigma\ \big]\ \bigg\}, 
\label{slice2}
\eea
where $t=t(u)$ and
\be R_{\le \oj}(t) = ( 1+i\frac{\sqrt{2\lambda}}{2} \sum_{j_1=0}^{j}\sum_{\omega_1(j_1)=M^{j_1}}^{M^{j_1+1}-1} C^{\omega_1(j_1)} t^{\omega_1(j_1)} \hat\sigma\ )^{-1}.\label{rc1}
\ee

Since $\bar\om_J=\om^J$ as sets, for each $\oj\in\bar\om_J$ there exist a unique $\og_j\in\om^J$ such that $\oj=\og_j$. So we can define the sliced interaction as $V_{\og_j}:=V_{\oj}$ and $
R_{\le\og_j}= ( 1+i\frac{\sqrt{2\lambda}}{2} \sum_{j_1=0}^j\sum_{\og=M^{j_1}}^{M^{j_1+1}-1} C^{\omega} t^{\omega} \hat\sigma\ )^{-1}$. Remark that since $t^{\og_j}=0$ for $\og_j\notin\om^J$, we have the following decomposition for the interaction term: 
\be V(\sigma, t)=\sum_{j =0}^{j_{\max}}\sum_{\oj\in\bar\Omega_j} V_\oj(\sigma, t)
=\sum_{\omega\in \Omega^J}V_{\omega}.
\ee
\subsubsection{Slicing the integrand}
Now each graph $G\in\cG(\Omega^J)$ is partitioned into the non-empty connected components $G_1,\cdots,G_n$ of $G$ with the corresponding index set of marked propagators $\Omega_a,\cdots,\Omega_n$, for which we have $\Omega_a\in\Omega^J$ and $\Omega_a\cap\Omega_b=\emptyset$ for $a\neq b$. The last property is also called the $hard\ core\ constraint$. Remark that the combinatorial weights of Feynman graph also factorize over connected components, namely
$c_G = \prod_a c_{G_a}$. 

We can write the result of the slice-testing expansion as:
\be  Z^{j_{max}}(\lambda)= \sum_{n=0}^{\infty}  \frac{1}{n!}   \int d\nu  (\sigma)   \sum_{\substack{\Om_1, \cdots , \Om_n;\Om_a \cap \Om_b= \emptyset, \ \forall a\neq b,\\ \Omega_a\cap\Omega_{stop}=\emptyset}} 
\bigl[ \prod_{a =1}^n \prod_{\omega\in \Om_a}  \int_0^1 dt^\omega  \bigr] \prod_{a=1}^n {\cI} (\om_a, \{t\}, \sigma), \label{testing21} 
\ee 
where
\be  {\cI} (\om_a, t, \sigma) = \sum_{G_a \in \cC\cG (\om_a) } c_{G_a}\bigl(  \prod_{\omega\in \om_a} [  e^{V_\omega(\{t\},\sigma) }]
A^R_{G_\alpha}(t,\sigma) \bigr) \label{testing22} ,
\ee
is the amplitude for the set of \emph{connected} graphs $\cC\cG (\om_a)\subset\cG (\om)$. The $1/n!$ in \eqref{testing21} comes from summing over the sequences $\om_1, \cdots ,  \om_n$ in \eqref{testing21}. The sum over $n$ is in fact finite due to the hardcore constraint, which forces all terms to be zero for $n\ge M^{j_{max}}+1$. $n=0$ corresponds to the factor 1 in the sum (the normalization of the free theory). 
We would like to factorize now the sum over the blocks $\Omega_a$. The obstacle is that the functions $\cI$ are still coupled through the integration w.r.t. the $\sigma$ variables, the hardcore constraints and the parameters $\{t\}$, which don't factorize the blocks $\Omega_a$. We shall discuss the first two issues in the next section and in the rest of this section we shall only focus on the factorization of the variables $\{t\}$. 

Let $\vec t_a$ be the list of $|\Omega_a|$ parameters $t^{\omega_j}$ for $\omega_j\in\Omega_a$ so that we can write the integration of $\cI$ w.r.t. $t$ as
\be\label{int1}
\int d\vec t_1\cdots d\vec t_n\ \prod_{a=1}^n{\cI_a}(\vec t_1,\cdots, \vec t_n; \Omega_a, \sigma,).
\ee
For each pair $\vec t_a$, $\vec t_b$ we introduce a pair interpolation parameters $x_{a,b}$
to simultaneously multiply the $\vec t_\bt$ dependence of ${\cI}_{a}$ and the $\vec t_a$ dependence of ${\cI}_{b}$. Then \eqref{int1} can be written as
\be\label{int2}
\int d\vec t_1\cdots d\vec t_n\ \prod_{a=1}^n{\cI_a}(x_{1,a}\vec t_1,\cdots,x_{a-1,a}\vec t_{a-1},\vec t_a,
x_{a,a+1}\vec t_{a+1},\cdots, x_{a,n}\vec t_n; \Omega_a, \sigma)\vert_{x_{a,b}=1\ \forall a,b=1,\cdots,n}.
\ee
The canonical way of factorizing the partition function into disconnected parts labeled by
forests is the standard BKAR forest formula \cite{BK, AR1}, which is a Taylor expansion formula with remainders:
\begin{theorem}[The BKAR Forest Formula.]\label{ar1}
Let $n\ge1$ be an integer, $I_N=\{1,\cdots, n\}$ be an index set and $\ell=(i,j), i\neq j$ be an unordered pair of $I_n$, also called a link on $I_n$. Let $\cP_n=\{\ell=(i,j), i, j\in I_n, i\neq j\}$ be the set of links of $I_n$, $\cF$ be the set of spanning forests over $I_n$ and $\cS$ be the set of smooth functions from $\mathds{R}^{\cP_n}$ to an arbitrary Banach space. Let ${\bf x}=(x_\ell)_{\ell\in\cP_n}$ be an arbitrary element of $\mathds{R}^{\cP_n}$ and $f\in \cS$, we have
\be
f({\bf 1})=\sum_{\cF}\Big(\int_0^1\prod_{\ell\in\cF} dw_\ell\Big)\Big(\prod_{\ell\in\cF}\frac{\partial}{\partial {x_\ell}}\ \Big)f[X^\cF(w_\ell)],
\ee
where the sum over $\cF$ runs over all forests with $n$ vertices, including the empty forest with no edge, ${\bf 1}\in \mathds{R}^{\cP_n}$ is the vector with every entry equals $1$, $X^\cF(w_\ell)$ is the vector ${(x_\ell)}_{\ell\in \cP_n}$ defined by $x_\ell= x_{ij}^\cF(w_\ell)$, the value at which we evaluate the derivations on $f$ and is defined as follows: $x_{ij}^\cF=1$ if $i=j$, $x_{ij}^\cF=0$ if $i$ and $j$ are not connected by a forest and $x_{ij}^\cF=\inf_{\ell\in P^{\cF}_{ij}}w_\ell$, where $P^{\cF}_{ij}$ is the unique path in the forest that connects $i$ and $j$. 
\end{theorem}

Then we perform the BKAR formula  w.r.t. the parameters $x_{a,b}$ and the above formula can be written as:
\bea\label{int3}
&&\sum_{\cF}\int d\vec t_1\cdots d\vec t_n\prod_{\ell=(a,b)\in\cF}\int_0^1 dw_{a,b} \prod_{\ell=(a,b)\in\cF}\frac{\partial}{\partial x_{a,b}}\\
&&\quad\prod_{a=1}^n{\cI_a}(x_{1,a}\vec t_1,\cdots,x_{a-1,a}\vec t_{a-1},\vec t_a,
x_{a,a+1}\vec t_{a+1},\cdots, x_{a,n}\vec t_n; \Omega_\al, \sigma)\vert_{x_{a,\bt}=X^{\cF}_{a,b}(\{w\})\ \forall a,b}\ .\nn
\eea
Consider a derivation $\frac{\partial}{\partial x_{a,b}}$ on $\cI_{a}$ for arbitrary $a,b$, we have 
\bea
&&\frac{\partial}{\partial x_{a,b}} I_a(\cdots,\vec t_a, \cdots, x_{a,b}\vec t_b)\nn\\
&&
=\vec t_b\cdot \frac{\partial}{\partial \vec t'_{b}} I_a(\cdots,\vec t_a, \cdots, \vec t'_b):=\sum_{i=1}^{|\Omega_b|}  t_{b_i} \frac{\partial}{\partial t'_{b_i}} I_a(\cdots,\vec t_a, \cdots, \vec t'_b)\label{aux2},
\eea
where $\vec t_b=( t_{b_1},\cdots, t_{b_{|\Omega_b|}})$ is the list of the interpolation parameters in $\vec t_b$ and
$\vec t'_\bt$ is the set defined by multiplying each member of the above set by $x_{a,b}$. The analysis is very similar to the slice-testing expansions (cf. \eqref{st0} and \eqref{partition}) so we don't repeat the explicit calculations here but give some general remarks. First of all it is easy to find that $|\Omega_b|$ marked propagators are attached to the graph $G_a$, as each derivation $\frac{\partial}{\partial t'_{b_i}} $ generates one marked propagator to $\cI_a$ in two ways: when it acts on the resolvents amplitude $A^R$ then a propagator $C^{\omega_{b_i}}$ together with a resolvent $R$ and a $c$-propagator $C(t)$
will be generated to $G_a$; when the derivation acts on the exponential, it will bring a new marked propagator $C^{\omega_{b_i}}$. The index set of the marked propagators is $\Omega_a\cup\Omega_b$.
Secondly, the graphs generated from the auxiliary expansions are not minimal in general, as a derivation in \eqref{aux2} could generate a connected amplitude such that two marked propagators bear the same indices.
One important difference from the slice-testing expansion is that the linear terms of the tadpole $T^{\omega_a}$ are not generated by the derivation, due to the fact that all the parameters $x_{a,b}$ associated with $T^{\omega_a}$ are equal to one.
The newly generated tadpoles are of the form $T^{\omega_a}T^{\omega_b}$. Since each of the newly generated tadpole is bounded by $\frac{1}{\omega_a \omega_b}$, they are not dangerous for the bound and can be bounded easily by the convergent factors 
gathered from the marked propagators. Another important difference from the slice-testing expansion is that each derivation $\frac{\partial}{\partial t_{b_i}}$ generate a term $RC^{\omega_{b_i}}\sigma RC^{\omega_a}\sigma$ from the exponential, which contains two $\sigma$ variables. We shall integrate out the $\sigma$ variables w.r.t. the Gaussian measure and this will generate combinatorial factors. But this is not dangerous, as each $\sigma$ variable is associated with an additional marked propagator $C^\omega$, which can be used to compensated these combinatorial factors (See Formula \eqref{boseg2} and Section 6.4 for the details.). 

The result of \eqref{int3} is a perturbation series indexed by forests $\cF$. Each connected component of $\cF$, noted by $\cT$, which is a tree, is indexed by a larger block of index sets $\Omega_\al^{\cM}=\Omega_{a_1}\cup\cdots\cup\Omega_{a_p}$ composed of $p$ previous blocks. We also call $\Omega_\al^{\cM}$ the modified index sets. In this way each vertex of a tree $\cT$ corresponds to a term $W_a$ and $\cT$ is a tree with $p$
vertices. Denoting the corresponding connected graphs by $\cC\cG^{\cM}$, the partition function result can be written as
\be  Z^{j_{max}}(\lambda)= \sum_{n=0}^{\infty}  \frac{1}{n!}   \int d\nu  (\sigma)   \sum_{\substack{\Om^{\cM}_1, \cdots , \Om^{\cM}_n;\Om^{\cM}_\alpha \cap \Om^{\cM}_\beta= \emptyset, \ \forall \alpha\neq\beta,\\ \Omega^{\cM}_\alpha\cap\Omega_{stop}=\emptyset}} 
 \prod_{\alpha =1}^n {\cI^{\cM}} (\om^{\cM}_\alpha, \sigma) ,
\ee 
where
\be  {\cI^{\cM}} (\om^{\cM}_\al, \sigma) =\sum_{\Pi}\sum_{\cT}\int dt_{\Omega^{\cM}}\int dw_\cT \sum_{G \in \cC\cG^{\cM} (\om^{\cM},\Pi,\cT )} c_{G}\bigl(  \prod_{\omega\in \om_\al^{\cM}} [e^{V_\omega(t,w, \sigma) }]
A^{R,M}_{G_\alpha}(t, w, \sigma) \bigr) ,
\ee
is the modified amplitude for the corresponding modified graph $\cC\cG^{\cM}$ resulted from the derivations w.r.t. $x_{\al,\bt}$, $\Pi$ is a partition of $\om_\al^{\cM}$ into disjoint unions $\om_{a_1}\cup\cdots\cup\om_{a_p}$; $A^{R,M}_{G_\alpha}(t,\sigma)$ is called the modified resolvent amplitude and has very similar expression than $A^R_{G^{\cM}}$, except that it contains now the quadratic terms of tadpoles;
The $w$ dependence means that all $t$ factors are multiplied by the weakening parameters $x^{\cT}_{a,b}(\{w\})$. 

Now we consider the factorization of the integration for the $\sigma$ variables and get rid of the hardcore constraints.
First of all recall that for a pair of Grassmann variables $\chi, \bar\chi$ one has
\be\label{grasr}
\chi^2=0=\bar\chi^2,\ \ \int d\chi=0=\int d\bar\chi,\ \int \chi d\chi=1=\int\bar\chi d\bar\chi.\ee

We can express the mutually disjointness of the index sets $\{\om_a\}$ in Formula \eqref{testing21} in terms of Grassmann integrals: 
\bea
\sum _{\om_\al,\om_\al\cap\om_\bt=\emptyset,\ \cup_\al \om_\al=\ I}\ 1=\int\prod_{\og\in I} d\bar\chi_\og d\chi_\og\ \sum_{\om_\al,\cup\om_\al=I}\ \prod_{\og\in\om_\al, \cup\om_\al=I}\chi_\og\bar\chi_\og,
\eea
since, if $\om^{\cM}_\al\cap\om^{\cM}_\bt\neq\emptyset$, the terms like $\chi_\rho^2$ or $\bar\chi_\rho^2$ would appear in the integrand for some $\rho\in\om^{\cM}_a\cap\om^{\cM}_\bt$, which are vanishing identically.

Using the the fact that $V=\int d\bar\chi d\chi e^{-\bar\chi V \chi}$, for any $V$ that is not a function of $\chi$ and $\bar\chi$, we can write the partition function as:
\bea
&&Z^{j_{max}}(\lambda) = \int d\nu  (\sigma)  d \mu_{\mathbb{I}_{\cS}}(\bar \chi ,\chi )   e^{ W}, \label{testing4}\\
&&W = \sum_{\om^{\cM} \subseteq \om^J}\sum_{\Pi}\sum_\cT\int dt_{\Omega^{\cM}}\int dw_\cT \sum_{G \in {\cC\cG^{\cM}}(\om^{\cM},\Pi,\cT) } 
 c_{G} \biggl(  \prod_{\omega\in\om} [ - \bar \chi_\omega  e^{V_\omega(t,w,\sigma) }  \chi_\omega ]
A^{R,M}_{G}(t,w,\sigma) \biggr)\label{testing5},\nn\\
\eea
where $d \mu_{\mathbb{I}_{I_\Lambda}}(\bar \chi ,\chi ) = \prod_{\omega=1}^{\omega_{max}} d\bar \chi_\omega d\chi_\omega \; e^{-\bar \chi_\omega \chi_\omega}$ is the standard normalized Grassmann 
Gaussian measure with covariance $\mathbb{I}_{\cS}$, which is the $(\Lambda+1)\times(\Lambda+1)$ dimensional identity matrix. The Grassmann integration implement automatically the hardcore constraints, and saturate the Grassmann pairs for $\omega_i\notin\om^{\cM}=\cup_\al\om^{\cM}_\al$.
Since $W^{\cM}$ contains potentially infinitely many vertices of type $V$, it is called an {\emph{exp-vertex}} \cite{RivW3}. 


\section{The Multiscale Loop Vertex Expansion, step four: the two level jungle formula}\label{multilve}
After this long preparation we shall compute in this section the vacuum correlation function $\log Z^{j_{max}}$, by first expand the partition function $Z^{j_{max}}$ into convergent perturbation series indexed by disconnected resolvent graphs. Then the perturbation series of $\log Z^{j_{max}}$ simply corresponds to the terms indexed by connected graphs.
It is well known that the fully expanded perturbation series labeled by Feynman graphs is divergent, due to the fact that Feynman graphs proliferate too fast, namely, the number of Feynman graphs of order $n$ in $\phi^4$ theory is proportional to $(2n)!\sim(n!)^2$, while the number of labeled trees with $n$ vertices is bounded by $O(n!)$, by Cayley's theorem. One doesn't really need to know the full information of the Feynman graphs to compute the partition function or the connected function; Information about connectedness properties would be enough. 
Since there are two different kinds of Gaussian integrations in the partition function \eqref{testing4}, one Bosonic and the other Fermionic, we shall expand the partition function as a convergent perturbation series such that each term is labeled by a two-level forest $\cJ_2=(\cF_1, \cF_2)$, which is a two-member sequence of forests such that $\cF_1$ are the Bosonic forests and $\cF_2$ are the $final$ forests formed by connecting certain trees in $\cF_1$ with $Fermonic$ lines. Clearly we have $\cF_1\subseteq\cF_2$. The canonical way of generating two-level forests is the two-level BKAR forest formula \cite{AR1},\cite{MLVE}, which can be obtained by using twice the
the one level BKAR forest formula (cf. Theorem \ref{ar1}), one for the Bosonic variables and another for the Fermionic variables. 

Now we derive the two-level forest formula for the partition function in \eqref{testing4}. Our derivation follows closely \cite{MLVE} and the interested reader is advised to look at \cite{MLVE} for the other aspects of the two-level forest formula. 

The first step is an analogue of the cluster expansions in statistical mechanics, which expand the partition function into a (new) power series that are indexed by forests. In order to do that we shall use the replica trick for the $\sigma$ variables, namely we assign different labels $\sigma_\al, \sigma_\bt, \cdots$ to the $\sigma$ variables in different vertices $W$ and we write $W(\sigma)^n=\prod_{a=1}^n W(\sigma_a)$.

Let $\cW = \{1, \cdots , n\}$ is the set of labels for the exp-vertices, we can write the partition function as:
\be\label{bof}
Z^{j_{max}}(\lambda)= \sum_{n=0}^\infty \frac{1}{n!}\int d\nu_{\cW}(\sigma, \bar \chi, \chi) \;  \prod_{\al=1}^n  W_\al  ( \sigma_\al, \bar \chi, \chi ) \; ,
\ee
where
\bea
W_\al (\sigma_\al, \bar \chi, \chi )=
\sum_{\om^{\cM} \subseteq \om^J}\sum_{\Pi}\sum_{\cT} \int dt_{\om^{\cM}}\int dw_{\cT}\sum_{G \in \cC\cG^{\cM}(\om^{\cM})} 
 c_{G} \biggl(  \prod_{\omega\in \om^{\cM}} [ - \bar \chi_\omega  e^{V_\omega(t,\sigma_\al) }  \chi_\omega ]
A^{R,M}_{G}(t,\sigma_\al) \biggr),\nn\\
\eea
and
\be
d\nu_{\cW}(\sigma,\bar\chi, \chi) = d\nu_{\bbone_\cW} (\{\sigma_\al\}) \;  d\mu_{\mathbb{I}_\cS  } (\bar \chi_\omega, \chi_\omega),
\ee
in which $\bbone_\cW$ is a $n$ by $n$ matrix with coefficients $1$ everywhere, reflecting the fact that the replicated Bosonic measure is completely degenerate, namely the Gaussian measure over the fields $\{\sigma_\al\}$ has covariance $\langle \sigma_{\al,mn},\sigma_{\bt,kl}\rangle=\delta_{ml}\delta_{nk}$. In order to factorize the block  $\bbone_\cW$ in the measure $d\nu$ we introduce the coupling parameters $x_{\al,\bt}=x_{\bt,\al}$, $x_{\al,\al}=1$, between the Bosonic vertex replicas.

It is useful two write the Gaussian measure as differential operators by using 
\be
\int d\mu_C(\phi) f(\phi)=e^{\frac{1}{2}\frac{\partial}{\partial\phi} C \frac{\partial}{\partial\phi}}\ f(\phi)\ \vert_{\phi=0},
\ee
where $d\mu_C(\phi)$ is the Gaussian measure with covariance $C$, then we can rewrite the partition function as:
\be
Z^{j_{max}}(\lambda) = \sum_{n=0}^\infty \frac{1}{n!} 
\Bigl[ e^{\frac{1}{2} \sum_{\al,\bt=1}^n x_{\al,\bt} 
\frac{\partial}{\partial \sigma_\al}\frac{\partial}{\partial \sigma_\bt} 
   + \sum_{\omega\in\om^J} \frac{\partial}{\partial \bar \chi_\og  } \frac{\partial}{\partial \chi_\og } } \; 
   \prod_{\al=1}^n W_\al  (\sigma_\al, \bar \chi, \chi )
   \Bigr]_{ \genfrac{}{}{0pt}{}{  \sigma, \chi,  \bar\chi  =0}{x_{\al,\bt}=1 }} \;.
\ee

\begin{figure}[!htb]
\centering
\includegraphics[scale=0.4]{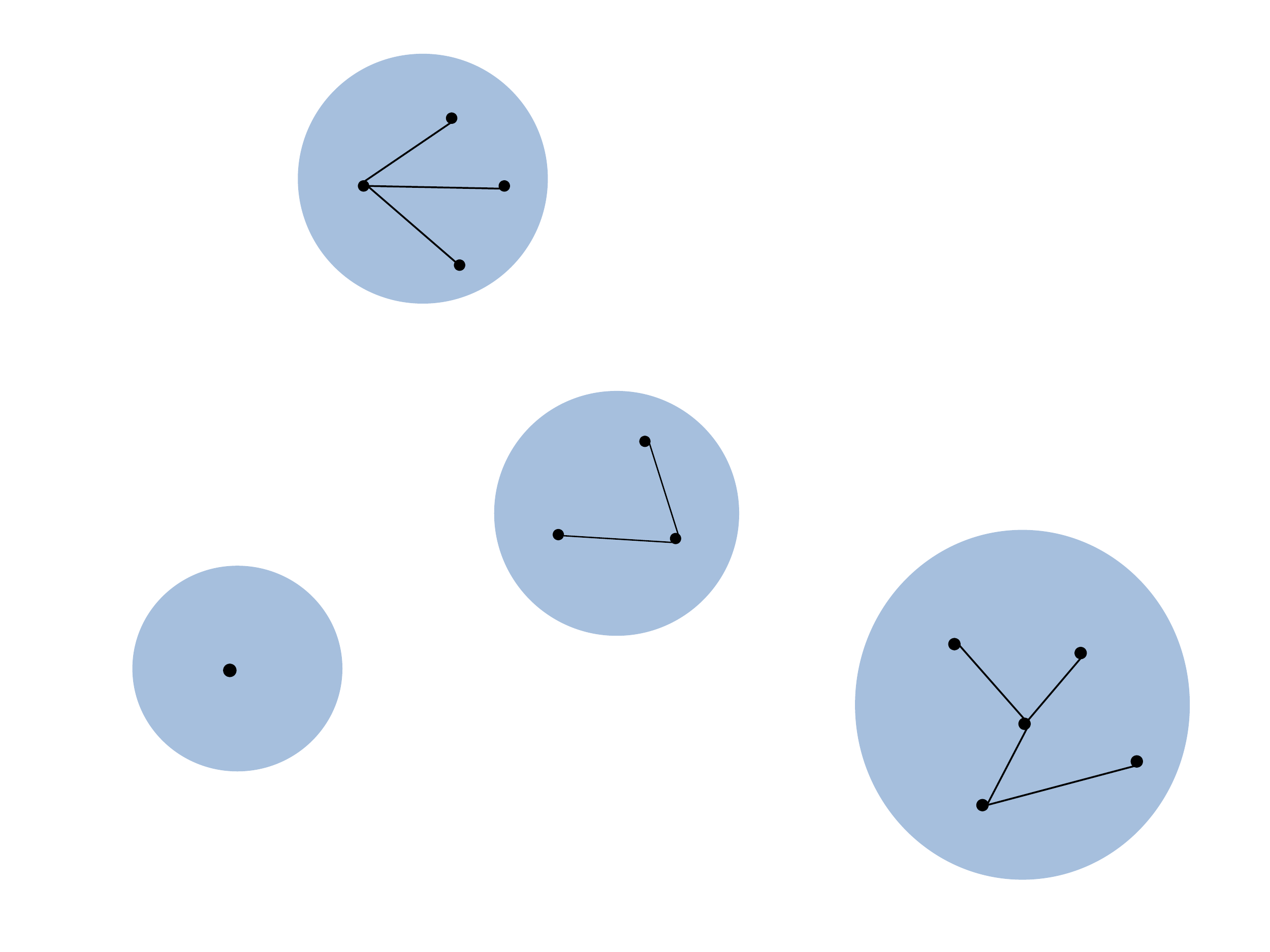}
\caption{A Bosonic forest. The configuration of the Blocks are trees, whose vertices correspond to the modified vertex $\{W_\al\}$.} \label{bosef}
\end{figure}

Then we apply the standard BKAR forest formula w.r.t. the parameters $\{x_{\al,\bt}\}$ and we have:
\bea\label{bosint0}
&&Z^{j_{max}}(\lambda) =  \sum_{n=0}^\infty \frac{1}{n!} \sum_{\cF_{B}} \int_{0}^1 \Bigl( \prod_{\ell_B\in \cF_B } dw_{\ell_B} \Bigr)\int \prod_{v\in \cV_{\cF_{B}}} d t_v\cdot
\\
&&\quad\cdot \Bigg[  e^{\frac{1}{2} \sum_{\al,\bt=1}^n X_{\al,\bt}(w_{\ell_B})  
   \frac{\partial}{\partial \sigma_{\al(\ell_B)}}\frac{\partial}{\partial \sigma_{\bt(\ell_B)}}
   +  \sum_{\al\in \cF_\cB}\sum_{\omega\in\om^{\cM}_\al} \frac{\partial}{\partial \bar \chi_\og} \frac{\partial}{\partial \chi_\og} }\nonumber\\
&&\quad
   \prod_{\ell_B=(\al,\bt) \in \cF_{\cB}} \Bigl(  
   \frac{\partial}{\partial \sigma_{\al(\ell_B)}}\frac{\partial}{\partial \sigma_{\bt(\ell_B)}}\   \Bigr)\ \; 
 \prod_{\al=1}^n  W_\al  (\sigma_{\al}, \bar \chi, \chi)\Bigg]_{ \sigma ,\chi , \bar\chi =0,\ x_{\al,\bt}
=X_{\al,\bt}(w_{\ell_B}) } \; ,\nonumber
\eea
where
\begin{itemize}

\item $\cF_{B}$ is a forest, called the Bosonic forest, with $n$ vertices such that each vertex corresponds to a term $W_\al$, $\al=\{1,\dots n\}$; the summation in the first line is over all Bosonic forests.

\item for each $\ell_{B}\in\cF_\cB$, $\al(\ell_\cB), \bt(\ell_\cB)$ are denoted as the ends vertices of $\ell_\cB$ and the weakening parameters $X_{\al\bt}(w_{\ell_\cB})$ are similarly defined as in Theorem \ref{ar1}, except that each vertex of the forest corresponds to a modified amplitude term $W_\al$, instead of $W_a$. 
\end{itemize}
The corresponding graph is shown in Figure \ref{bosef}, in which each block represents a connected graph.

Remark that an analogue of this formula has been discussed in \cite{MLVE} for the construction of a vector model, which is simpler in that renormalization is not needed.

The Gaussian integrations over $\sigma$ are still not completely factorized in that the variables $\sigma$ are coupled with 
the Grassmann variables $\bar\chi, \chi$, which are not factorized. So we repeat the procedure in the first step: first of all we introduce the replica Fermionic fields $\chi^{\cB}_\og$ for the blocks in $\cF_B$  and the coupling parameters $y_{\cB\cB'}=y_{\cB'\cB}$ for each pair of blocks $\cB, \cB'\in\cF_\cB$. 
The last step applies (once again) the forest formula, 
this time for the coupling parameters $y_{\cB,\cB'}$ and generates a set of forests $\cF_F$. 
Let $L_{F}$ be a generic Fermionic edge in the Fermionic forest, which connects two blocks and let $\cB(L_F), \cB'(L_F) $ 
the end blocks of $L_F$, we have:
\bea
&&  Z^{j_{max}}(\lambda) = \sum_{n=0}^\infty \frac{1}{n!} \sum_{\cF_{B}}\sum_{\cF_F} \int_{0}^1 \Bigl( \prod_{\ell_B\in \cF_B } dw_{\ell_B}\prod_{L_F\in\cF_F} dw_{L_F}\Bigr)\nonumber\\
&&\cdot  \Bigg\{ e^{\frac{1}{2} \sum_{\al,\bt=1}^n X_{\al,\bt}(w_{\ell_B})  
   \frac{\partial}{\partial \sigma_{\al(\ell_B)}}\frac{\partial}{\partial \sigma_{\bt(\ell_B)}}
   +  \sum_{\cB,\cB'}Y_{\cB\cB'}(w_{L_F})\sum_{\al\in \cF_\cB}\sum_{\omega\in\om^J} 
\frac{\partial}{\partial \bar \chi^\cB_\og} 
\frac{\partial}{\partial \chi^{\cB'}_\og} }\nonumber\\
&&\quad\quad\quad\cdot\prod_{\ell_B=(\al,\bt) \in \cF_{B}} \Bigl(  
   \frac{\partial}{\partial \sigma_{\al(\ell_B)}}
\frac{\partial}{\partial \sigma_{\bt(\ell_B)}}\ \Bigr)\ \nn\\
&&\cdot\prod_{L_F\in\cF_F}\Bigl(\sum_{\al\in \cF_\cB}
\sum_{\omega\in\om^{\cM}_\al}\ \big[\ \frac{\partial}{\partial \bar \chi_\og^{\cB(L_F)}} 
\frac{\partial}{\partial \chi^{\cB'(L_F)}_\og}  
+\frac{\partial}{\partial \bar \chi^{\cB'(L_F)}_\og} 
\frac{\partial}{\partial \chi_\og^{\cB(L_F)}}\ \big]\ \Bigr)\cdot\nonumber\\
&&\quad\quad\quad\ \quad\cdot\prod_{\al=1}^n  W_\al  (\sigma_\al, \bar \chi, \chi)
\Bigg\}_{ \sigma ,\chi , \bar\chi =0 } \; .
\eea
Organizing the Fermionic forests and Bosonic forests into two-level jungles and we obtain the two-level jungle formula:
\be\label{twol}
Z^{j_{max}}(\lambda)  =  \sum_{n=0}^\infty \frac{1}{n!}  \sum_{\cJ} 
 \;  \int dw_\cJ  \;  \int d\nu_{ \cJ} 
\ \partial_\cJ   \Big[ \prod_{\cB} \prod_{\al\in \cB} W_{\al}   (   \sigma_\al , \chi^{ \cB } , \bar \chi^{\cB}  )
  \Big] \; ,
\ee
where
\begin{itemize}
\item $\cJ$ is a two level jungle, which is defined as an ordered pair $\cJ = (\cF_\cB, \cF_F)$ of two (each possibly empty) disjoint forests on $\cW$, such that $\cF_B$ is a forest over the vertices $a\in \cV_{\cF_\cB}$, $\cF_F$ is a forest over the connected components of $\cF_\cB$  and
the union $\bar \cJ= \cF_B \cup \cF_F $ is still a forest on $\cW$. The forests $\cF_B$ and $\cF_F$ are called the Bosonic and Fermionic components of $\cJ$, respectively. The sum runs over all two-level jungles $\cJ$.
 
\item  $\int dw_\cJ$ means integrals over the parameters $w_\ell$, one for each edge $\ell \in \bar\cJ$, namely
$\int dw_\cJ  = \prod_{\ell\in \bar \cJ}  \int_0^1 dw_\ell  $.
There is no integration for the empty forest since by convention an empty product is 1. A generic variable $w_\cJ$
is therefore made of $\vert \bar \cJ \vert$ parameters $w_\ell \in [0,1]$, one for each $\ell \in \bar \cJ$.

\item 
\be \partial_\cJ  = \prod_{\genfrac{}{}{0pt}{}{\ell_B \in \cF_B}{\ell_B=(\al,\bt)}} \Bigl(
\frac{\partial}{\partial \sigma_\al}\frac{\partial}{\partial \sigma_\bt} \Bigr)
\prod_{\genfrac{}{}{0pt}{}{\ell_F \in \cF_F}{\ell_F=(\al',\bt') } } \sum_{\omega_{\ell_F} =0}^{\omega_{max}} \Big(
\frac{\partial}{\partial \bar \chi^{\cB(\al')}_{\omega_{\ell_F} } }\frac{\partial}{\partial \chi^{\cB(\bt')}_{\omega_{\ell_F} }}+  \frac{\partial}{\partial \bar \chi^{ \cB(\bt') }_{\omega_{\ell_F} } } \frac{\partial}{\partial\chi^{\cB(\al')}_{\omega_{\ell_F}  }} \Big) \; 
\ee
generate both Boonic lines and Fermionic lines. Remark that a Fermionic line $\ell_F$ connects
two Bosonic blocks $\cB$ and $\cB'$ on the vertices $\al'\in\cB$, $\bt'\in\cB'$.

\item The measure $d\nu_{\cJ}$ has covariance $ X (w_{\ell_B}) $ on Bosonic variables and $ Y (w_{\ell_F}) \otimes \mathbb{I}_\cS  $  
on Fermionic variables:
\be
\int d\nu_{\cJ} F = \biggl[e^{\frac{1}{2} \sum_{\al,\bt=1}^n X_{\al,\bt}(w_{\ell_B })  \frac{\partial}{\partial \sigma_\al}\frac{\partial}{\partial \sigma_\bt} 
   +  \sum_{\cB,\cB'} Y_{\cB\cB'}(w_{\ell_F})  \sum_{\omega \in \om^J}
   \frac{\partial}{\partial \bar \chi_{\omega}^{\cB} } \frac{\partial}{\partial \chi_{\omega}^{\cB'} } }   F \biggr]_{\sigma = \bar\chi =\chi =0}\; .
\ee

\item  $X_{\al,\bt} (w_{\ell_B} )$  is the infimum of the $w_{\ell_B}$ parameters for all the Bosonic edges $\ell_B$
in the unique path $P^{\cF_B}_{\al \to \bt}$ from $a$ to $b$ in $\cF_B$. This infimum is set to zero if such a path does not exists and 
to $1$ if $\al=\bt$. 

\item  $Y_{\cB\cB'}(w_{\ell_F})$ is the infimum of the parameters $w_{\ell_F}$ for all the Fermionic
edges $\ell_F$ in any of the paths $P^{\cF_B \cup \cF_F}_{\al\to \bt}$ from some vertex $\al\in \cB$ to some vertex $\bt\in \cB'$. 
This infimum is $0$ if there are no such paths, is equal to $1$ if such paths exist but do not contain any Fermionic edges.
\end{itemize}

\begin{figure}[!htb]
\centering
\includegraphics[scale=0.4]{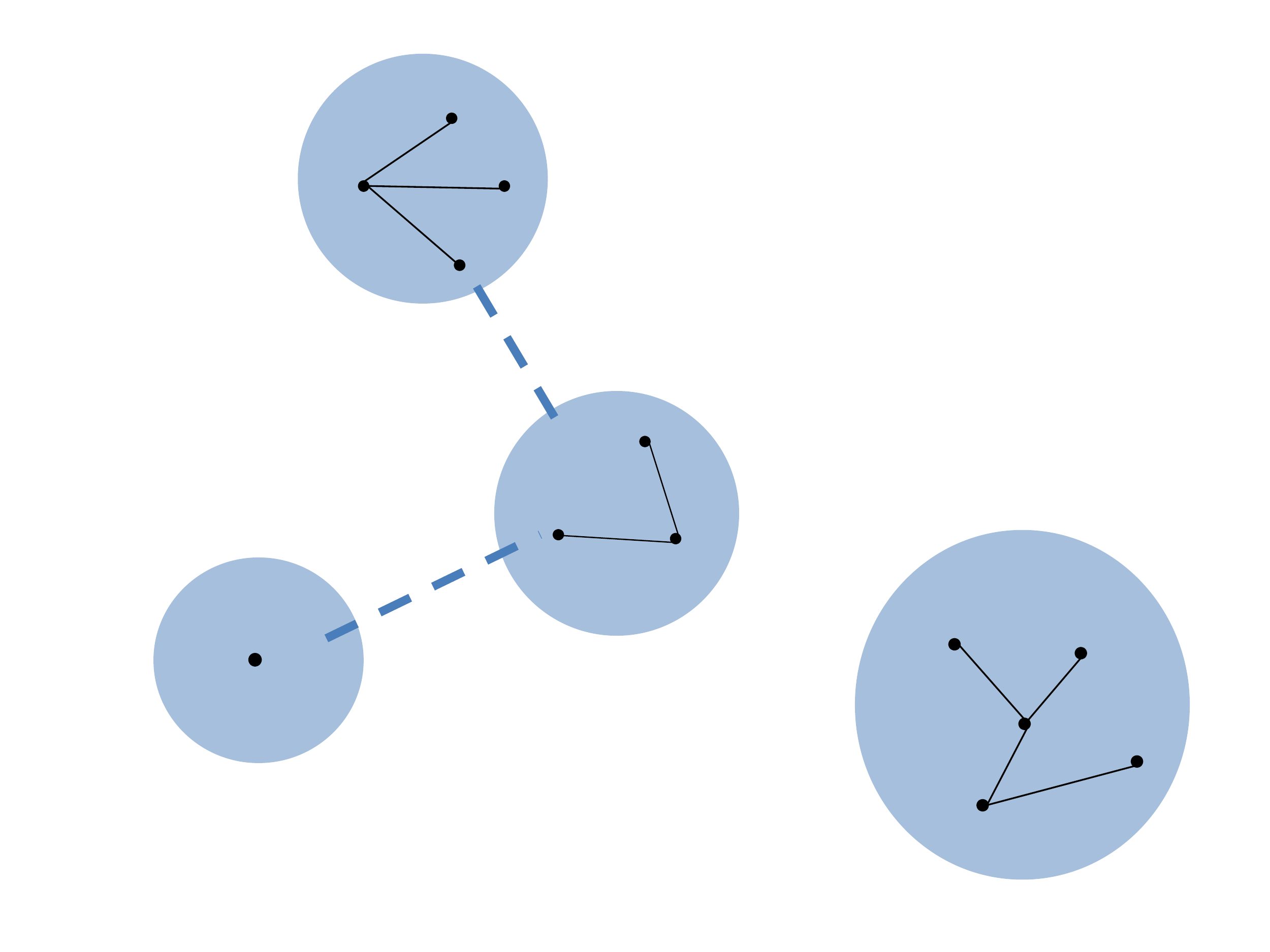}
\caption{A Fermionic forest, which is made of two trees.}\label{twotree}
\end{figure}

The resulting graph are two-level forests made of two level trees $\cJ^\cT$, which is a connected graph with both Bosonic lines and Fermionic lines such that the vertices together with the Bosonic lines form the Bosonic blocs.
A two level tree made of one single Bosonic bloc (without Fermionic or Bosonic edges) is allowed. Figure \ref{twotree} shows an example of a MLVE forest made of two trees.
Recall the following Lemma about the combinatorial properties of the two level trees. 

\begin{lemma}\label{labeled}
The number of labeled two-level trees over $n\ge1$ vertices is equal to $2^{n-1}n^{n-2}$.
\end{lemma} 
\begin{proof}
This is a well known result in graph theory and can be generalized to the case of arbitrary labeled $m$-level trees, $m\ge1$, over $n\ge1$ vertices: the number of labeled $m$-level trees over $n$ vertices is given by $m^{n-1}n^{n-2}$. Indeed there are $m$ kinds of edges of different nature (if the edges are of the name nature, e.g. all edges are Bosonic, 
we can label them with $m$ colors) in a $m$-level tree so that each edge has $m$ choices. Since there are $n-1$ edges in the tree and the total number of labeled trees is $n^{n-2}$, the result comes. A constructive estimate about the number of labeled $2$-level trees is given \cite{MLVE} and the interested readers are advised to consult that paper.
\end{proof}

\vskip.5cm

Now that the assignments for the slices, the integration measure and the integrand are now 
factorized over the connected components of $\cJ$, the logarithm of $Z$ is easily computed as exactly the same sum but restricted 
to two-levels spanning trees, denoted by $\cJ^{\cT}$, on $\cW= [1, \cdots , n]$. We have:
\bea \label{treerep}  
\log Z^{j_{max}}(\lambda)=  \sum_{n=1}^\infty \frac{1}{n!}  \sum_{\cJ^{\cT}} \;   \int dw_{\cJ^\cT}   \int d\nu_{ \cJ^{\cT}}  \  \partial_{\cJ^\cT}   \Big[ \prod_{\cB} \prod_{\al\in \cB}   \Bigl(     W_{\al}   (   \sigma_\al , \chi^{ \cB } , \bar \chi^{\cB}  )\Bigr)    \Big]  .\nonumber\\
\eea

The main result is the convergence of this representation uniformly in $j_{max}$ for $\lambda$ in the Cardioid domain, 
allowing to perform in this domain the ultraviolet limit of the theory:
\begin{theorem} \label{thetheorem} Let $0<\rho <1$ be a fixed small constant. The series \eqref{treerep} is absolutely convergent, uniformly in $j_{max}$, for $\lambda$ in the cardioid domain $\cC ard_\rho:=\{\lambda\in \mathds{C}:|\lambda|<\rho\cos^2(\frac12\arg(\lambda))\}$.
Its ultraviolet limit $\log Z (\lambda) = \lim_{j_{max}  \to \infty}  \log Z^{j_{max}}(\lambda)$ is therefore well-defined and
analytic in $\lambda$ in the cardioid domain; furthermore it is the Borel sum
of its perturbation series in powers of $\lambda$. 
\end{theorem}

\section{Proof of Theorem \ref{thetheorem}}\label{maintech}
In this section we shall prove the Borel summability for perturbation series of $\log Z^{j_{max}}$ in the cardioid domain. The key steps to prove this theorem is to get the optimal bounds for the Fermionic integral and the Bosonic integrals. While the former is essentially algebraic and the procedure for getting the bound is very similar to \cite{RivW3}, the latter is new due to the linear divergence of the vacuum tadpole term $\Pi^\Lambda$ (see \eqref{glasses}).
We shall start with the bounds for the Grassmann integrals.

\subsection{Grassmann Integrals}
First of all consider a fixed forest $\cF_F$ with $k$ edges: $\{\ell_i=(\al_i, \bt_i), i=1\cdots k\}$. Using the basic properties of the Grassmann integrals ( see Formula \eqref{grasr} ) we can easily find that: 
\bea\label{eq:grassmaint}
\int \biggl[ \prod_{\cB} \prod_{\omega}(d\bar \chi^{\cB}_{\omega}d\chi^{\cB}_{\omega}) \biggr]
e^{- \sum_{\omega=0}^{\omega_{max}}\bar\chi^{\cB}_{\omega} Y_{\cB\cB'}(w_{\ell_F})\chi^{\cB'}_{\omega}}  
\prod_{\genfrac{}{}{0pt}{}{\ell_F \in\cF_F}{\ell_F=(\al,\bt)}} 
\Big(\chi^{\cB(\al)}_{\omega_{\ell_F} } \bar  \chi^{\cB(\bt)}_{\omega_{\ell_F}}+ 
\chi^{\cB(\bt) }_{\omega_{\ell_F}}\bar \chi^{\cB(\al) }_{\omega_{\ell_F} } \Big) \nonumber
 \\ =   
 \Bigl(\prod_{\cB} \prod_{\al,\bt\in \cB} \bbone( \om_\al \cap \om_\bt =\emptyset) \Bigr)
 \Bigl( {\bf Y }^{\hat \bt_1 \dots \hat \bt_k}_{\hat \al_1 \dots \hat \al_k}  + 
 {\bf Y }^{\hat \al_1 \dots \hat \bt_k}_{\hat \bt_1 \dots \hat \al_k}+\dots + {\bf Y }_{\hat \bt_1 \dots \hat \bt_k}^{\hat \al_1 \dots \hat \al_k}   \Bigr) \; ,
\eea 
where \be
{\bf Y }^{\hat \al_1 \dots \hat \bt_k}_{\hat \bt_1 \dots \hat \al_k}=\int\big(\prod_id\bar\chi_i d\chi_i \big)e^{-\sum_{i,j}\bar\chi_i Y_{ij}\chi_j}\prod_{i=1}^k\bar\chi_{\al_i}\chi_{\bt_i}\ ,
\ee
is the minor of $Y$ with the lines $\al_1\cdots \al_k$ and columns $\bt_1\cdots \bt_k$ deleted.
There are precisely $2^k$ such terms in the sum, as
there are exactly $2^k$ ways to exchange the vertices $\al_i$ and $\bt_i$ of each $\ell_i=(\al_i, \bt_i)\in F$.
The factor $\prod_{\cB} \prod_{\al,\bt\in \cB} \mathds{I}( \om_\al \cap \om_\bt =\emptyset) $
means \emph{the hard core constraint inside each block}, ensures the disjointness of the slices in each block.

Since the matrix $Y$ is positive semi-definite with each element smaller or equal to one, we can easily find that each minor is bounded:
\be
\Big{|}  {\bf Y }^{\hat \al_1 \dots \hat \bt_k}_{\hat \bt_1 \dots \hat \al_k} \Big{|}\le 1 \; .
\ee
So the Grassmann integral is bounded by $O(1)\cdot 2^k$. 


\subsection{Bosonic Integrals}
The main problem is now the evaluation of the Bosonic integrals in \eqref{treerep}, for which the two-level trees $\cJ^\cT$ restrict to the bosonic trees
$\cT_\cB$ in certain bosonic block $\cB$. So it is sufficient to bound separately this integral in each fixed block $\cB$. Let the resolvent graphs associated with that block be ${\cal G}(\cB)$ and the index sets of refined indices be $\om(\cB) = \cup_{\al \in \cB}\  \om_\al$. Let the Bosonic Gaussian measure restricted to $\cB$ be $d\nu_\cB(\sigma) $, the bosonic integral reads:
\be
\int d \nu_\cB(\sigma)\  F_\cB(\sigma) = \biggl[ e^{\frac{1}{2} \sum_{\al,\bt\in \cB} X_{\al,\bt}(w_{\ell_B }) 
 \frac{\partial}{\partial \sigma_\al}\frac{\partial}{\partial \sigma_\bt}} F_\cB(\sigma) \biggr]_{\sigma =0},
\ee
where $F_\cB(\sigma)$ is obtained by evaluating the action of the derivatives on the $exp-vertices$:
\bea \label{manyder}  F_\cB &=&  \prod_{\al\in \cB}\bigl[ \prod_{e \in E^\al_\cB} \bigl( \frac{\partial}{\partial \sigma_e} \bigr)  W_{\al}\bigr]\nonumber\\
&=&  \prod_{\al\in \cB} \prod_{e \in E^\al_\cB} \bigl( \frac{\partial}{\partial \sigma_e} \bigr)      \biggl[\prod_{\og\in \om^{\cM}_\al} \bigl[ \int_0^1 dt_\og  e^{V_\og(t,\sigma_\al) }  \bigr] A_{G_a}(t,\sigma_\al)   \biggr]\ ,
\eea
where $\om^{\cM}_\al$ is the modified set of refined scaling indices of vertex $\al$, $E^\al_\cB$ runs over the set of all edges in $\cT_{\cB}$ ending at the vertex $\al$, hence $\vert E^\al_\cB \vert$ is equal to the
degree of coordinations of the tree $\cT_{\cB}$ at vertex $W_\al$, denoted by $d_\al( \cT_{\cB} )$. We call the newly generated graph $G^{\cM}_\cB$. 
The derivatives $\big(\ \prod_{e \in E^\al_\cB}\ \frac{\partial}{\partial \sigma_e}\ \big)$ act either on the amplitudes $A^{\cM}_{G_\al}$ or generate new loop vertices 
from the exponential $\prod_{\omega\in \om^{\cM}_\al} e^{V(\sigma_\al)}$. There are two cases to be considered: when the block $\cB$ is reduced to a single vertex $\al$, there is no derivative to compute and the integrand reduces simply to the one of the slice-testing expansion; when $\cB$ has more than one vertex, then each vertex $\al\in \cB$ is touched by at least one derivative. We only consider the second case.

\begin{itemize}
\item the exponential $\prod_{\al\in\cB}\prod_{\omega \in \om^{\cM}_\al} e^{V_\omega (\sigma_a)}$ doesn't disappear since the exponential function is its own derivative,

\item the derivatives which act on the graph integrand $A_{G_a}(\sigma_a) $ must act on the $R$ resolvent factors and create therefore new propagators sandwiched by resolvents,
through $\frac{\partial}{\partial \sigma} R =  -i\frac{\sqrt{2\lambda}}{2} R C R$ (see \eqref{opeq1}),

\item each derivative acting on the exponential $e^{V_\omega(\sigma)}$ generates either a term of the type $ i\frac{\sqrt{2\lambda}}{2} (R-1)C^\og$, a counter terms $T^\og$, or a term proportional to $RCRC^{\omega}\sigma$, which contains a $\sigma$ variable in the numerator (see \eqref{slice2}).  Similar to the discussion of the auxiliary expansions, the counter-term $T^\omega$ is bounded by $1/\omega<1$; In last case, we shall integrate out these $\sigma$ variable by Gaussian integration. Since each $\sigma$ variable is associate with a marked propagator $C^\omega$, the combinatorial factors from Gaussian integrations can be eventually bounded (see the next section).
\end{itemize}

We can write the Bosonic integrand as:
\bea\label{eq:bosogauss}
 \int dw_\cB\ \int d \nu_\cB  F_\cB =\sum_{\cT_\cB}[\ \prod_{\ell\in\cT_\cB}\int_0^1 dw_{\ell}\ ] \int d\nu_\cB(\sigma)\prod_{\al\in\cB}\prod_{\omega\in\Omega_\al} e^{ V_\omega(\sigma_\al)}  \sum_{G_\cB \in \cG (\cB)} A_{G_\cB} (\sigma),
\eea 
where $\cG(\cB)$ is the set of resolvent graphs after the derivation and $A_{G_\cB}$
is the amplitude for a member $G_\cB\in\cG(\cB)$.

It is useful to collect the terms that are linear in the $\sigma$ variables, which are generated from the sliced interaction terms $V_\omega$ by derivations w.r.t. the interpolation parameter $t$ in the auxiliary expansions (cf. Formula \eqref{aux2}) or by derivations w.r.t. the $\sigma$ variables in the two-level forest formula. An interesting property of these terms is that each $\sigma$ variable always comes with a marked propagator $C^\omega$. 
So we denote these terms as $\cP(\sigma,C^\omega)$. Remark that $\cP(\sigma,C^\omega)\sim (\sigma C^\omega)^p$ is a monomial of $\sigma C^\omega$ with even power $2\le p\le 2|B|$. Formula \eqref{eq:bosogauss} can be written as
\bea\label{boseg2}
\int d \nu_\cB(\sigma)\prod_{\al\in\cB}\prod_{\omega\in\Omega_\al} e^{ V_\omega(\sigma_\al)}   \sum_{G^\cR_\cB \in \cG (\cB)}  A^{\cR}_{G^\cR_\cB} (\sigma)\cP(\sigma,C^\omega)
 \; ,
\eea 
where $ A^{\cR}_{G^\cR_\cB} (\sigma)$ is defined as the remainder terms in $A_{G_\cB} (\sigma)$ with corresponding resolvent graphs $G_\cB^\cR$.

The merit of such identification is as follows. The graphs $G_\cB$ resulted after the auxiliary expansions and the two level forest formulas are not minimal in general due to the derivations w.r.t. the sliced interaction potential $V_\omega$. These derivatives generate marked propagators to a connected resolvent graph such that the indices of the newly generated marked propagators may coincide with that in the resolvent graph. We call these terms the minimal-violation terms, the collection of which is exactly $\cP(\sigma, C^\omega)$. So after
identifying the terms $\cP(\sigma, C^\omega)$ the resolvent graphs $G^\cR_\cB$ corresponding to the remainder terms $ A^{\cR}_{G^\cR_\cB} (\sigma)$ are still minimal.

Due to the positivity of the integration measure $d\nu_\cB$ we can use the H\"older inequality for the Bosonic integrations:
\bea  \label{CS}  && \sum_{G_\cB \in \cG (\cB)}   \Bigl{|}\int dw_\cB\  \int d \nu_\cB(\sigma) \prod_{\al\in\cB}\prod_{\omega\in\Omega_\al} e^{ V_\omega(\sigma_\al)}  A_{G^\cR_\cB} (\sigma) \Bigl{|}\\
&\le& \; \sum_{G_\cB \in \cG (\cB)} \Bigr{(} \int d \nu_\cB(\sigma)\   \vert  A^{\cR}_{G^\cR_\cB} (\sigma)  \vert^2  \Bigr{)}^{1/2}  \Bigr{(} \int d \nu_\cB(\sigma)\  \vert\prod_{\al\in\cB}\prod_{\omega\in\Omega_\al} e^{ V_\omega(\sigma_\al)}\vert^4   \Bigr{)}^{1/4}\cdot\nonumber\\
&&\quad\quad \quad\cdot \Bigr{(}\int d \nu_\cB(\sigma)\ \cP^4(\sigma, C^\omega)\Bigr{)}^{1/4}\ .
\eea
In the following we shall write $A^{\cR}_{G^\cR_\cB}$ as $A_{G_\cB}$ and note ${G^\cR_\cB}$ as $G_\cB$, just for simplicity.

\subsection{Non-Perturbative Bounds I: Bounds for the Resolvent Operators}

\begin{figure}[!htb]
\centering
\includegraphics[scale=0.45]{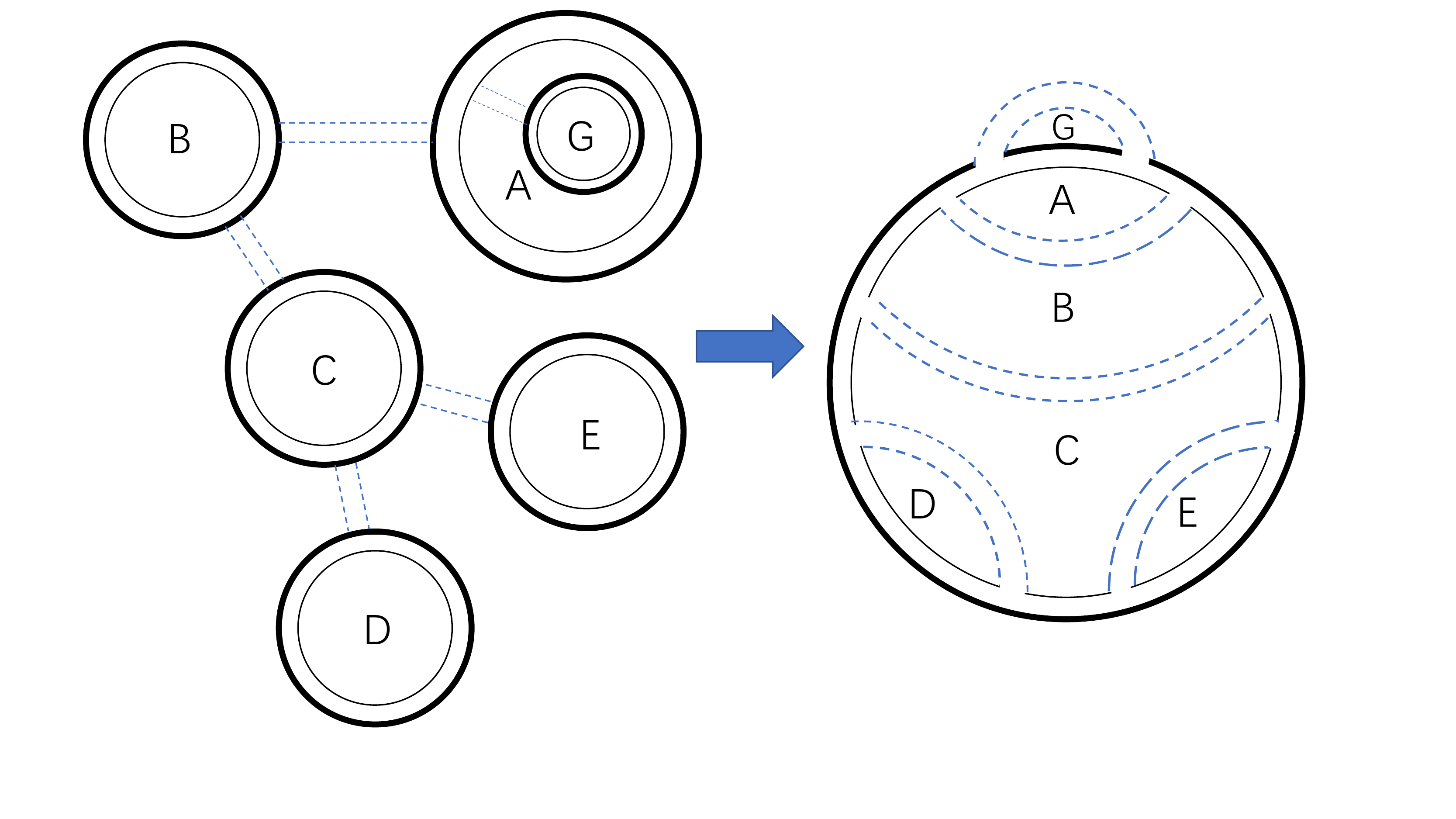}
\caption{A planar LVE graph in the direct representation and the dual representation. The ribbons on the l.h.s. are the resolvent propagators in which the $c$-propagators are not distinguished with the resolvents. The dotted lines are the $\sigma$ propagators which are attached to both borders of the ribbons. The graph on the r.h.s. is the dual graph, in which each region enclosed by the $\sigma$ propagator correspond to a loop vertex on the l.h.s.. If any loop vertices $A,\cdots, E$ contains non-planar graphs the corresponding dual graph is constructed in the same way.  
} \label{dual}
\end{figure}

In this subsection we explain how to bound, for a fixed graph $G_\cB$ (which is minimal) the factor 
\be I= \bigl{(}\int d \nu_\cB(\sigma )   \vert  A_{G_\cB} (\sigma)  \vert^2   \bigr{)}^{1/2} 
\ee
using the Cauchy-Schwarz inequality \eqref{CS}.
This is a non-perturbative problem, since the resolvents in the $A_{G_\cB} (\sigma)$, if expanded in power series of $\sigma$
and integrated out with respect to $ d \nu_\cB  $, would lead to infinite divergent series of Feynman graphs. Hence we shall
use the norm bound \eqref{resobound} to get rid of these resolvents. This part follows closely \cite{RivW3} and the interested reader is advised to look at \cite{RivW3} for more details.  
We shall work on the dual representation of for the graphs generated in MLVE, defined as follows.
\begin{definition}[Dual Graph of MLVE]
For any tree generated by MLVE one can define the dual representation for the graph, also called the dual tree, as follows: the dual tree graph is represented as a single big circle divided by dotted lines representing the $\sigma$ propagator. Each region enclosed by the $\sigma$ propagators in the dual graph represents a loop vertex in the direct representation, see Figure \ref{dual}. The cyclic ordering for the resolvents is the same as in the direct representation. 
\end{definition}
Remark that due to the cyclic ordering of the various objects (eg. the resolvents, the c-propagators, the $\sigma$ propagators) in the ribbon graph, planar or not, the dual representation always exist and is uniquely defined given the ribbon graph in the direct representation. 

For each graph in the dual representation we can defined its complex conjugate graph:
\begin{definition}[Complex Conjugate Graph]
A graph $\bar G_\cB$ is called the complex conjugate graph for the resolvent graph $G_\cB$ if $\bar G_\cB$ is isomorphic
to $G_\cB$ as a combinatorial object and if each resolvent in $\bar G_\cB$ is the complex conjugate of the corresponding one in $G_\cB$. Then $G_\cB\cup\bar G_\cB$ is considered as a (non-connected) graph with twice the number of propagators and vertices of $G_\cB$.
\end{definition}

Let $A_{\bar G_\cB} (\sigma)$ be the amplitude of $\bar G_\cB$, we can write square of the amplitude as
\be I^2 = \int d \nu_\cB   A_{G_\cB} (\sigma)  A_{\bar G_\cB} (\sigma),
\ee
where $A_{\bar G_\cB} (\sigma)$ is the amplitude for the complex conjugate graph. 


Before proceeding it is important to observe that $A_{G_\cB}$ contains still many renormalized tadpoles in the form of $\Tr\  (R-1)\tilde C$. In order to get the correct bounds we write each $R-1$ as $R-1=-i\frac{\sqrt{2\lambda}}{2}C\hat\sigma R$ and contract all the $\sigma$ variables using integration by parts, starting one arbitrary $\sigma$ variable in the dual graph and follow the cyclic ordering. Each step of the contraction decreases the number of $\sigma$ fields by two if they contract on themselves or by one if they contract to $R$. At the end of this step we obtain a new set of resolvents graph in which no tadpoles are presented but many new resolvents are generated. Let the set of the newly generated $\sigma$ propagators be $\gamma$ and we still use $G_\cB$ to represent the resulting graph of order $n$. The amplitude for the resulting graphs can be written as:
\bea\label{rcs1}
A_{G_\cB}=[(-\lambda)^{|\Omega_\cB|}]\Big\{\sum_{\cT\in\cB} \Tr\ [\prod_{\ell\in\cT(\cB)}R(\sigma)C\ R(\sigma)C_{\og}]\ \Big\} ,
\eea
where $\cT(\cB)$ means a spanning tree over vertices in $\cB$ and $\Omega_\cB$ is the set of refined scaling indices in $\cB$ with cardinality $|\Omega_\cB|$.

\begin{figure}[!t]
\begin{center}
{\includegraphics[width=9cm]{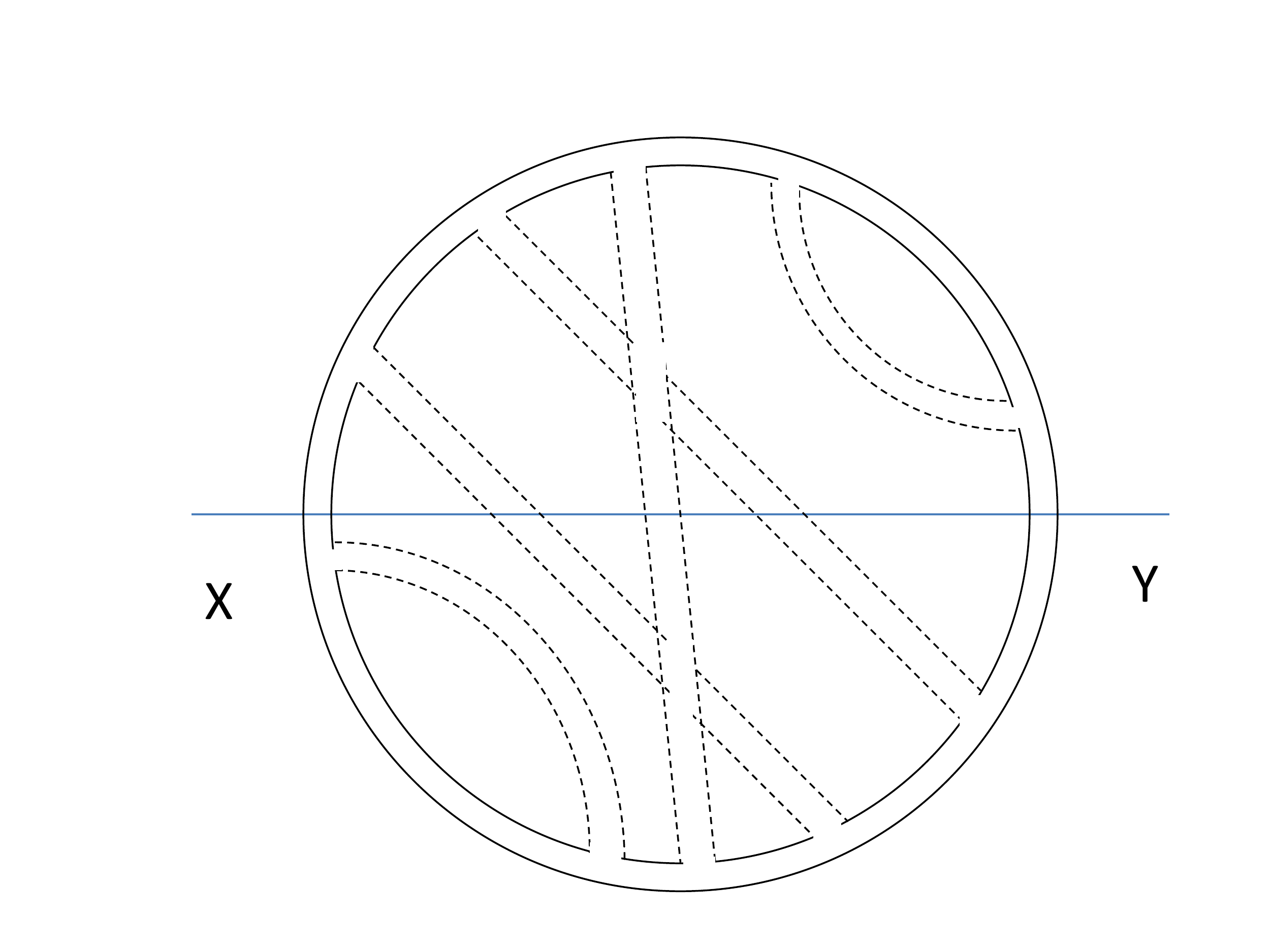}}
\end{center}
\caption{A resolvent graph with a balanced cut. Here $\sigma$-propagators are pictured as dotted lines.}
\label{cs1}
\end{figure}

In the next step we shall get rid of the resolvents $R(\sigma)$, by using the fact that each resolvent is bounded by $1/\cos(\arg\lambda /2)$ in the cardioid domain (see Lemma \ref{keybound}). This is not trivial and we shall rely on the technique of recursive Cauchy-Schwarz (CS) inequalities. \cite{RivMag, RivW3}. It is more convenient to perform the recursive CS equality in the dual representation of the graphs of MLVE. The key step of performing the CS inequality is to choose a balanced cut:
\begin{definition}[Balanced Cut for the Dual Graph]
Let $G$ be a resolvent graph in the dual representation. A balanced cut with ends $X$ and $Y$ for $G$ is a partition of the $G$ into two pieces $H_t$ and $H_b$ such that $H_t$ and $H_b$ contains equal number of resolvents (up to one unit if the initial number of resolvent propagators is odd). Here $H_t$ is called the upper chain and $H_b$ is called the lower chain. $X$ and $Y$ are called the ends of the chains, each made of a half resolvent propagator. 
\end{definition}
\begin{remark}
Remark that although the $\sigma$ propagators can attach to either the inner border or the outer border of the ribbon, we still have the cyclic ordering for the resolvents and c-propagators. So we can choose the balanced cup as above no matter to which border the $\sigma$ propagators are attached.
\end{remark}
Figure \ref{cs1} shows one example of such balance. Remark that the $\sigma$ propagators have no reason to occur at 
symmetric positions along the top and bottom chains.

Let $(G,\cR)$ be a graph in the dual representation with the set of resolvents $\cR$ and
$m$ $\sigma$ propagators. Balanced cuts for a $(G,\cR)$ with $\cR \not = \emptyset$ can be obtained in many different ways. 
A nice way to define such cuts is to  first go to the direct representation and select a spanning tree of $\sigma$ propagators of $G$. Turning around the tree provides a well defined cyclic ordering for the resolvent propagators and pure propagators of the graph (jumping over the $\sigma$-propagators not in the tree). Then we go back to the dual representation.
We shall choose the balanced cuts by first \emph{contracting all marked propagators along the cycle}, and selecting an antipodal\footnote{Or almost antipodal if the number of resolvents is not even; this can happen only at the first
CS step.} pair $(X, Y)$
among the resolvent propagators left in that cycle. We then
cut the cycle across that pair (see Figure \ref{cs1}). 

\begin{figure}[!t]
\begin{center}
{\includegraphics[width=7cm]{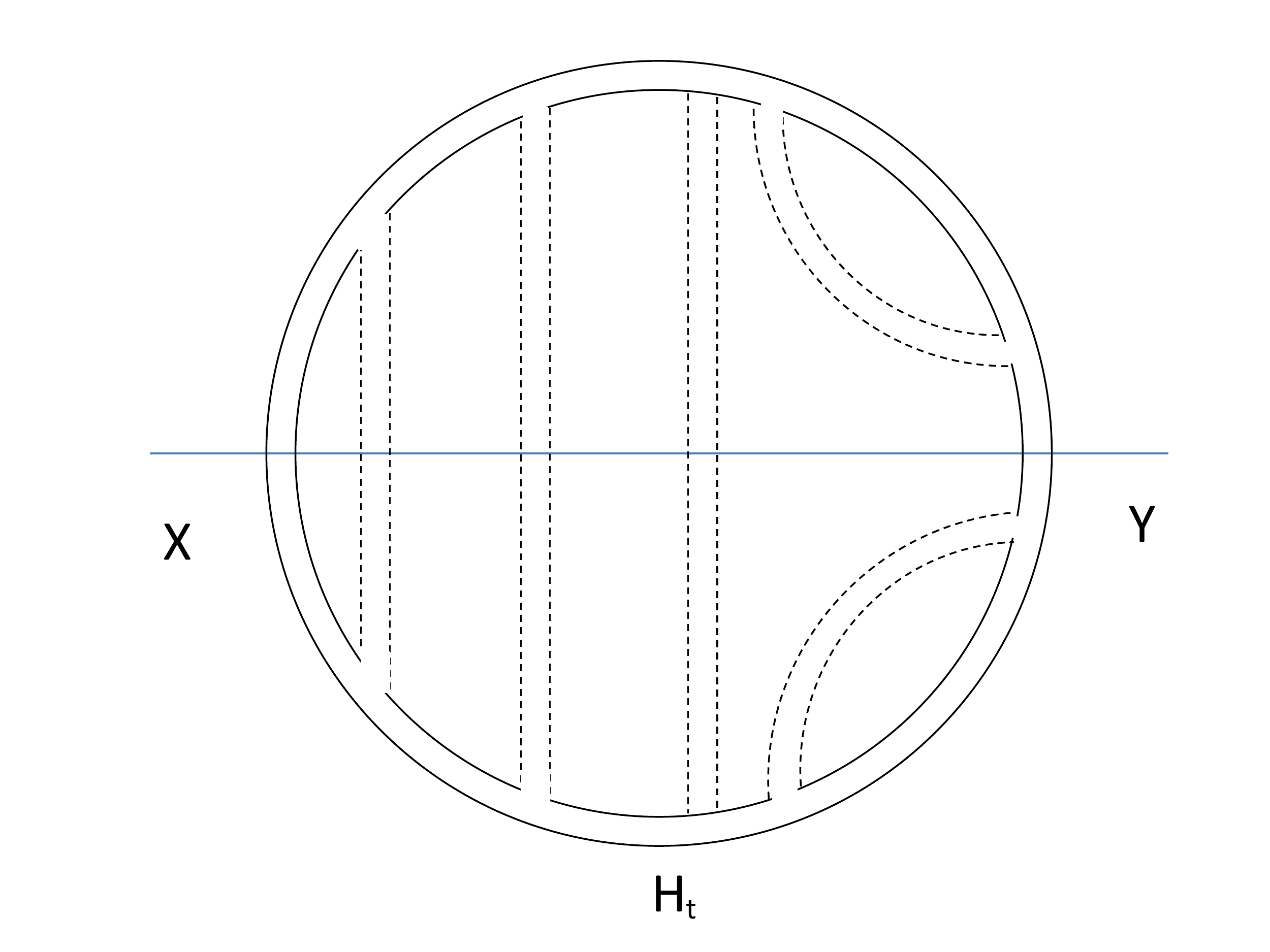} \hskip.5cm\includegraphics[width=7cm]{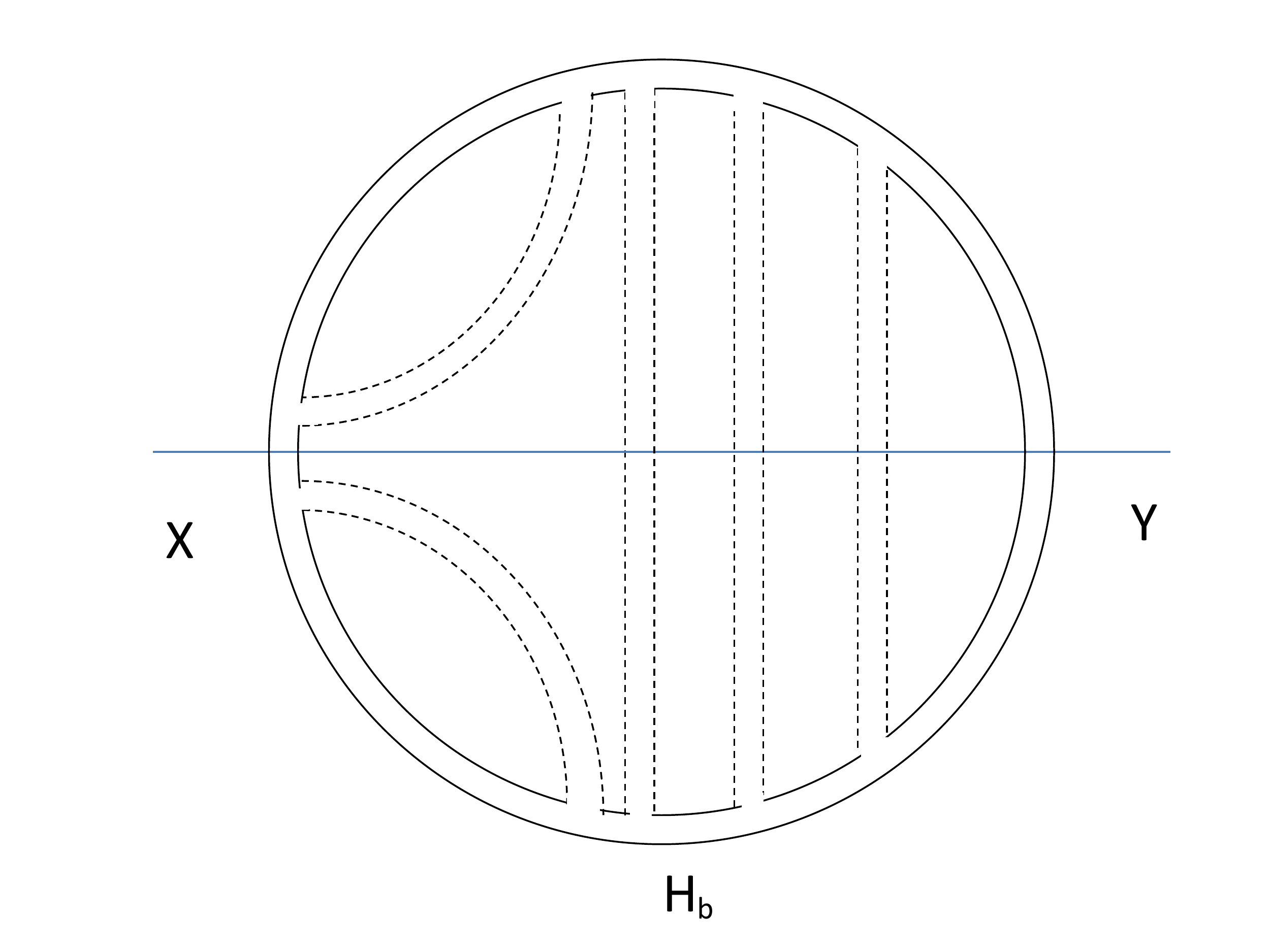}}
\end{center}
\caption{The top and bottom mirror graphs for the balanced cut of Firure \ref{cs1}.}
\label{cs2}
\end{figure}

To any such balanced cut is associated a Cauchy-Schwarz (CS) inequality. It bounds the
resolvent amplitude $A_{G_\cB}$ (see \eqref{rcs1}) by the geometric mean of the amplitudes of the two graphs
$G_t = H_t \cup \bar H_t$ and $G_b = H_b \cup \bar H_b$. These two graphs are
obtained by gluing $H_t$ and $H_b$ with their mirror image along the cut. Remark that in this gluing the $\sigma$ propagators crossing the cut are 
fully disentangled: in $G_t$ and $G_b$ they no longer cross each other, see Figure \ref{cs2}. Remark also
that the right hand side of the CS inequality, hence the bound obtained for $A_G$, is a priori different for different balanced cuts.

Now the two half propagators $X$ and $Y$ crossed by the balanced cut
at the end of the top and bottom chain, which had values $RC$ in the amplitude $A_G$,  can be replaced by two \emph{ordinary} 
propagators $C$ in the amplitudes of $G_t$ and $G_b$, loosing simply a factor $\cos^{-2} (\phi/2)$ from the norm of $R$. 
More precisely, let $\cR - \{X \cup Y\}$ be the set of resolvents such that the two half resolvent $(RC)$ related at $X$ and $Y$ are replaced by $C$, respectively. Then we have


\begin{lemma} \label{CSstep}
For any balanced $X-Y$ cut
\be  \vert A_{G,\cR} (\sigma) \vert   \le \frac{1}{\cos^2 (\arg\lambda /2)} \sqrt{A_{G_t, \cR - \{X \cup Y\}}(\sigma)}\sqrt{A_{G_b, \cR - \{X \cup Y\}}(\sigma)}
\ee
uniformly in $\sigma$.
Hence we have cleaned the two resolvent propagators $X$ and $Y$ crossed by the cut.
\end{lemma}
\prf\ 
Each $CS$ inequality is simply obtained by writing
\be <H_t , \cO H_b  > \le \Vert \cO \Vert  \sqrt{<H_t H_t >} \sqrt{<H_b H_b >} \label{tensocs}
\ee
in the tensor product of $2+q$ Hilbert spaces corresponding to the two c-propagators associated to the ends of the cuts and the $q$ $\sigma$-propagators
which cross the cut. 
We symmetrize first the operators $RC$ of the two cut propagators, writing them as $C^{1/2}B C^{1/2} $  with $ B = C^{-1/2}R C^{1/2} $.
The operator $\cO = B \otimes \Pi \otimes B$ 
in  \eqref{tensocs} is the tensor product of the two end operators $B$ and of a permutation operator $\Pi$ for the remaining
$H^{\otimes q}$ tensor product of the $q$ crossing $\sigma$-propagators. Therefore
$\Vert \cO\Vert \le \Vert B \Vert ^2  \Vert \Pi \Vert $. Any permutation operators has eigenvalues which are roots of unity, hence has norm bounded by 1, and 
$\Vert  B \Vert = \Vert R \Vert \le 1/\cos (\arg\lambda/2)$ (see Lemma \ref{keybound}).
\qed


Starting with a full resolvent graph, the inductive Cauchy-Schwarz inequalities of \cite{magnen} consist in iterating lemma \ref{CSstep} until all resolvent operators are bounded and no resolvents are left. In this way we can therefore reach a bound 
made of a geometric mean of $2^m$ ordinary perturbation amplitudes for an initial resolvent graph of of order $m$. 
To understand the result of the induction, let us observe that
\begin{itemize}

\item Only at the first step the number of resolvent propagators can be odd. In that case 
we choose an almost antipodal pair (antipodal up to half a unit): 
but at all later stages the mirror gluing creates an even number of resolvents and we can choose truly antipodal pairs. 

\item The result of $m$ complete inductive layers of CS steps applied to a starting graph $G$ of order $m$ is a family $F_m^\cC(G)$ of $2^m$ graphs,
which depends on the inductive choices, noted by $\cC$, of all the balanced cuts of the induction. Each layer of graphs and can be pictured to stand at the leaves of a rooted binary tree, with the initial graph
$G$ standing at the root.

\item Although the graphs in the family $F_q^\cC(G)$ may have very different orders, they all have the same number of resolvents (up to one at most, if the initial
number of resolvents was odd).

\item No matter which inductive choice $\cC$ is made, every 
c-propagator $\ell$ of the initial graph $G$ gets finally copied into exactly $2^m$  c-propagators in the union of all graphs
of $F_m^\cC(G)$. Notice that all these copies have the same set of refined scaling indices than the initial propagator. But they are not evenly distributed among the members of the family.

\end{itemize}

This is summarized in the following lemma.

\begin{lemma}
For any choice $\cC$ of $m$ recursive cuts we have
\be  \vert A_{G} (\sigma) \vert   \le  \bigl[ \prod_{G' \in F_n^\cC(G)} \vert A_{G'}  \vert \bigr]^{2^{-m}},
\ee
uniformly in $\sigma$. The amplitudes $A_{G'}$ are computed with coupling constants $\rho$ instead of $\vert \lambda \vert$.
\end{lemma}

Remark that no additional combinatorial factor is generated in the course of removing the resolvents since the mirror graphs are generated symmetrically and the cuts are chosen by following the cyclic ordering. The amplitudes $A_{G'}$ have no resolvent factors any more, hence are ordinary perturbative amplitudes no longer depending on $\sigma$. 

\subsection{Non-Perturbative Bound II: Bounding the polynomials $\cP(\sigma,C^\omega)$}
In this part we shall consider the bound for integration 
\be
\Bigr{(}\int d \nu_\cB(\sigma)\ \cP^4(\sigma, C^\omega)\Bigr{)}^{1/4},\label{np20}
\ee
where $\cP$ is a monomial of $(\sigma C^\omega)$ of degree $2\le p\le 2(|\cB|-1)$ resulted from the auxiliary expansions and the forests formula. We have

\be\label{np2a}
\Bigr{(}\int d \nu_\cB(\sigma)\ [C^\omega\sigma]^{4p} \Bigr{)}^{1/4}
\le |C^\omega|^{|p|}\Bigr{(}\int d \nu_\cB(\sigma)\ \sigma^{4p} \Bigr{)}^{1/4}.
\ee

Then \eqref{np2a} is bounded by
\be
O(1)^{p}M^{-jp}(\frac p2)!\le O(1)^{p}[(\frac{p}{2e})^{\frac12}M^{-j}]^{p}.
\ee
Due to the fact that $p\le M^{2j}$ for any scale $j$, summing over $p$ for the above formula is bounded by $O(1)$ so that \eqref{np20} is bounded by $O(1)$.

\subsection{Non-Perturbative Bound III: The Remaining Interaction}
In this subsection we bound of the Gaussian integration for the interacting potential in \eqref{CS}. First of all we have
\bea
\int d \nu_\cB \ \vert\prod_{\al\in\cB}\prod_{\omega\in\Omega_\al}  e^{ V_\omega (\hat\sigma)}\ \vert^4 =\int d \nu_\cB \ [\ \prod_{\al\in\cB}\prod_{\omega\in\Omega_\al} \vert\ e^{  V_\omega(\hat\sigma) }\ \vert\ ]^4.
\eea 

We shall first consider the bound for $\vert e^{V_\og(\hat\sigma)}\vert$, for which we have the following lemma:

\begin{lemma}\label{boundv}
For $\lambda$ in the cardioid domain $\cC ard_\rho$ defined in Theorem \ref{thetheorem} and $\og\in I_j$
we have:
\bea \vert \exp V_\og (\hat\sigma)   \vert &\le&  \exp\ \bigl[\ 0(1)\,   \rho +  \frac1\omega\ \vert \lambda \vert^{1/2}  \sin ( \phi/2) \ {\rm Tr} \, \hat\sigma  + \rho\ {\rm Tr}  \bigl( C_{\le \og} \hat\sigma C_\og   \hat\sigma \bigr)\ \bigr].\nonumber\\
\label{boundvj}
\eea
\end{lemma}
\prf   
Using \eqref{slice2} we write, putting ${\rm Arg}\; \lambda = \phi $, we have:
\bea
&&\vert \exp V_\og (\hat\sigma)  \vert  \le 
\exp \Re \bigl( 
\int_0^1 dt^\og \Big[ \lambda\sum_{p\in I_j}(\sum_{\og_1=M^{j}+1}^\og T_p^{\og_1}(t)\ ) T_p^{\og} \nonumber\\  
&&\quad\quad +\ i \frac{\sqrt{2\lambda}}{4} \Tr\ T^{\og}\hat\sigma
 - i \frac{\sqrt{2\lambda}}{4}\Tr\  [ R_{\le \og}(t) -1 ]  C_{\og} \hat\sigma \Big]  \bigr)\nonumber\\
&&\quad\quad\le  \exp \bigl(  O(1)\;  \bigl( \rho  +  \vert \lambda \vert^{1/2}\sin ( \phi/2) \ \frac1\omega  \ {\rm Tr}  \, \hat\sigma  + \vert \lambda \vert
 \vert{\rm Tr} \bigl(   C_\og \hat\sigma  R_{\le \og} C_{\le \og} \hat \sigma   \bigr)\vert \bigr)
\nonumber \\
&&\quad\quad\le \exp \bigl(  O(1)\;  \bigl( \rho +  \vert \lambda \vert^{1/2}\sin ( \phi/2)\  \frac1\omega \    {\rm Tr} \, \hat\sigma  + \rho\;    {\rm Tr}  \bigl(  C_{\le \og}^{1/2} \hat\sigma C_\og \hat\sigma C_{\le \og}^{1/2}  \bigr) \bigr)
\nonumber \\
&&\quad\quad= \exp \bigl(  O(1)\;  \bigl( \rho+  \vert \lambda \vert^{1/2}\sin ( \phi/2)\  \frac1\omega 
\  {\rm Tr}\, \hat\sigma  + \rho \;    {\rm Tr}  \bigl(  C_{\le \og} \hat\sigma C_\og \hat\sigma  \bigr)\bigr).  \label{noper1} 
\eea

From the first to second line we used the fact that $ ( R_{\le \og}(t) -1 )  C_{\og} \hat\sigma  =   -\frac{\sqrt{2\lambda}}{4}i  R_{\le \og} C_{\le \og} \hat \sigma  C_\og \hat\sigma $ and the bounds $|T^\og_p|\le O(1)\frac1\omega$, $\sum_{p\in I_j}\ 1\le O(1) M^j$ (cf. Section 4.1).
We also used the fact that for a positive semi-definite Hermitian operator $A$ and a bounded operator $B$ we have $\vert  \Tr\  A B \vert \le \Vert B \Vert\ \Tr A $. 
Indeed if $B$ is diagonalizable with eigenvalues $\mu_i$, computing the trace in a 
diagonalizing basis we have $\vert \sum_i   A_{ii} \mu_i \vert \le \max_i \vert \mu_i \vert \sum_i   A_{ii}  $;
if $B$ is not diagonalizable we can use a limit argument. 
We can now remark that for any Hermitian operator $L$ we have, if $\vert {\rm Arg}\; \lambda \vert = \vert \phi\vert  < \pi$, 
$\Vert ( 1- i \sqrt \lambda L )^{-1}\Vert \le \frac{1}{\cos (\phi /2)}$. 
We can therefore
apply these arguments to $A = C_{\le \og}^{1/2}  \hat\sigma C_\og\hat\sigma C_{\le \og}^{1/2} $ (which is Hermitian positive)
and $B = C_{\le \og}^{-1/2} R_{\le \og} C_{\le \og}^{1/2}$.
Indeed 
\be \Vert B \Vert =  \Vert R_{\le \og} \Vert  = \Vert ( 1- i \sqrt  \lambda C_{\le \og}^{1/2} \hat\sigma C_{\le \og}^{1/2} )^{-1}\Vert \le  \frac{1}{\cos (\phi /2)}. 
\label{resobound}
\ee
We conclude since $ \frac{\vert \lambda \vert}{\cos^2 (\phi /2)} < \rho$ in the cardioid domain.  \qed
\vskip.5cm
Let $J(\cB)$ be the set of coarse scaling indices for the marked propagators in the block $\cB$. In order to derive the bound in a more explicit way we shall identify the ordered refined scaling index $\og\in\om_a\subset\om(\cB)$ with the corresponding index
$\oj\in\om_j$, for each $\al\in\cB$. So we have:
\be\label{sca}
\sum_{a\in B}\sum_{\og\in \om_a}\ 1=\sum_{j\in J(\cB)}\sum_{\oj\in\om_j}\ 1.
\ee
As a corollary of the above lemma we have:
\begin{corollary}\label{bound2}
For $\lambda$ in the cardioid domain $\cC ard_\rho$ defined in Theorem \ref{thetheorem}
we have
\bea \vert \exp\ [ \ \sum_{\oj\in \om_j}  V_\oj (\hat\sigma)\ ] \vert &\le&  \exp \bigl( 0(1)\,   \rho \,  M^j +  \vert \lambda \vert^{1/2}  \sin ( \phi/2)     {\rm Tr} \, \hat\sigma  + \rho\ {\rm Tr}  \bigl(\sum_{\og\in\om_j}  C_{\le \og} \hat\sigma C_\og   \hat\sigma \bigr)] \bigr).\nonumber\\
\label{boundvj}
\eea
\end{corollary}

We have the following bound for the first integration in the Cauchy-Schwartz inequality \eqref{CS}:

\begin{theorem}[Bosonic Integration]\label{BosonicIntegration}
Let $\rho $ be a fixed positive constant that is small enough, one has
\be
\big[\int d \nu_\cB \ \vert\prod_{\al\in\cB}\prod_{\omega\in\Omega_\al}  e^{ V_\omega (\hat\sigma)}\ \vert^4\ \big]^{1/4}\le e^{ O(1) \rho \sum_{j \in J(\cB)}  M^j }.  \label{rhobound}
\ee
\end{theorem}

\prf  \ 
From Lemma \ref{boundv} we know that it would be enough to prove that
\bea
&&\big(\ \int d \nu_\cB\prod_{\al\in\cB}\prod_{\omega\in\Omega_\al} \exp\big[\ 4 O(1)\,   \rho + \frac4\omega \vert \lambda \vert^{1/2}  \sin ( \phi/2)     {\rm Tr} \, \hat\sigma_\al  + 4\rho\ {\rm Tr}  \bigl( C_{\le \og} \hat\sigma_\al C_\og   \hat\sigma_\al \bigr)\ \big]\ \big)^{1/4}\nn\\
&&\quad\quad\le e^{ O(1) \rho \sum_{j \in J(\cB)}  M^j }\ .
\eea

Then it is easy to find that the first term (cf. Corollary \ref{bound2}) in the above formula gives precisely the bound $\exp [ O(1) \rho \sum_{j \in J(\cB)}  M^j] $. 
So it remains to check that
\be\Bigr{(}\int d\nu_\cB \prod_{\al\in\cB}\prod_{\omega\in\Omega_\al}e^{\ [\ \frac4\omega \vert \lambda \vert^{1/2}\sin ( \phi/2) {\rm Tr}\, \hat\sigma_\al +4\rho \; {\rm Tr} ( C_{\le \og} \hat\sigma_\al C_\og\hat\sigma_\al ) ]}\ \Bigr{)}^{1/2} \le 
e^{ O(1) \rho \sum_{j \in J(\cB)} M^j }.\label{bi1}
\ee
The idea of proving the above formula is to collect the quadratic terms in $\sigma$ and perform the Gaussian integration w.r.t. the $\sigma$ variables.

So we can rewrite \eqref{bi1} as
\be
\int d\nu_\cB  \prod_{\al\in\cB}\prod_{\omega\in\Omega_\al}e^{\ [\ \frac4\omega \vert \lambda \vert^{1/2}\sin ( \phi/2) {\rm Tr}\, \hat\sigma_\al +4\rho \; {\rm Tr} ( C_{\le \og} \hat\sigma_\al C_\og\hat\sigma_\al ) ]}:=
\int d\nu_\cB  \; e^{ \; \frac{1}{2} < \sigma , \bQ \sigma >  + < \sigma, \bP >}\ ,
\ee
where the inner product $< \sigma , \bP>$ is defined as
\be  
 < \sigma , \bP> =\sum_{\al\in\cB}  < \sigma_\al , P^\al>=
 O(1)|\lambda|^{1/2}\sin(\phi/2)\sum_{\al \in \cB}\sum_{\og \in \om_\al}\ \frac1\omega\ \Tr\ \sigma_\al
 \label{niceequ2}.
\ee
So that $\bP=(P^\al)$, $\al=1,\cdots, {|\cB|}$, can be considered as a vector in which
\be\label{pa1}
P^\al= O(1)|\lambda|^{1/2}\sin(\phi/2)\sum_{\og \in \om_\al}\frac1\omega\ .
\ee

In order to evaluate the summation in the above formula we shall pass to the summation over the scaling indices $\{\omega(j)\in\om_j, j\in J(\cB)\}$. See Formula \eqref{goodtra}.

The matrix $\bQ$ is defined by the following equation:
\bea  < \sigma , \bQ\sigma >  &=&  \sum_{\al \in \cB}\sum_{\omega\in\om_\al} < \sigma_\al , Q^\og \sigma_\al > 
= \sum_{\al \in \cB}\sum_{\omega\in\om_\al}
O(1)  \rho  {\rm \ Tr}  \bigl(  C_{\le \og}   \hat\sigma_\al C_{\og}  
 \hat\sigma_\al \bigr)\nn\\
&=&\sum_{\al \in \cB}\sum_{\omega\in\om_\al}
O(1)  \rho  {\rm \ Tr}  \bigl(  C^{1/2}_{\le \og}   \hat\sigma_\al C_{\og}  
 \hat\sigma_\al C^{1/2}_{\le \og} \bigr)\ .
 \label{niceequ1}
\eea
Using the fact that (cf. \eqref{int}) $\hat\sigma_{mn,pq}=\sigma_{np}\delta_{mq}+\sigma_{qm}\delta_{np}$,
the trace in above formula reads:
\bea\label{kernq}
&&\Tr\ C^{\le\omega}\hat\sigma C^{\omega}\hat\sigma=
\sum_{mnkl}(C^{\le\omega}\hat\sigma)_{mn,kl}(C^\omega\hat\sigma)_{lk,nm}\nn\\
&&=\sum_{mnkl}(C^{\le\omega}_{mn}\sigma_{nk}\delta_{lm}+C^{\le\omega}_{mn}\sigma_{lm}\delta_{nk})
(C^{\omega}_{lk}\sigma_{kn}\delta_{lm}+C^{\omega}_{lk}\sigma_{ml}\delta_{nk})\nn\\
&&=2\sum_{mnk}\sigma_{nk}C^{\le\omega}_{mn}C^{\omega}_{mk}\delta_{mm}\sigma_{kn}+
2\sum_{pq}\sigma_{mm}
C^{\le\omega}_{mp}C^{\omega}_{qn}\sigma_{nn}\nn\\
&&=:\sum_{mn,kl}\sigma_{nm}Q^\omega_{mn,kl}\sigma_{lk},
\eea
we have
\be\label{kq2}
Q_{mn,kl}=O(1)\rho \ \big(2 \delta_{ml}\delta_{nk}\sum_{r} C^{\le\omega}_{rl}C^{\omega}_{rk}+ 
2\delta_{mn}\delta_{kl}\sum_{pq} C^{\le\omega}_{mp}C^{\omega}_{ql}\ \big).
\ee

\begin{figure}[!htb]
 \centering
 \includegraphics[scale=0.45]{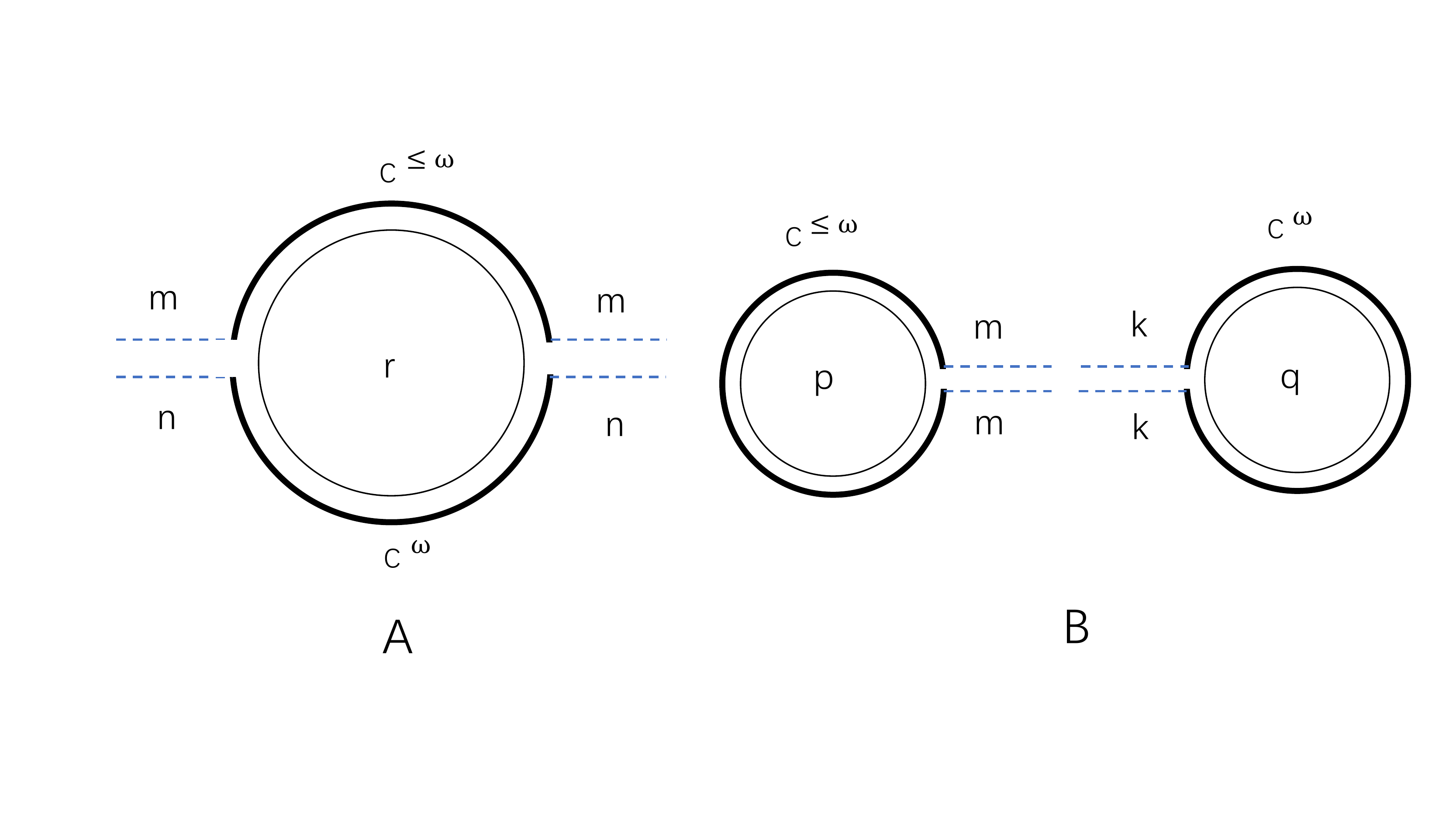}
 \caption{The graphic representations for the kernel functions in \eqref{kq2}. We have drawn explicitly the intermediate fields $\sigma$, which correspond to the half dash lines. } \label{qkn}
 \end{figure}

First of all it is easy to find that $\bQ$ is a positive and symmetric matrix. What's more, it is diagonal in the replica space generated by an orthogonal basis $\{e_\al\}$, $\al\in\cB$. 
It is useful to look at the graphical representation of the above two terms. While the first term in\eqref{kq2} can be represented by Graph A of Figure \ref{qkn}, the second term corresponds to Graph B of Figure \ref{qkn}.

We have the following lemma concerning the kernel $Q$:
\begin{lemma}
Uniformly in $j_{max}$ we have the following bound for $Q^\og$:
\bea
\Vert Q^\og \Vert  &\le& O(1) \rho  M^{-j} \label{bonnenorm},\\
|{\rm Tr}\; Q^\og |&\le&  O(1)  \rho .\label{bonnetrace}
\eea
\end{lemma}
\begin{proof}
Using the refined scaling indices $\omega(j)\in I_j$ we have $Q^\og=Q^\oj$ and
formula \eqref{bonnenorm} follows directly by using the bound for $C^\oj$ and $C^{\le \oj}$  (cf. Section \ref{slice1}). Indeed for the first term in \eqref{kq2} we have
\bea
&&\sum_{r\in I_j} C^{\le\omega(j)}_{rl}C^{\omega(j)}_{rk}= \sum_{r\in I_j}(\frac{1}{r+l+1}\mathds{1}_{r+l\le\omega(j)})
(\frac{1}{r+k+1}\mathds{1}_{r+k=\omega(j)})\nn\\
&&\le\frac{1}{\omega(j)+1}\sum_{r\in I_j}\frac{1}{r+l+1}\mathds{1}_{r+l\le\omega(j)}=\frac{1}{\oj+1}T^j\le\frac{ O(1)}{\omega}\ .
\eea
Since $\omega\in I_j= [M^j, M^{j+1}-1]$ we have 
\be
\sum_{r\in I_j} C^{\le\omega(j)}_{rl}C^{\omega(j)}_{rk}\le O(1)M^{-j}.
\ee
And the second term in \eqref{kq2} is bounded by (cf. \eqref{bd1}, \eqref{bd1a})
\be
|T^\omega(j)| |T^{\le\omega(j)}|\le\frac{O(1)}{\omega(j)}\le O(1)M^{-j}.
\ee
Thus we have proved Formula \eqref{bonnenorm}.
In the same way we can derive formula \eqref{bonnetrace}, by using the bound for $C^\oj$ and $C^{\le \oj}$ and for $T^{\le \oj}$, $T^\oj$ (cf. Section \ref{slice1}). We have
\be
|{\rm Tr}\; Q^\oj|= O(1)\rho|\sum_{m,n,p\in I_j}C^{\le \og}_{mn,nm}C^{ \og}_{mp,pm}|=O(1)\rho\sum_{m\in I_j}T^{\le \omega(j)}_m T^{\omega(j)}_m\le O(1)\rho.
\ee
\end{proof}
The two terms of \eqref{kq2} are presented in Figure \ref{qkn}, in which we also keep the $\sigma$ fields explicit. It is easy to see that the two graphs have the same amplitude after contracting the $\sigma$ fields w.r.t. the Gaussian measure, as both of them correspond to the vacuum tadpole graphs with one marked propagator but no resolvents,
which also implies that the two terms in \eqref{kq2} have the same bounds.

The covariance $\bX$ of the Gaussian measure $d\nu_\cB$ (cf. \eqref{bosint0}) is a symmetric matrix on the big space $\bV$, which is the tensor product 
of the identity in matrix spaces with the matrix $X_{a\bt} (w_{\ell_B} )$ in the replica space. Since the matrix $\bX=(X_{\al\bt})$ commutes with $\bQ$ (to remember the $\bQ$ is diagonal in the replica indices $\al$), we can perform the Gaussian integration over $\sigma_\al$, we have
\be
\int d\nu_\cB  \; e^{\frac{1}{2}  <\sigma, \bQ \sigma > + <\sigma ,\bP > }  = e^{\frac{1}{2} <\bP , \bX (1-\bA)^{-1}  \bP > }\ [ \det(  1-\bA)  ]^{-1/2},
\ee
where $\bA \equiv \sqrt\bX \bQ \sqrt\bX = \bX \bQ$.

\begin{lemma}
The following bounds hold uniformly in $j_{max}$: 
\bea
\vert\Tr \; \bA\vert &\le&  O(1)  \rho  \,  \sum_{j \in J(\cB)}  M^j  ,\label{bonnegtrace}\\
\Vert \bA \Vert  &\le& O(1) \rho  \label{bonnegnorm}.
\eea
\end{lemma}

\prf\quad We shall use again the identity $\sum_{\al\in\cB}\sum_{\og\in\om_\al}1=\sum_{j\in J(\cB)}\sum_{\oj\in\om_j}1$, we can write $\bQ = \sum_{j \in J_\cB}\sum_{\oj\in\om_j} Q^\oj $. Since $\bQ$ is diagonal in replica space, we have: 
\bea
\vert\Tr \; \bA \vert=\vert \sum_{a \in \cB} \sum_{\og \in \om_a}\Tr \bX Q^\og\vert
&=&
  \vert\sum_{j \in J_\cB}\sum_{\oj\in\om_j} \Tr \bX Q^\oj\vert\nonumber\\ 
&\le& \sum_{j \in J_\cB}\sum_{\oj\in\om_j}\vert{\rm Tr}\; Q^\oj \vert\le  O(1)  \rho  \,  \sum_{j \in J(\cB)}  M^j ,
\eea
where in the third inequality we have used the fact that $X_{\al\al}=1$ and in the last inequality we used \eqref{bonnetrace} and the fact that $|\om_j|\le O(1) M^j$. Furthermore by the triangular inequality in \eqref{niceequ1} and using \eqref{bonnenorm}
\bea
\Vert \bA \Vert  \le   \sum_{a \in \cB}  X_{aa} (w_{\ell_B} )  \sum_{j \in J_a}\sum_{\oj\in\om_j} \Vert  Q^\oj \Vert &=& \sum_{a \in \cB} \sum_{j \in J_a}  O(1)  \rho  M^{-j}\nonumber\\
&\le& O(1)   \rho.
\eea
\qed

We can now complete the proof of Theorem \ref{BosonicIntegration}. 
Since $ | \Tr \bA^n | \le | \Tr \bA |  \Vert \bA \Vert^{n-1} $, by \eqref{bonnegnorm} for $\rho$ small enough the series $|\sum_{n=1}^\infty  \Vert \bA \Vert^{n-1} /n |$ converges and is bounded by $ 2$. So we have:
\bea 
&&[ \, \det (1 - \bA) ]^{-1/2} 
= e^{\frac{1}{2}  \sum_{n=1}^\infty (\Tr \bA^n)/n  }\nonumber\\
&& \quad\quad \le e^{\frac{1}{2}  | \Tr \bA |\ \sum_{n=1}^\infty  \Vert \bA \Vert^{n-1} /n } \le e^{\vert\Tr \bA \vert} \le e^{ O(1)  \rho  \sum_{j \in J(\cB)}  M^j }  .
\eea
Moreover, passing to the index set $(J(\cB), \Omega_j)$ we have (c.f. \eqref{pa1}):
\bea
\sum_{\al\in\cB}\ P^\al&=& O(1)|\lambda|^{1/2}\sin(\phi/2)\sum_{\al\in\cB}\sum_{\og \in \om_\al}\frac1\omega=O(1)|\lambda|^{1/2}\sin(\phi/2)\sum_{j\in J(\cB)}\sum_{\oj \in \om_j}\frac1\oj\nn\\
&\le& O(1)|\lambda|^{1/2}\sin(\phi/2)\sum_{j\in J(\cB)} O(1)\le O'(1)|\lambda|^{1/2}\sin(\phi/2)|J(\cB)|,
\eea
where $O'(1)$ is another constant of order 1 and we shall not distinguish it with $O(1)$. In the second line of the above formula we have used the fact that $\om_j\subset I_j=\{M^j,\cdots, M^{j+1}-1\}$ (cf. \eqref{slice0}) so that $\sum_{\oj \in \om_j}\frac1\oj\le O(1)$.
So we have
\bea
e^{\frac{1}{2} <\bP , \bX (1-\bA)^{-1}  \bP > }  
&\le& e^{\frac{1}{2} \Vert (1-\bA)^{-1}  \Vert <\bP , \bX \bP > }\le e^{\sum_{\al,\bt\in\cB}
X_{\al\bt} (w_{\ell_B} ) P^\al P^\bt}
\nn\\
&\le&   e^{( \vert\sum_{\al \in \cB} P^\al \vert)^2}\le e^{ O(1)  \rho\  | J(\cB)|^2} 
\le e^{ O(1)\rho \sum_{j\in J(\cB)}{M^j}},
\label{goodtra}
\eea
where in the fourth inequality we used the fact that $|\lambda|\sin^2(\phi/2)\le\rho$ and $  X_{a\bt} (w_{\ell_B} ) \le 1$ $\forall \al,\bt\in\cB$ and $|J(\cB)|^2\le \sum_{j\in J(\cB)}{M^j}$ for the last inequality. 


\subsection{The Combinatorial factors and the non-perturbative Bounds}
\label{finalsubsec}
Now we gather all the convergent factors and combinatorial factors for an arbitrary fixed connected graph $G$ of order $n$. We write the summing over refined scaling indices $\og\in\om_\cB$ in \eqref{CS} as \eqref{sca} and gather the factors scale by scale, from $j_{max}\in J(G)$ to $j_{min}\in J(G)$, to get the final bound. Remark that since the  nonperturbative bound from the vacuum tadpole is $e^{\lambda\Pi^\Lambda}\sim e^{\lambda M^{j_{max}}}$, we can simply choose $j_{min}=j_{max}-1$.
\begin{itemize}
\item Let $n_{G,j_{max}}$ be the number of marked propagators of coarse scale $j_{max}$ in graph $G$.
The number of marked propagators at scale $j$, $j=j_{max}$ or $j_{max}-1$, is bounded by $n_{G,j_{max}}$\ ;

\item The total number of minimal resolvent graphs with $n_{G,j_{max}}$ marked propagators is bounded by $4^{n_{G,j_{max}}}n_{G,j_{max}}!$\ ;

\item The nonperturbative bound for each scale $j_{max}$ is $e^{\rho O(1) n_{G,j_{max}}}$ (see Formula \eqref{boundvj}). Each counter-term at scale $j$ generated in the MLVE is bounded by $O(1)$, hence we have the bound
$O(1)^{n_{G,j_{max}}}$\ ;

\item Since from each marked propagator of scale $j_{max}$ we gain the convergent factor $M^{-j_{max}}$, we gain in total the convergent power-counting factor $ M^{-j_{max}n_{G,j_{max}}}$. This convergent factor is good enough to compensate both the combinatorial factor and the nonperturbative bound:
\bea
&&e^{\rho O(1) n_{G,j_{max}}}4^{n_{G,j_{max}}}(n_{G,j_{max}}!) M^{-{j_{max}}n_{G,j_{max}}}\nn\\
&&\quad\quad\quad\quad\sim e^{\rho O(1) n_{G,j_{max}}} (\frac{4n_{G,j_{max}}}{e})^{n_{G,j_{max}}} M^{-{j_{max}}n{G,j_{max}}}\le 1,
\eea
as long as $n_{G,j_{max}}\le aM^{j_{max}}$. This condition can be certainly verified since at each scale $j$ we generate no more than $aM^j$ marked propagators and we can choose $a<\frac{e}{4}$;

\item The summation $\sum_\Pi\sum_\cT\sum_{\cJ^\cT}\ 1$, where the first two sums are from the auxiliary expansions and the third sum runs over all two-level trees, is bounded by $O(1)^nn!$, which is partially canceled by $\frac{1}{n!}$ from the perturbation expansion of $\log Z$. 
Indeed the auxiliary expansions together with the two-level forest expansions can be considered as a $three$-level forest expansion. From Lemma \ref{labeled} we know that the number of labeled three-level trees over $n$ vertices is bounded by $3^{n-1}n^{n-2}$.  
So we find that the amplitude of the MLVE graph at any order $n$ is bounded by polynomials $(O(1)\rho)^n$.

\end{itemize}
Remark that we have considered both the planar graphs and the non-planar graphs. The amplitude of a non-planar graph is less divergent than planar ones. They are dangerous only because their number can be much larger than the planar ones if the theory is fully expanded. So we have to impose the stopping rule to control the total number of expanded graphs. In the above we proved that the perturbation series of $\log Z^{j_{max}}$ at any order is bounded by polynomial. We shall consider the remainder terms in the next subsection. 

\subsection{Borel summation}
The Borel summation technique determines a unique analytic function from a divergent perturbation series.
\begin{definition} \cite{Sok, Riv}
A family $F_j(\lambda)$ of functions is called Borel summable in $\lambda$ uniformly in $j$
if
\begin{itemize}
\item each $F_j$ is analytic in a disc $D_\lambda=\{\lambda\vert \Re\lambda>0,  {\rm Arg} \lambda\in(-\pi/2, \pi/2)\}$,
\item each $F_j$ admits an asymptotic power series $\sum{a_{j,k}\lambda^k}$, which is the Taylor expansion with remainder terms around $\lambda=0$,
\be
F_j(\lambda)=\sum_{k=0}^{n-1}a_{j,k}\lambda^k+R_{j,n}(\lambda),
\ee 
such that the bound 
\be
\vert R_{j,n}\vert\le A_j K^n n!\vert\lambda\vert^n
\ee
holds uniformly in $n$ and $K$ is a constant independent of $j_{max}$.
\end{itemize}
\end{definition}

Finally we have the theorem:
\begin{theorem}
The perturbation series is Borel summable \cite{Sok, MLVE} in the Cardioid domain $Card_\rho=\{\lambda\vert\ |\lambda|<\rho\cos^2[({\arg}\ \lambda)/2] \}
\subset D_\lambda$, uniformly in $j_{max}$. 
\end{theorem}

\begin{proof}
In the previous section we proved the polynomial bounds for the perturbation series of arbitrary order $n$ and now we consider the remainder term. This corresponds to further Taylor expanding the vacuum correlation functions that are labeled by trees to completely graphs by adding loop lines. This is called the mixed expansion \cite{gur13}. Here we don't repeat it here and the interested is advised to consult \cite{gur13} for more details.

Remark that an easier way to obtain the correct combinatorial factors for the remainder term is to go back to the direct representation and contract all the fields $\phi$ by Wick's rule. In this way we can easily obtain the bound $n!(O(1)\rho)^n$ for the remainder term and prove the Borel summability of the perturbation series. 
\end{proof}

\section{Conclusions and Perspectives}
In this paper we have proved that the perturbation series of the vacuum correlation functions of the two-dimensional Grosse-Wulkenhaar model is Borel summable, with the method of Loop vertex expansions. Remark that this method is also suitable for the construction of the general $2n$-point Schwinger functions. In fact one can introduce in the partition function \eqref{part0} a term $\Tr\ \phi J$, where $J$ is an $(\Lambda+1)\times(\Lambda+1)$ matrix, called the Schwinger source matrix. We can still integrate out the terms linear and quadratic in $\phi$ so that Formula \eqref{lv1} is replaced by:
\be\label{lvap}
Z(\lambda, J)=\int d\nu(\sigma)\ e^{\frac{ i\sqrt{2\lambda}}{2}\ \Tr\ T^\Lambda\hat\sigma-
\frac{1}{2}\Tr\log[1+i\frac{\sqrt{2\lambda}}{2} C\hat\sigma\ ]+\frac{1}{2}\lambda \Pi^\Lambda+\frac{1}{2}\Tr\ \hat J R\hat J},
\ee
where $\hat J=J\otimes I+I\otimes J^t$ and $\hat R=(\frac{1}{1+i\frac{\sqrt{2\lambda}}{2} C\hat\sigma})$ is the resolvent matrix. Deriving the above formula w.r.t. $\{J_{mn},\ m,n=0\cdots\lambda\}$ for $2n$-times one obtains a formal expression for the $2n$-point Schwinger functions (See also \cite{MagRiv} for the construction of $2n$-point functions for the commutative $\phi^4$ model and \cite{fabien, gur13} for the tensor models, with the method of loop vertex expansions). Performing the slice-testing expansions followed by the 3-level forest expansions one can obtain the expression for the $2n$-point connected correlation functions, which is similar to \eqref{treerep} but with derivations w.r.t. the source matrix elements $\partial_{ J_{mn}}$ involved. Then using the method in Section \ref{maintech} one can prove that the perturbation series for the connected $2n$-point correlation functions is Borel summable in the Cardiod domain.  

Remark that although this method might not be applicable to the $4$-d Grosse Wulkenhaar model, which is a just renormalizable model in which all the two-point and four-point functions are divergent, this method could be applicable to the construction of $3$-dimensional Grosse-Wulkenhaar model with one commutative coordinate and two non-commutative coordinates. This is left to future publications \cite{gw3}.

\section{Appendix: Second Order Expansions}
In the appendix we shall consider the second order slice-testing expansion.
In the previous section we have calculated
\bea
\cA_1&=&\int d\nu(\sigma)\ e^{V(\sigma, t)}\big\{-\lambda\tr\ [C^{\omega(j)}(R-1)]\circ \tr\ [C(t)(R-1)]\nonumber\\
&+&{\rm non-planar\ terms}\big\}.\label{od1}
\eea
We shall forget the contributions from the non-planar graphs as their amplitudes are not divergent. The second order expansion reads:
\bea
\cA_2&=&\int d\nu(\sigma)\ dt^{\og(j_1)}dt^{\og(j_2)}\frac{d}{dt^{\og(j_2)}} \frac{d}{dt^{\og(j_1)}}\ e^{V(\sigma, t)}=\int dt^{\og(j_2)}\frac{d  }{dt^{\og(j_2)}}\cA_1\nonumber\\
&=& \int d\nu(\sigma)\ dt^{\og(j_1)}dt^{\og(j_2)}e^{V(\sigma, t)}\big\{-\lambda\ \tr\ \frac{d}{dt^{\og(j_2)}}\ [C^{\omega(j_1)}(R-1)]\circ \tr\ [C(t)(R-1)]\nonumber\\
&-&\lambda\ \tr\ [C^{\omega(j_1)}(R-1)]\circ \tr\ \frac{d}{dt^{\og(j_2)}}\ [C(t)(R-1)]\nonumber\\
&-&\lambda\ \tr\  [C^{\omega(j_1)}(R-1)]\circ \tr\ [C(t)(R-1)]\ \frac{d}{dt^{\og(j_2)}}V(t, \sigma)\ \big\}.
\eea
Using the same method as the the first order slice-testing expansion, including using the flipping symmetry ( see Figure \ref{duality} ) for the tadpoles and after some lengthy but straightforward calculation we have:
 \begin{figure}[!htb]
 \centering
 \includegraphics[scale=0.25]{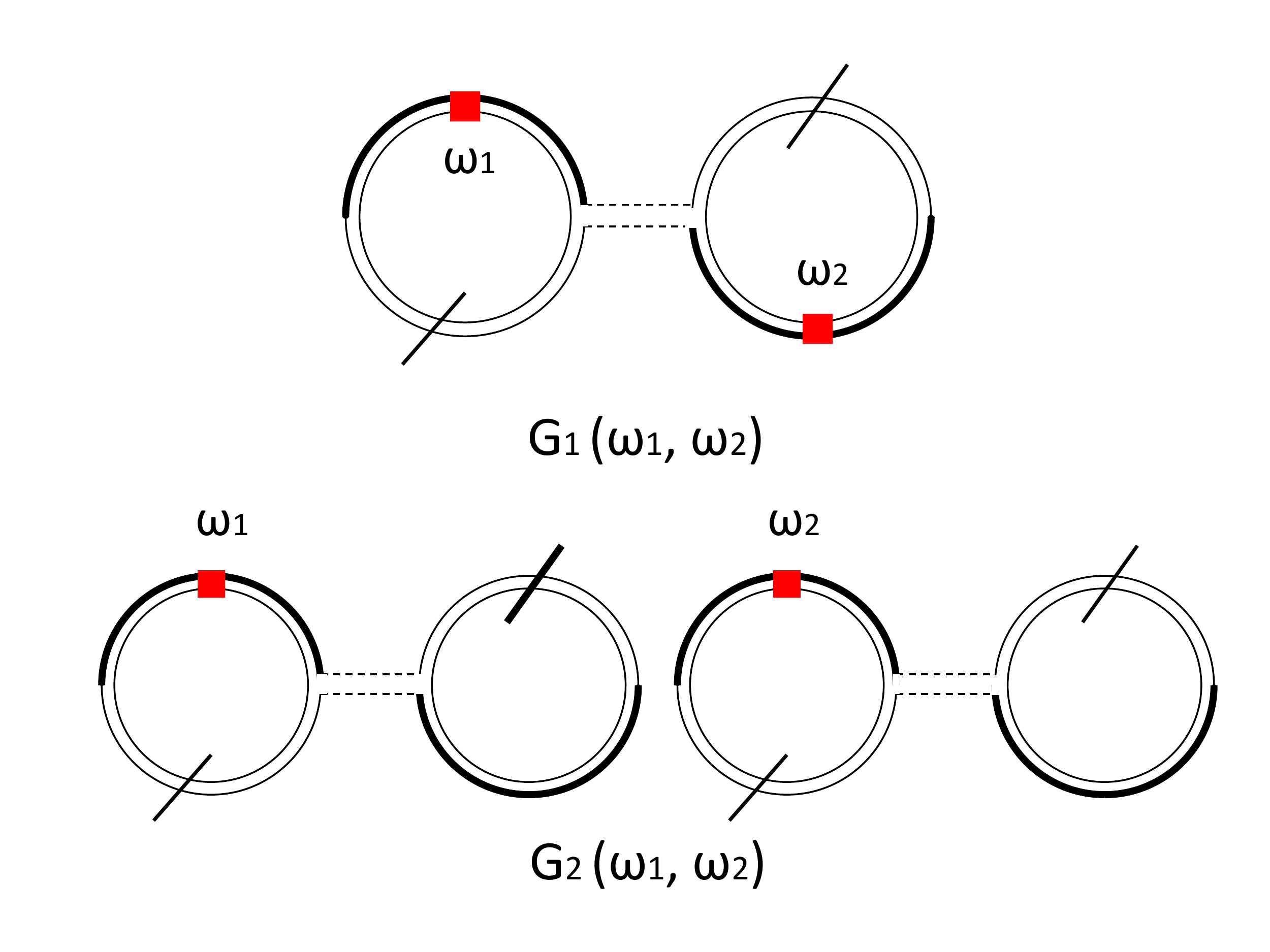}
 \caption{The graphs for the second order slice-testing expansions. We have only shown the case such that all the $\sigma$ propagators are hooked to the outer border of the ribbon graphs.} \label{gsa}
 \end{figure}

 \begin{figure}[!htb]
 \centering
 \includegraphics[scale=0.3]{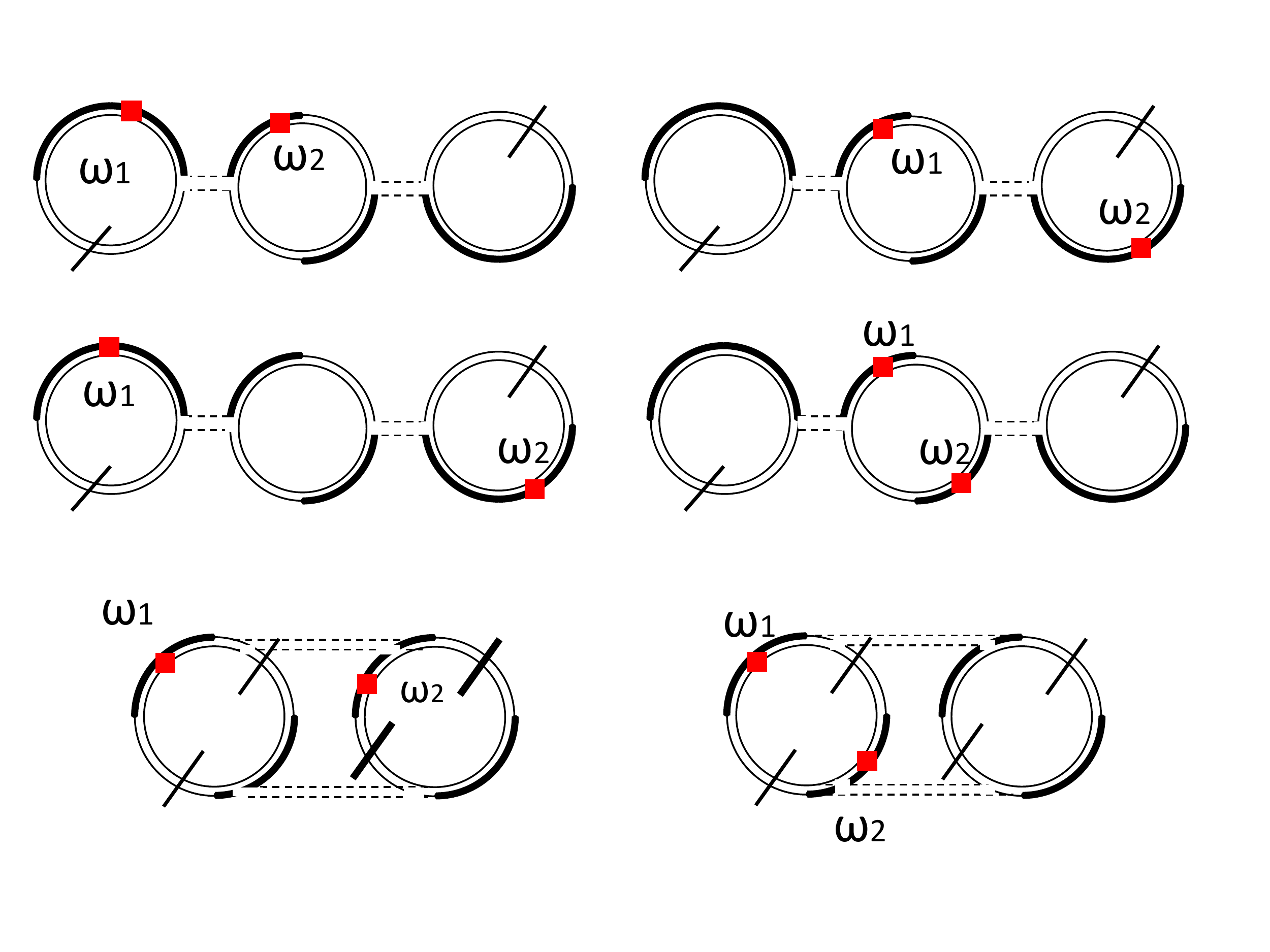}
 \caption{The graphs for the second order slice-testing expansions.We have only shown the case such that all the $\sigma$ propagators are hooked to the outer border of the ribbon graphs.} \label{gsb}
 \end{figure}
\bea
\cA_2&=&\int d\nu(\sigma)dt^{\og(j_1)}dt^{\og(j_2)}\ e^{V(\sigma, t)}\ \big\{\nonumber\\
&&\ 2\lambda^2\ \big[\ \sum_{mn}\tr[(R-1)C^{{\og(j_1)}}]_{mm}\circ \tr[RC(t)RC^{\og(j_2)}]_{mm,nn} \circ\tr[(R-1)C(t)]_{nn}\nonumber\\
&+&\sum_{mn}\tr[(R-1)C^{\og(j_2)}]_{mm}\circ \tr[RC(t)RC^{\og(j_1)}]_{mm,nn}\circ \tr[(R-1)C(t)]_{nn}\nonumber\\
&+&\sum_{mn}\tr[(R-1)C(t)]_{mm}\circ \tr[RC^{\og(j_1)}RC^{\og(j_2)}]_{mm,nn}\circ \tr[(R-1)C(t)]_{nn}\nonumber\\
&+&\sum_{mn}\tr[(R-1)C^{\og(j_1)}]_{mm}\circ \tr[RC(t)RC(t)]_{mm,nn}\circ \tr[(R-1)C^{\og(j_2)}]_{nn}  \  \big]\nonumber\\
&+& \lambda^2\sum_{mn,pq} \big[\ \ [RC^{\og(j_1)}RC^{\og(j_2)}]_{mn,pq} [RC(t)RC(t)]_{qp,nm}\nonumber\\
&+&[RC^{\og(j_1)}RC(t)]_{mn,pq} [RC^{\og(j_2)}RC(t)]_{qp,nm}\ \big]\nonumber\\
&+&\lambda^2\ \tr\ [(R-1)C^{\og(j_1)}]\circ\tr [(R-1)C(t)]\times \tr\ [(R-1)C^{\og(j_2)}]\circ\tr [(R-1)C(t)]\nonumber\\
&-&\lambda \tr[(R-1)C^{\og(j_1)}]\circ \tr[(R-1)C^{\og(j_2)}]\ \big\},\label{second}
\eea
where $[\tr A]_{nn}$ means the partial trace introduced in Section 4 (see Definition \ref{traceprod} ); similarly we can define $\tr A_{mm,nn}:=\sum_k\sum_q A_{mk,km;nq,qm}$. The product $\circ$ is defined in Formula $\eqref{circ}$;
The product $\times$ means that
the terms are not connected (see Graph $G_2(\og_1, \og_2)$ in Figure \ref{gsa}). The terms in the last line of \eqref{second} correspond to Graph
$G_1(\og_1, \og_2)$ in Figure \ref{gsa} while the terms in the previous line correspond to Graph $G_2(\og_1, \og_2)$ in Figure \ref{gsa}. The first four terms correspond to the first four graphs in Figure
\ref{gsb} while the two following two terms correspond to the last two terms in Figure \ref{gsb}.
Here for simplicity we have only shown the graphs such that all the $\sigma$ propagators are hooked to the outer border of the planar graphs. The other half such that all the $\sigma$ fields are hooked to the inner border of the ribbon graph has been omitted. 

\medskip
\noindent{\bf Acknowledgments} The author is very grateful to Prof. Vincent Rivasseau for many useful discussions and Dr. J.T. Zang for a simpler proof of Cayley's theorem for multi-level trees. He is very grateful to the referee for his excellent work, which has made this paper much more readable. Part of this work has been finished during the author's stay at the Shanghai Center for Mathematical Science (SCMS), Fudan university and he author is very grateful to Prof. Xiaoman Chen and Guoliang Yu for their hospitality. The author is partially supported by NSFC(11701121) and research funding from HIT.

\end{document}